\documentclass[a4paper, 10pt, twocolumn, superscriptaddress]{revtex4-2}     
\usepackage{amsmath}
\usepackage{times}
\usepackage{graphicx}
\usepackage{setspace}
\usepackage{txfonts}
\usepackage{pxfonts}
\usepackage{xcolor}
\usepackage{comment}
\usepackage{adjustbox}
\usepackage[toc,page]{appendix}
\usepackage{array}
\newcolumntype{x}[1]{>{\centering\arraybackslash\hspace{0pt}}p{#1}}
\newcolumntype{R}[1]{>{\raggedleft\let\newline\\\arraybackslash\hspace{0pt}}m{#1}}
\newcolumntype{L}[1]{>{\raggedright\let\newline\\\arraybackslash\hspace{0pt}}m{#1}}
\usepackage{ulem}
\usepackage{natbib}
\usepackage{nameref}
\usepackage{soul}

\let\emph\textit

\usepackage{hyperref}
\begin{document}


\title{A robust method for fitting degree distributions of complex networks}
\author{Shane Mannion}
	\affiliation{MACSI, Department of Mathematics and Statistics, University of Limerick, Limerick, V94 T9PX, Ireland}
\author{P\'adraig MacCarron}
	\affiliation{MACSI, Department of Mathematics and Statistics, University of Limerick, Limerick, V94 T9PX, Ireland}

\begin{abstract}
{This work introduces a method for fitting to the degree distributions of complex network datasets, such that the most appropriate distribution from a set of candidate distributions is chosen while maximizing the portion of the distribution to which the model is fit. Current methods for fitting to degree distributions in the literature are inconsistent and often assume \textit{a priori} what distribution the data are drawn from. Much focus is given to fitting to the tail of the distribution, while a large portion of the distribution below the tail is ignored. It is important to account for these low degree nodes, as they play crucial roles in processes such as percolation. Here we address these issues, using maximum likelihood estimators to fit to the entire dataset, or close to it. This methodology is applicable to any network dataset (or discrete empirical dataset), and we test it on over 25 network datasets from a wide range of sources, achieving good fits in all but a few cases. We also demonstrate that numerical maximization of the likelihood performs better than commonly used analytical approximations. In addition, we have made available a Python package which can be used to apply this methodology.}
\end{abstract}

\maketitle

\section{Introduction}
\label{section:intro}
Although complex networks have been studied for around a quarter of a century, there is still a lack of consistency in determining one of their core properties -- namely, the degree distribution. 
One of the early observations about complex networks was that their degree distributions differed from those of the random graph models which had been studied for many years before. Specifically it was observed that complex networks had right-skewed distributions, and Barab\'asi and Albert claim that these right tails often followed power-law distributions~\cite{barabasiscaling}. Their work appears to show that different types of networks have different power-law exponents and these exponents are often in the range $2 \le \gamma \le 3$. 

However, Amaral et al.~\cite{Amaral} argue that power-law distributions are not so common and suggest that for social networks other distributions are sometimes more appropriate. In addition they define three classes of small-world networks: power-law or `scale-free' networks, `broad-scale' networks (power-law distributions with sharp cut-offs), and `single-scale' networks, which have distributions that decay quickly, such as exponential or Gaussian distributions. The distribution of the network determines which of these classes it belongs to. 

More recently, Broido \& Clauset~\cite{broido2019scale} argue that scale-free networks are rare. They introduce a strict procedure for identifying power-laws as well as several weaker regimes. This method, however, tends to be more focused on the tail of the distribution, as even the strongest criteria they outline only requires a power-law to be a good fit to the 50 highest-degree nodes and the method requires comparing to many simulated graphs which are generated under strict conditions. It is more concerned with whether or not evidence can be found to support or rule out a power-law distribution than being able to robustly identify the most appropriate distribution. 

Knowing the distribution of a graph is important as it allows 
us to model and estimate other properties of the network structure. Similarly, it allows us to build general models for the formation of complex networks such as the preferential-attachment model~\cite{barabasiscaling}. The preferential-attachment model is still one of the most popular complex network formation models, though it requires the degree distribution to follow a power law, or at least with slight deviations from it. 

The question of what distribution a given network's degree distribution follows remains a heavily researched topic (see~\cite{Itojapan, broido2019scale}). In addition, some controversy exists around the prevalence of power-law distributions in empirical networks~\cite{broido2019scale,Voitalov,Holmerare}. 

Despite this, current methods for fitting degree distributions are inconsistent. Often still, the method  of least squares is used to determine the parameter of a distribution, despite work highlighting the issues with this approach~\cite{goldstein2004problemsfitting}. The method of least squares has several drawbacks, for example, in the case of a power-law distribution, one would plot the distribution on a log-log scale and use least squares to determine the slope of the relatively straight line through the data. This assumes \textit{a priori} that the data is a power-law however, and data appearing as a relatively straight line on a log-log scale is not a sufficient condition to rule out other possibilities. Additionally, it is not applicable to data that is less easy to linearize. 

A better method is described in the work of Clauset \textit{et al.}~\cite{Clauset2009}. This uses the method of maximum likelihood to estimate the parameters of the distribution. This method is intended for general empirical datasets. A problem with using it to specifically fit degree distributions is that it focuses on fitting to the tail of the distribution, and in doing so can ignore huge amounts of the data available, as we will discuss in Section~\ref{section:powerlaw}. Nodes with a low degree are important for  percolation processes~\cite{newman2001random} and community identity in social networks~\cite{mehrabi2019community}, choosing a high degree cut-off will miss these. 

In this paper we 
build on the method of Clauset \textit{et al.}~\cite{Clauset2009} in order
to identify the most appropriate distribution for the degree distribution of a network from a set of candidate distributions and apply this methodology to a variety of network datasets. We do this while taking minimal approximations (for example using a minimum value of $k_{\rm min}$), which are discussed in more detail throughout this paper. The methods are applicable to any network dataset and do not rely on choosing things such as $k_{\rm min}$ or the distribution by eye.

The rest of this paper proceeds as follows. In the next section we cover important background information concerning some basics of network science. In Section~\ref{section:powerlaw} we cover power-law distributions, methods of estimating power-law exponents and maximum likelihood estimators. Section~\ref{sec:comparison} compares the estimates for the power-law exponent covered in the previous section. In Section~\ref{section:other} we introduce the distributions which we use in our methodology before discussing model selection and goodness of fit in Section~\ref{section:modelselection}. Section~\ref{section:methodology} describes in detail exactly how our methodology works. An in-depth analysis of the results we obtain and discussion around them are found in Sections~\ref{section:results} and~\ref{discussion}.

A Python package for the implementation of these methods is included in Appendix~\ref{appendix:A}.

\section{Background}
\label{section:bg}
A network is defined as an ordered set of $N$ nodes and $M$ edges between them. In this work we are concerned with unweighted and undirected networks, however the methods described can be easily applied to weighted degree distributions and directed networks. Nodes and edges can represent many things in different networks. For example, nodes may represent people in a social network and the edges represent friendship ties between them.

The \textit{degree} of a node is the number of edges connected to it. The distribution of the degrees in a network gives some important characteristics related to its structure. The \textit{degree distribution} $p_k$ is the fraction of vertices in a network with degree $k$,
\begin{equation}
p_k = \frac{1}{N}\sum_{i=1}^N\delta_{k_i k}, \label{eqn:pk} 
\end{equation}
where $k_i$ is the degree of node $i$ and $\delta_{k_i, k}$ is the Kronecker delta function which is one if $k_i = k$ and zero otherwise. We are assuming there are no nodes of degree zero in the network. We use $p_k$ to denote probability mass functions (for discrete distributions) and $p(k)$ to denote probability density functions (for continuous distributions).

For complex networks, degree distributions are often found to have a positive or right skew~\cite{Newmanstructure}, which is also referred to as being heavy-tailed. These heavy tails are frequently due to a small number of vertices having very large degrees. Such highly connected nodes are often referred to as \emph{hubs}. In contrast, Erd\H{o}s-R\'enyi graphs tend to lack hubs due to having a small standard deviation in their degree when compared to complex networks. 

\begin{figure*}[t!]
\begin{center}
\includegraphics[width=0.5\textwidth]{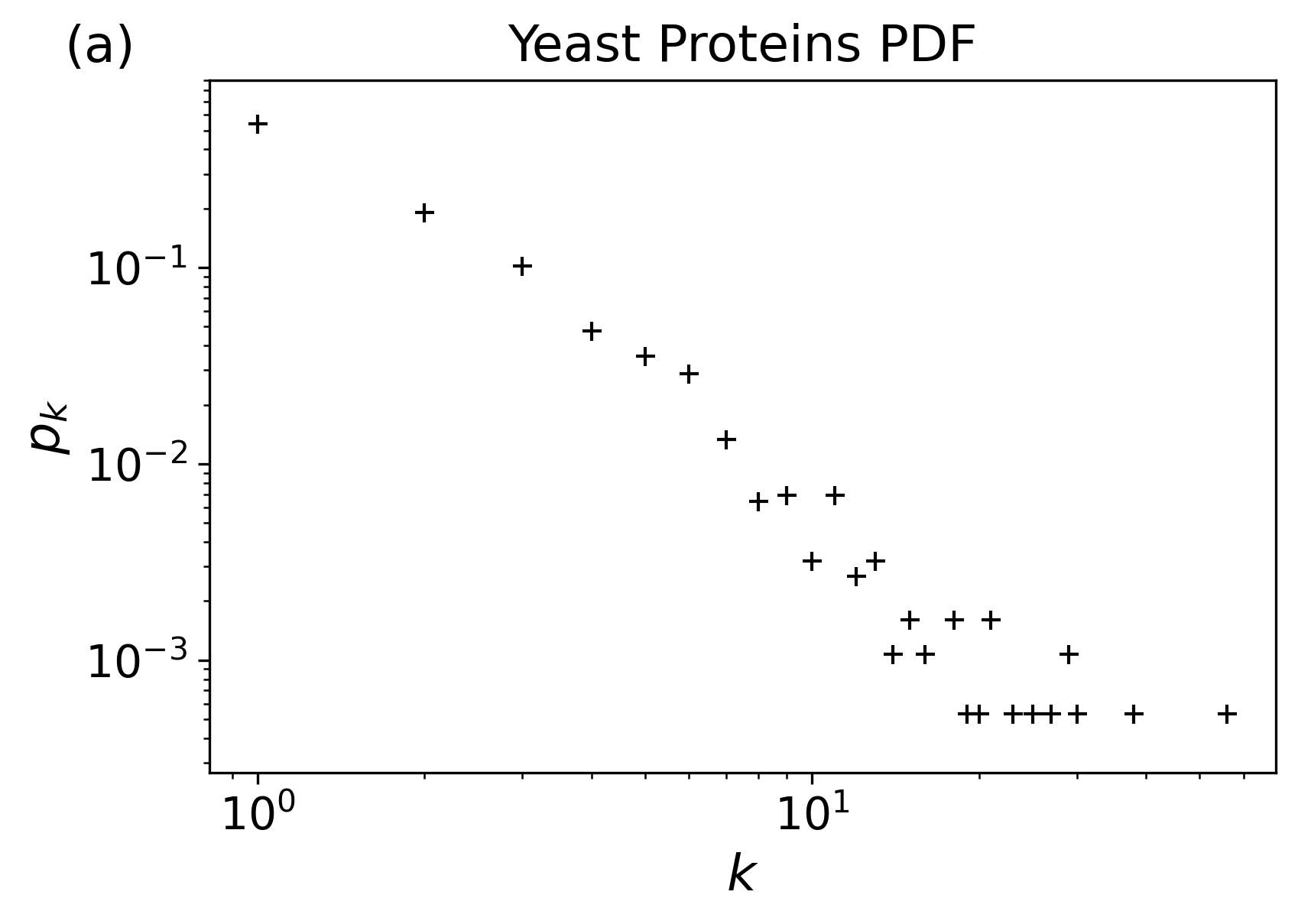}\includegraphics[width=0.5\textwidth]{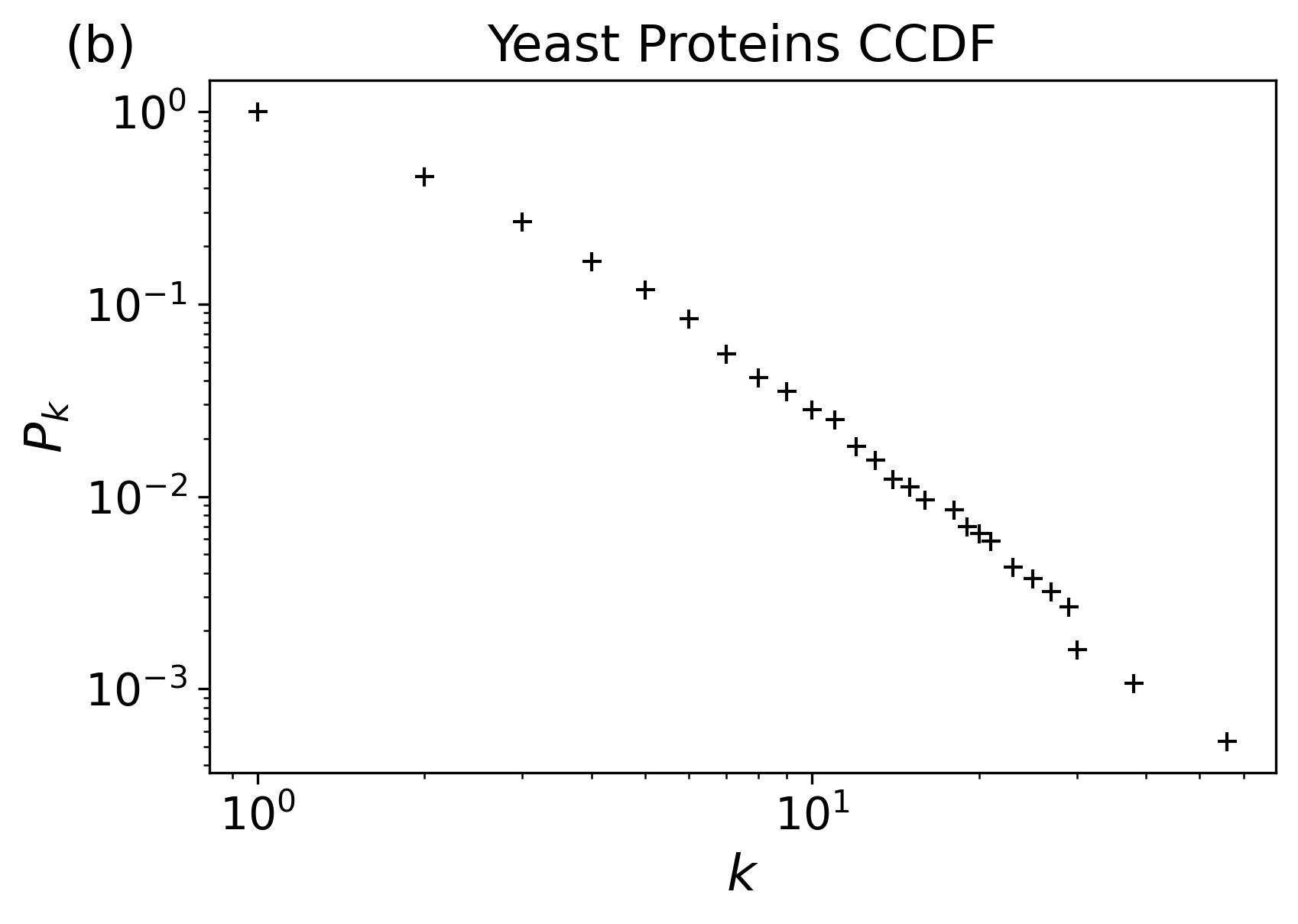} 
\caption{Degree distribution for yeast protein interactions on a log-log scale (a) and the complementary cumulative degree distribution, again on a log-log scale.}
\label{fig:yeast}
\end{center}
\end{figure*}

In Figure~\ref{fig:yeast} we display the degree distribution for protein interactions in yeast, data taken from~\cite{Jeongprotein}. The distribution has a noisy tail due to many instances of only one vertex with a specific (large) degree. To reduce the noise in the tail we often consider instead the complementary cumulative distribution function (CCDF) for the dataset~\cite{Newmanstructure}. This is the probability that a node has a degree greater than or equal to $k$ and is denoted $P_k$:
\begin{equation}
P_k = \sum_{j = 0}^{\infty} p_{k+j} = \sum_{q=k}^{\infty} p_q.\label{eqn:cumulativepk}
\end{equation}

\section{Power-Law distributions}
\label{section:powerlaw}
One of the most common and thoroughly studied distributions observed in complex networks is the power-law distribution~\cite{Strogatz2001,AlbertBarabasi2002}. A power-law distribution has the form
\begin{equation}
p_k \sim k^{-\gamma}, \label{eqn:powerlaw}
\end{equation}
where the exponent of the power-law, $\gamma \ge 1$. This is the most basic definition of a power-law, though the criteria for what is classified as a power-law is a much debated topic in network science (see ~\cite{broido2019scale} for a summary of different definitions). As discussed in Section~\ref{section:intro}, this behaviour is usually found only in the tail of the distribution beginning at some value $k_{\rm min}$. If we take logarithm of Equation~(\ref{eqn:powerlaw}),
\begin{equation}
\log p_k \sim \log{k^{-\gamma} \sim -\gamma \log{k}}, \label{eqn:linearisedpl}
\end{equation}
we can see that the relationship between the log of $p_k$ and the log of $k$ is linear. Hence, when a power-law is displayed on a log-log scale, it will appear linear.

Therefore, one way in which we can obtain an estimate for the exponent $\gamma$ is by performing a least squares regression on Equation~(\ref{eqn:linearisedpl}). In this section we look at this and other methods of estimating the power-law exponents. First, we must normalize Equation~(\ref{eqn:powerlaw}). We start with
\begin{equation}    
\sum_{k=1}^{k_{\rm min} - 1} \rho_k + \sum_{k=k_{\rm min}}^{k_{\rm max}} p_k = 1, \label{eqn:rhokandpk}
\end{equation}
where $\rho_k$ is the distribution below $k_{\rm min}$. Our equation therefore becomes
\begin{equation}
    p_k = \frac{(1 - \Delta)}{\zeta(\gamma, k_{\rm min})} k^{-\gamma},
    \label{eqn:normalisedpl}
\end{equation}
where 
$$
\Delta \equiv \sum_{k=0}^{k_{\rm min} - 1}\rho_k \;\;{\rm and } \;\;\zeta(\gamma, k_{\rm min}) \equiv \sum_{n=0}^{\infty}(k_{\rm min} + n)^{-\gamma}.
$$
Here, $\zeta(\gamma, k_{\rm min})$ is the Hurwitz-zeta function. In the case where the entire distribution follows a power-law and $k_{\rm min} = 1$, the denominator is just the Riemann zeta function $\zeta(\gamma)$. The complementary cumulative distribution function for the full degree distribution is given by
\begin{equation}
    P_k = \sum_{q=k}^{\infty}p_q = (1-\Delta)\frac{\zeta(\gamma, k)}
    {\zeta(\gamma, k_{\rm min})}.
    \label{eqn:cumulativepl}
\end{equation}

\subsection{Continuous Approach}
\label{section:continuous}
Often when fitting, the approximation is made that the distribution is continuous instead of made up of discrete integers. We now consider the continuous regime to introduce an approximation for the exponent of a power-law before returning to the discrete case. 
In the continuous case, Equation~(\ref{eqn:powerlaw}) can be normalized by
$$
\int_{0}^{k_{\rm min}}\rho(k)\,dk + \int_{k_{\rm min}}^{\infty}p(k)\,dk = 1,
$$
where $\rho(k)$ is some unknown distribution below $k_{\rm min}$. The normalized distribution is then
\begin{equation}
p(k) = (1-\delta)\frac{\gamma - 1}{k_{\rm min}}\left(\frac{k}{k_{\rm min}}\right)^{-\gamma}, \label{eqn:continuouspl}
\end{equation}
where $\delta \equiv \int_{0}^{k_{\rm min}}\rho(k)\, dk$ is treated as constant. We can then integrate Equation~(\ref{eqn:continuouspl}) to determine the cumulative distribution function,
$$
P(k) = \int_{k_{\rm min}}^{\infty} p(q)\,dq = (1 - \delta)\left(\frac{k}{k_{\rm min}}\right)^{-\gamma + 1}.
$$
We can see from this that the cumulative distribution function also follows a power-law but with an exponent differing by one  to the original. As with Equation~(\ref{eqn:linearisedpl}), a least squares fit to the cumulative distribution can be used to determine an estimate for the exponent. This estimate will usually provide a better fit than a fit to the probability density function as the cumulative distribution is less noisy in the tail~\cite{Newmanstructure}. 

\subsection{Moments of The Distribution}
\label{section:moments}
If we make the assumption that the distribution can be used for the entire data (i.e., set $k_{\rm min} = 1$) and thus $\delta \rightarrow 0$, the probability density function Equation~(\ref{eqn:continuouspl}) is 
\begin{equation}
p(k) = (\gamma - 1)k^{-\gamma}. \label{eqn:contpldelta0}
\end{equation}
Since the probability that $k = 1$ is bounded by unity, we must have $\gamma < 2$. The first moment of Equation~(\ref{eqn:contpldelta0}) is
\begin{equation}
\langle k \rangle = (\gamma - 1)\frac{1}{2-\gamma}\left[k^{2 - \gamma}\right]^{\infty}_1. \label{eqn:plfirstmoment}    
\end{equation}
If $\gamma \le 2$, then $\langle k \rangle$ will diverge when we evaluate this integral (for the discrete case it will be a large finite value). As a result, we should use $k_{\rm min} > 1$ to remain consistent with $k=1$ implying $\gamma < 2$.
Empirically in complex networks we often observe that the mean degree is often low compared to the size of the network. Hence, choosing a high value of $k_{\rm min}$ could lead to the distribution only being fitted to a small number of nodes. Indeed, in the case of 1,000 simulated true power-law distributions 
with $\gamma=2.5$, 918 of these have an average degree less than three. 

It may be also worth considering the using the modal degree as the value for $k_{\rm min}$ as the distribution likely decays beyond this point. Clauset \textit{et. al}~\cite{Clauset2009} provide various methods for estimating $k_{\rm min}$, however, they assume large values of $N$, and so these methods may not be applicable to smaller datasets. We aim to provide a methodology to fit to datasets of any size. Methods for choosing $k_{\rm min}$ will be discussed in section \ref{section:modelselection}.
For now we continue in the continuous regime. If $\gamma > 2$ in Equation~(\ref{eqn:plfirstmoment}) then we can obtain an estimate for $\gamma$ from the mean,
\begin{equation}
    \langle k \rangle = \frac{\gamma - 1}{\gamma - 2} \implies \gamma = \frac{2\langle k \rangle - 1}{\langle k \rangle - 1}. \label{eqn:plsecondmoment}   
\end{equation}
We will use this as one of the estimates for the exponent.
The same procedure can be repeated for the
second moment and we obtain
$$
\langle k^2 \rangle = \int_{K_{\rm min}}^{N-1}k^2p(k)\,dk \sim \frac{1}{3-\gamma}\left[k^{3 - \gamma}\right]_{k_{\rm min}}^{N}.
$$
As $N$ goes to infinity, one observes that the second moment diverges when $\gamma \le 3$. Therefore, if the degree distributions follow a power-law with $\gamma \le 3$, they would be expected to have a large value of $\langle k^2 \rangle$.
\subsection{Maximum Likelihood Estimators}
\label{section:mle}

From the continuous regime, we have three estimates for the exponent, 1) a least squares fit to the log-log probability mass function, 2) a least squares fit to the log-log complementary cumulative distribution function (exponent will be different by one) and 3) an estimate from the first moment of the degree distribution.
We now turn our attention to the main focus of this paper which is using maximum likelihood estimators to obtain estimates for distribution parameters. 

In general, the likelihood $\cal L$ is the probability that an independent and identically distributed dataset of $N$ observations were drawn from a model $p_k$ with parameter $\theta$,
\begin{equation}
    {\cal L} (\theta \vert k) = \prod_{i}^{N}p_\theta(k_i). \label{eqn:generallnl}
\end{equation}
 
Here $k_i$ is the degree of node $i$ and $N$ is the number of nodes in the network. As mentioned above, the  complementary cumulative distribution in Equation~(\ref{eqn:cumulativepk}) is often used to reduce noise in the tail of the probability distribution. Fits are then often made to the cumulative distribution $P_k$ rather than the original distribution $p_k$.
It has been suggested that applying the method of least squares to cumulative distributions is unsuitable for empirical data~\cite{Clauset2009,Edwards2007}. This is in part due to the data being discrete and therefore the continuous approach to the complementary cumulative distribution may not be appropriate (particularly for small datasets). Network data will always contain dependencies between the observations, violating one of the assumptions underlying maximum likelihood estimation~\cite{Gerlachtesting}. Despite this, MLE still shows good results as we see below and shown in Ref.~\cite{broido2019scale}.

Furthermore, the estimates for the parameters obtained using a least squares fit can be inaccurate, and the error in the parameter value can be difficult to estimate~\cite{Clauset2009}. Instead, maximum likelihood estimators (MLEs) are said to provide better measurements for the parameters as no estimator has lower asymptotic error in the large sample size as the MLE ~\cite{Clauset2009}.

Sticking with the continuous regime for now, the likelihood corresponding to the power-law from Equation~(\ref{eqn:continuouspl}) is
\begin{equation}
    {\cal L}(\gamma \vert k) = \prod_{i = 1}^{N}(1 - \Delta)\frac{\gamma - 1}{k_{\rm min}}\left(\frac{k_i}{k_{\rm min}}\right)^{-\gamma}. \label{eqn:contpllnl}
\end{equation}
Taking the logarithm we obtain
\begin{equation*}
    \ln{\cal L} = N\ln(1-\Delta) + N\ln(\gamma - 1) - N\ln{k_{\rm min}} - \gamma \sum_{i=1}^{N} \ln{\frac{k_i}{K_{\rm min}}}.
\end{equation*}
We can obtain a value for $\Delta$ from the data. This does not affect the maximisation, however, it is important to have an estimate for $\Delta$ when displaying the fit to the distribution. Typically, $\Delta$ or $\sum_{k=1}^{k_{\rm min} - 1} \rho_{k}$ as in Equation~(\ref{eqn:rhokandpk}), is ignored in works such as ref.~\cite{Clauset2009} or any other methodology that focuses on the tail of the distribution. Here we aim to account for it as fully as possible, though it does not affect the maximization.

We set $\partial(\ln{\cal L}/\partial \gamma) = 0$ and solve analytically for an estimate for the power-law exponent. We obtain
\begin{equation}
    \hat{\gamma} = 1 + N\left[\sum_{i = 1}^{N}\ln\frac{k_i}{k_{\rm min}}\right]^{-1},
    \label{eqn:gammahat} 
\end{equation}
where $\hat{\gamma}$ denotes an estimate from the data rather than the true value. If a power-law is a good model for the data then $\hat{\gamma} \approx \gamma$.
Our tests show however, that this estimate performs relatively poorly and this will be discussed in the next section (see Tables \ref{tab:1}, \ref{tab:comparisontablen100}, \ref{tab:comparisontablen10000}).

In addition, complex network data, even when relatively large, is rarely a complete set of integers and there may be large gaps between consecutive degrees. 
It is therefore much more accurate to look at the discrete case from Equation~(\ref{eqn:normalisedpl}). Taking the log likelihood of this we get
\begin{equation}
    \ln{\cal{L}} = N\,\ln(1 - \Delta) - N\,\ln{\zeta(\gamma, k_{\rm min})} - \gamma\,\sum_{i=1}^{N}\ln{k_i}.
    \label{eqn:pllnl}
\end{equation}
This can be numerically maximized to obtain an estimate for $\gamma$.
Additionally, Clauset \textit{et al.}~\cite{Clauset2009} find an approximation for the log likelihood in Equation~(\ref{eqn:pllnl}) as
\begin{equation}
    \hat{\gamma} = 1 + N\left[\sum_{i=1}^{N}\ln{\frac{k_i}{k_{\rm min} - \frac{1}{2}}}\right]^{-1}. \label{eqn:clausetapprox}
\end{equation}
However, the convenience of this estimate comes at the cost of accuracy for smaller networks, particularly with small $k_{\rm min}$ as we show later.
We compare all of these estimates in Section~\ref{sec:comparison}.

\section{Comparison of Estimates}
\label{sec:comparison}
In the previous section we outlined four ways of obtaining estimates for the power-law estimate $\gamma$. In this section we will compare these four estimates along with estimates obtained by performing least squares on the log-transformed PDF and CCDF of the data.

We denote as $\gamma_{\rm c}$ the estimate obtained from the continuous version of the distribution as in Equation~(\ref{eqn:gammahat}), $\gamma_{\rm mle}$ as the maximisation of the discrete discrete log likelihood in Equation~(\ref{eqn:pllnl}), and $\gamma_{\rm d}$ as the discrete approximation from Equation~(\ref{eqn:clausetapprox}).

Additionally we define $\gamma_{\rm pdf}$ to be the value obtained by performing least squares regression on log-log plot of the probability density function, and $\gamma_{\rm ccdf}$ to be the value obtained by performing least squares regression on the log-log plot of the complementary cumulative distribution. Lastly, $\gamma_{\langle k \rangle}$ is the estimate calculated from the first moment of the distribution as described by Equation~(\ref{eqn:plsecondmoment}).

For this we generated synthetic power-law distributions of varying sizes, all with a parameter $\gamma = 2.5$. We generate these using the NumPy package's built-in random number generators. For each value of $N$, we generated 1,000 distributions and calculated each estimate. The average estimates as well as their standard distributions for $N = 1,000 $ are shown below in Table~\ref{tab:1}. Additional tables for graphs of size $N=100$ and $N=10,000$ can be found in Appendix~\ref{appendix:B}. We see that in this case, with small $k_{\rm min}$, the numerical maximisation $\gamma_{\rm mle}$ performs best when compared to other estimates. 

In Ref.~\cite{Clauset2009}, the authors say that the continuous estimate, $\gamma_{\rm c}$ is less accurate than the discrete estimate and no easier to calculate and so it should not be used. Our results here agree with this statement as we can see the continuous estimate is inaccurate even at large values of $k_{\rm min}$. Instead, they suggest using the discrete approximation, however, we observe here that the numerically maximized discrete MLE performs better than the approximation.

As we can see here the least squares fit to the CCDF, $\gamma_{\rm ccdf}$, is highly consistent under variation of $k_{\rm min}$, however, Clauset et al.~\cite{Clauset2009} also point out flaws with this method, such as the underlying assumptions being invalid. The fit to the PDF, $\gamma_{\rm pdf}$, is always less accurate than that to the CCDF, on account of the PDF having a noisier tail~\cite{newman2005pareto}. In addition, as mentioned in Sections~\ref{section:intro} and~\ref{section:mle}, there are drawbacks and limitations to applying the method of least squares to cumulative distributions.

The discrete estimate, $\gamma_{\rm d}$, that Clauset \textit{et al.}~\cite{Clauset2009} recommend using is accurate at $k_{\rm min} = 3$ and above, however, at this point already over 90\% of the points will be missed. In Ref.~\cite{Clauset2009}, it is stated that this method works well only for $k_{\rm min} \gtrapprox 6$. With this we would run into a more extreme version of this problem, and our tests show that it begins to become less accurate at this point.

The method calculated using the first moment of the distribution, $\gamma_{\langle k \rangle}$ performs poorly for all values of $k_{\rm min}$ and we recommend that it should not be used. 

We can see from Table~\ref{tab:1} that in the case of small $k_{\rm min}$, the numerical maximisation of the log likelihood $\gamma_{d}$ performs best of all the estimates compared here. This is not surprising, given that all other methods are approximations, however, we believe the increased accuracy of maximum likelihood estimators justifies the relative increase in difficulty of their implementation. There are several other advantages that MLEs have over the other estimates as well, as mentioned in Section~\ref{section:mle}.

Furthermore, none of the other estimators here are applicable to any distribution in the way that MLEs are, as the estimates $\gamma_{\rm c}$ and $\gamma_{\rm d}$ are unique to power-laws, and least squares regression is only applicable if the data can be easily linearized. The method of moments can be applied to some other distributions (e.g. exponential), however, its performance is still poor. Finally, all of these other estimates assume the distribution \textit{a priori}, whereas MLEs allow for easy comparison of multiple distributions using information criteria. For these reasons, we argue that maximum likelihood estimators should be the standard approach for fitting distributions to network data.

Indeed, this is a clarification needed when discussing the work of Clauset et al.~\cite{Clauset2009}; the methodology they discuss is applied to empirical data, but not necessarily network distributional data, which are often small and incomplete, especially in the case of poor sampling. It should be noted as well that the above paper is more concerned with determining if a power-law distribution is an appropriate fit for data than it is with testing many distributions to find the most appropriate fit. In the succeeding sections we will show all of the common distributions that one should consider when fitting a distribution to network data as well as equations for the log likelihood to be maximized, and some of the pitfalls that should be avoided in plotting these distributions against empirical data to ensure correct fitting.

\begin{table*}[t]
\renewcommand{\arraystretch}{1.2}

\caption{Values for each power-law exponent estimate discussed in section~\ref{sec:comparison} obtained on 1,000 datasets each of size $N=1,000$ with true parameter value $\gamma = 2.5$. Values shown are means with standard deviations in brackets. $\hat{\gamma}_c$ is the continuous estimator for the exponent from eq.~(\ref{eqn:gammahat}), $\gamma_{\rm pdf}$ 
and $\gamma_{\rm ccdf}$ are the values obtained by least squares regression of the pdf and ccdf respectively. $\gamma_{\rm d}$ is the value obtained by the discrete approximation eq.~\ref{eqn:clausetapprox}, $\gamma_{\langle k \rangle}$ is the value obtained by the method of moments and $\gamma_{\rm mle}$ is the value obtained by maximum likelihood estimation. Best estimates for each value of $k_{\rm min}$ are highlighted in bold.}
\label{tab:1}
\begin{tabular}{|l|x{2.2cm}|x{2.2cm}|x{2.2cm}|x{2.2cm}|x{2.2cm}|x{2.2cm}|}
    \hline
$k_{\rm min}$ & $\gamma_{\rm c}$ & $\gamma_{\rm pdf}$ & $\gamma_{\rm ccdf}$ & $\gamma_{\rm d}$ & $\gamma_{\langle k \rangle}$ & $\gamma_{\rm mle}$ \\
    \hline
     1   & 4.47\,\,(0.23)                                  & 2.57\,\,(0.28)                             &  2.53\,\,(0.23)                              & 2.02\,\,(0.02)                                  &3.15\,\,(0.23)                                        & \textbf{2.50\,\,(0.05)}                             \\ \hline
2                                   & 3.28\,\,(0.21)                                  & 2.34\,\,(0.29)                             & 2.45\,\,(0.26)                              & 2.38\,\,(0.07)                                  & 2.29\,\,(0.06)                                        & \textbf{2.50\,\,(0.10)}                              \\ \hline
3                                   & 3.00\,\,(0.24)                                   & 2.18\,\,(0.30)                              & 2.43\,\,(0.30)                               & 2.46\,\,(0.13)                                  & 2.17\,\,(0.04)                                        & \textbf{2.52\,\,(0.14)}                            \\ \hline
4                                   & 2.88\,\,(0.29)                                  & 2.04\,\,(0.32)                             & 2.43\,\,(0.34)                              & \textbf{2.50\,\,(0.18)}                                   & 2.12\,\,(0.03)                                        & 2.53\,\,(0.19)                            \\ \hline
5                                   & 2.81\,\,(0.32)                                  & 1.89\,\,(0.30)                              & 2.41\,\,(0.36)                              & \textbf{2.51\,\,(0.22)}                                  & 2.09\,\,(0.02)                                        & 2.53\,\,(0.23)                            \\ \hline
6                                   & 2.77\,\,(0.35)                                  & 1.74\,\,(0.28)                             & 2.37\,\,(0.38)                              & \textbf{2.53\,\,(0.26)}                                  & 2.07\,\,(0.02)                                        & 2.54\,\,(0.26)                            \\ \hline
7                                   & 2.75\,\,(0.38)                                  & 1.62\,\,(0.25)                             & 2.37\,\,(0.39)                              & \textbf{2.54\,\,(0.29)}                                  & 2.06\,\,(0.02)                                        & 2.56\,\,(0.29)                            \\ \hline
8                                   & 2.74\,\,(0.42)                                  & 1.52\,\,(0.25)                             & 2.37\,\,(0.43)                              & \textbf{2.55\,\,(0.33)}                                  & 2.06\,\,(0.02)                                        & 2.56\,\,(0.34)                            \\ \hline
9                                   & 2.73\,\,(0.48)                                  & 1.42\,\,(0.27)                             & 2.35\,\,(0.48)                              & 2.57\,\,(0.38)                                  & 2.05\,\,(0.02)                                        & \textbf{2.57\,\,(0.37)}                            \\ \hline
10                                  & 2.76\,\,(0.54)                                  & 1.37\,\,(0.27)                             & 2.37\,\,(0.50)                               & 2.60\,\,(0.44)                                   & 2.05\,\,(0.01)                                        & \textbf{2.60\,\,(0.42)}                             \\ \hline
\end{tabular}
\end{table*}

\section{Other Distributions}
\label{section:other}
Additionally we looked at the following distributions; (a) truncated power-law distributions, (b) exponential distributions, (c) stretched exponential distributions, (d) Weibull distributions, (e) Poisson distributions, (f) normal (Gaussian) distributions and (g) lognormal distributions. Most of these distributions are for continuous variables, but here we discretize and normalize them in order to obtain their corresponding log likelihoods. We choose these as they are common in formation processes and the complex network literature, e.g.~\cite{Amaral, Clauset2009, ebel2002dynamics}.

\subsection{Truncated Power-Law Distributions}
The second class of network described by Amaral \textit{et al.}~\cite{Amaral} contains a power-law regime followed by a sharp cut-off. Albert \& Barab{\'a}si~\cite{AlbertBarabasi2002} provide a list of exponents and cut-offs for a range of complex networks.
A truncated power-law, or a power-law with exponential cut-off has the form
\begin{equation}
    p_k \sim k^{-\gamma}{\rm e}^{-k/\kappa}. \label{eqn:truncpl}
\end{equation}
A truncated power-law is shown for the collaboration network of the condensed matter arXiv from 1995-1999 in Ref.~\cite{Newmanscientific}. The log likelihood then is
\begin{equation}
    \ln{\cal L} = N\ln(1-\Delta) + \frac{Nk_{\rm min}}{\kappa} 
    - N\ln{Z(k_{\rm min})} - \sum_{i=1}^{N}\left(\gamma\ln{k_i} + \frac{k_i}{\kappa}\right), \label{eqn:truncpllnl}
\end{equation}
where $Z(k_{\rm min}) \equiv \sum_{m = 0}^{\infty} (x + m)^{-\gamma}{\rm e}^{-m/\kappa}$. We then maximize this numerically.
Details of obtaining Equation~(\ref{eqn:truncpllnl}) from Equation~(\ref{eqn:truncpl}) are shown in Appendix~\ref{appendix:A}.
\subsection{Exponential and Weibull Distributions}
The final class of network described by Amaral \textit{et al.} are so-called single-scale networks characterized by fast-decaying tails. Although they are rarely used, these classifications provide a useful ordering by complexity of the different types of distributions.

We first look at the exponential distribution which is given by
\begin{equation}
p_k \sim {\rm e}^{-k/\kappa}. \label{eqn:exponential} 
\end{equation}
Its log likelihood is 
\begin{equation}    
\ln{\cal L} = N\ln(1-\Delta)
+ N\,\ln\left(1 - {\rm e}^{-1/\kappa}\right)
-\frac{1}{\kappa}\,\sum_{i=1}^{N}(k_i - k_{\rm min}). \label{eqn:exponentiallnl}
\end{equation}
The exponential distribution can be generalized to a stretched exponential distribution which has been found to be a good fit for some empirical datasets, such as the jazz band network in Ref.~\cite{gleiserjazz}, however, the stretched exponential \& Weibull distributions are 
related with
the stretched exponential being the CCDF of the Weibull distribution. 
Hence we only include the Weibull here.

The Weibull distribution is given by
\begin{equation}
    p_k \sim \left(\frac{k}{\kappa}\right)^{\beta - 1}{\rm e}^{(k/\kappa)^\beta}. \label{eqn:weibull}
\end{equation}
As we can see when $\beta = 1$, the Weibull simplifies to the exponential distribution. When $\beta = 2$, it is closely related to the normal distribution (below). Its log likelihood is
\begin{multline}
    \ln{\cal L} = N\ln(1 - \Delta) - 
    N \ln\left(\sum_{m=k_{\rm min}}^{\infty}\left(\frac{m}{\kappa}\right)^{\beta - 1}{\rm e}^{(-m/\kappa)^\beta}\right) \\
    + \sum_{i = 1}^N\ln\left((k_i/\kappa)^{\beta - 1}\right)
    + \sum_{i=1}^{N}\left(\frac{k_i}{\kappa}\right)^\beta. \label{eqn:weibulllnl}
\end{multline}
Details of this calculation can once again be found in Appendix~\ref{appendix:A}.
\subsection{Normal \& Lognormal Distributions}
The normal distribution (or Gaussian) is given by
\begin{equation}
    p_k \sim {\rm e}^{-\frac{(k-\mu)^2}{2\sigma^2}},
\end{equation}
where $\mu$ is the mean and $\sigma^2$ is the variance. This distribution is included here as it is found to fit some social networks \cite{Amaral}, however, given the skewed nature of network degree sequences it is unlikely to be chosen in most cases. The log likelihood is given by
\begin{equation}
    \ln {\cal L} = N\ln(1-\Delta) - N\ln \left(\sum_{m=k_{\rm{min}}}^{\infty} {\rm e}^{-\frac{(m-\mu)^2}{2\sigma^2}} \right) - \sum_{i=1}^{N} \frac{(k_i-\mu)^2}{2\sigma^2}.
\end{equation}
The lognormal distribution then is given by 
\begin{equation}
	p_k \sim \frac{1}{k} {\rm e}^{-\frac{(\ln k-\mu)^2}{2\sigma^2}}. \label{eqn:logn}
\end{equation} 
The log likelihood is given by
\begin{multline}
    \ln{\cal L} = N\ln(1 - \Delta) - N\ln\left(\sum_{m=k_{\rm min}}^{\infty}\frac{1}{m}{\rm e}^{-\frac{(\ln(m) - \mu)^2}{2\sigma^2}}\right) \\ - \sum_{k=k_{\rm min}}^{N}\ln(k_i) - \sum_{k=k_{\rm min}}^N\frac{(\ln(k_i) - \mu)^2}{2\sigma^2}. \label{eqn:lognlnl} 
\end{multline}    
\subsection{Poisson Distribution}
The last distribution we test for these network datasets is the Poisson distribution. It is given by
\begin{equation}
    p_k = \frac{\lambda^k}{k!}{\rm e}^{-\lambda}. \label{eqn:poisson}
\end{equation}
The log likelihood then is 
\begin{multline}    
    \ln{\cal L} = N\ln(1 - \Delta) - \ln\left(1 - {\rm e}^{-\lambda}\sum_{m = 0}^{k_{\rm min} - 1} \frac{\lambda^m}{m!}\right)\\ 
    -N\lambda + \ln{\lambda}\sum_{i=1}^N k_i - \sum_{i=1}^N \ln(k_i!). \label{eqn:poissonlnl}
\end{multline}
Again, details of obtaining this equation as well as those for the normal and lognormal distributions can be found in Appendix~\ref{appendix:A}. As with the normal distribution, a Poisson is an unlikely candidate for most network datasets given their skewed nature, however, both are included here for completeness given their ubiquity in other empirical datasets and scientific settings \cite{lyon2014normal, johnson2005poisson}.

\section{Model selection, Choosing $k_{\rm min}$ , and Goodness of Fit}
\label{section:modelselection}
As discussed in Section~\ref{sec:comparison}, there are several advantages to using maximum likelihood estimation. Most importantly is the ability to compare the likelihoods of different distributions in order to choose the best distribution for a given dataset.
In order to compare different models we use the \textit{Akaike information criterion with correction for finite sample sizes} (${\rm AIC_c}$)~\cite{Akaike,Burnham2004} and the \textit{Bayesian information criterion} (BIC)~\cite{Schwarz}. The ${\rm AIC_c}$ is given by
\begin{equation}
    {\rm AIC_c} = -2\ln{\cal L}(\theta | k_i) + 2n_{\theta} 
    + \frac{2n_{\theta}(n_{\theta} + 1)}{N - n_{\theta} - 1}, \label{eqn:AICc}
\end{equation}
and the BIC is given by
\begin{equation}
    {\rm BIC} = -2\ln{\cal L}(\theta | k_i) + n_{\theta}\ln N.
    \label{eqn:BIC}
\end{equation}
In each equation $n_{\theta}$ is the number of parameters in the model, and $N$ is the sample size. This ${\rm AIC_c}$ 
gives greater penalty for number of parameters, and the penalty is diminished with a larger sample size. Anderson \& Burnham~\cite{Burnham2004} recommend always using the ${\rm AIC_c}$ instead of the regular AIC (which is the same as Equation~\ref{eqn:AICc} but without the last term) as in the case of large $N$ the correction tends to zero.

We also make use of AIC weights. For a given set of $m$ models for a given dataset, the AIC weights are given by
\begin{equation}
    W_i = \frac{{\rm e}^{-\Delta_{i}/2}}{\sum_{1}^{m}{{\rm e}^{-\Delta_{i}/2}}},
\end{equation}
where $\Delta_{i} = {\rm AIC_c}_i - {\rm min}({\rm AIC_c})$. These values represent the relative likelihood of the models and allow us to determine if there is support for more than one model. The values of the weights will sum to 1, and the closer a value is to one the more support for that model there is. Usually the weights are accompanied by some cut-off, below which we say a model is not supported.

Since the ${\rm AIC}_c$ and the BIC both depend on $N$, we can only compare two distributions that start at the same value of $k_{\rm min}$. For this reason, the model and $k_{\rm min}$ must be chosen in tandem. It is important to remember that to capture the true distribution of a dataset, it should be a priority that $k_{\rm min}$ be minimized. How exactly we achieve this is discussed in Sections~\ref{section:methodology} and~\ref{section:results}.

Goodness-of-fit tests for this type of analysis leave much to be desired. Two common means of assessing goodness-of-fit are (i) Kolmogorov-Smirnov tests, and (ii) Q-Q plots, both of which have their drawbacks. Descriptions of how these tests work can be found in Refs.~\cite{massey1951kolmogorov} and~\cite{kratz1996qqplot} respectively.

First, with network datasets, we are often dealing with sample sizes of the order of 10,000 nodes or more. These sample sizes are in contrast to the work done by Goldstein \textit{et al.}~\cite{goldstein2004problemsfitting}, who find success using the KS test on sample sizes of the order of 1,000 nodes. The critical value for a KS test at a 5\% level of significance in the limit of large $n$ is given by $1.36/\sqrt{n}$, where $n$ is the sample size or in our case the number of nodes. Our tests find that the KS test is simply too restrictive, and will tell the user that even the best of our fits are poor.

Quantile-quantile or Q-Q plots (shown in Figure~\ref{fig:qqplots}) are useful in many cases but since they require visual interpretation, they are not suitable in a situation where automation is desired. Furthermore they are sensitive to outliers or extreme values, which are common across all of our datasets.
\section{Fitting}
\label{section:methodology}
Here we outline the methodology used to choose distributions for datasets based on the methods discussed already. The datasets to which this methodology was applied can be found in Table~\ref{tab:networkdata}. For a given degree sequence, the methodology works as follows:

(i)	Beginning with $k_{\rm min} = 1,$ we find the parameter values that maximize the log likelihood functions of each of the distributions discussed above.

(ii) The ${\rm AIC}_c$ and AIC weights for each distribution are calculated. The user may choose to use the ${\rm AIC}_c$ or the BIC to choose the best distribution or both.

(iii)	We increase $k_{\rm min}$ by one and repeat steps (i) and (ii). Once  the same distribution is chosen for three consecutive $k_{\rm min}$ values we choose this distribution for the dataset. In the case of small datasets (here chosen to  be fewer than 2,500 nodes), the same distribution need only be chosen for two consecutive $k_{\rm min}$ values. The value 2,500 was chosen based on our datasets; below this value, requiring the same distribution to be chosen three times simply resulted in cutting off too much of the data below $k_{\rm min}$.

(iv) We use the determined distribution and parameters to fit a curve to the CCDF of the degree sequence for visual purposes. At this point, both the CCDF and the PDF should be inspected, as we find that looking at just the CCDF is insufficient to determine if we have obtained a good fit or not. Other goodness-of-fit metrics such as a Q-Q plots can be used. If one determines that the fit obtained is poor, then $k_{\rm min}$ should be increased and the algorithm repeated. We find that we can best determine the degree distribution when using all available information. We will explore this further in the results section.

(v) We then bootstrap the dataset 1,000 times and fit the same distribution to our bootstrapped samples in order to obtain means, standard deviations, and other desired summary statistics for the parameters of our fitted distributions. These summary statistics are shown in Table~\ref{tab:networksummary}.

When fitting we must account for data below $k_{\rm  min}$. This is done using our empirical CCDF. If, for example, we have $k_{\rm min} = 3$, then we multiply the values in our fitted CCDF by $\sum_{k=0}^{3} p(k) \equiv P(3)$. This ensures that the fitted curve begins at the correct point.

There are some limitations for what can be done given a network dataset. Many networks are poorly sampled or not a complete sample. Consider the Enron email dataset~\cite{enronsource}, here emails between employees were recorded but not emails to external addresses, i.e., contacts outside of the organisation would not add to an individual’s degree. Therefore, we do not have the actual degrees of nodes, and so conclusions from the dataset about the degree distribution cannot be extended to the network as a whole. In this case we may just want to fit to the tail and not the full distribution so a high value of $k_{\rm{min}}$ may be appropriate.

In contrast to this is the power-grid network studied in Ref.~\cite{wattsstrogatz}. Here, the degree is the number of physical connections to a generator, a transformer, or a substation and so the degree of a node is correct to the underlying network. In this case, a high $k_{\rm{min}}$ may result in a different distribution being chosen for the tail and not be representative of the full network.

Keeping these limitations in mind, this methodology can be applied to any distribution of discrete integers. In the following section we will see how it performs on a wide variety of network types.

\section{Results}
\label{section:results}
\begin{figure*}[t!]
\begin{center}
\includegraphics[width=0.32\textwidth]{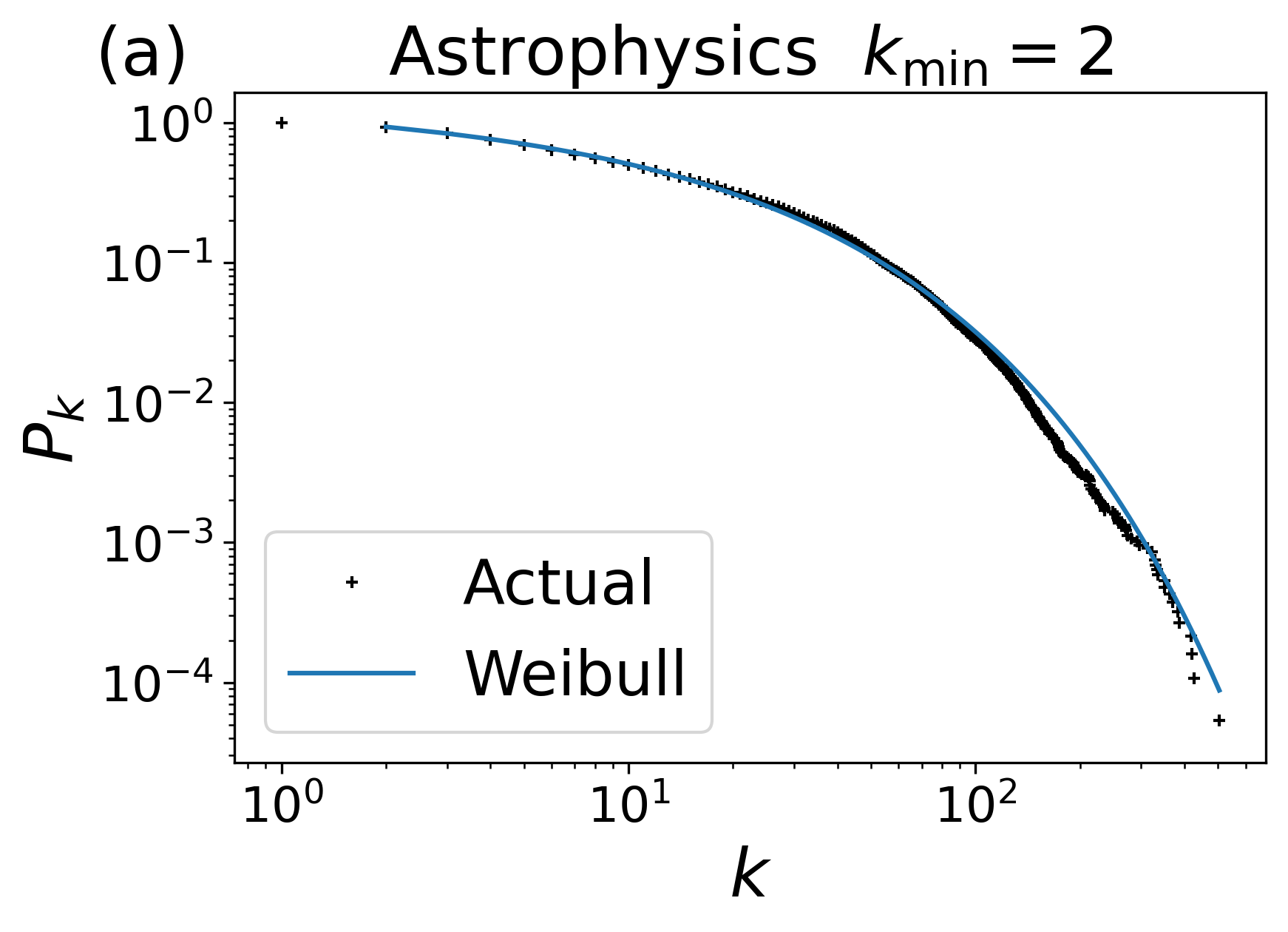}
\includegraphics[width=0.32\textwidth]{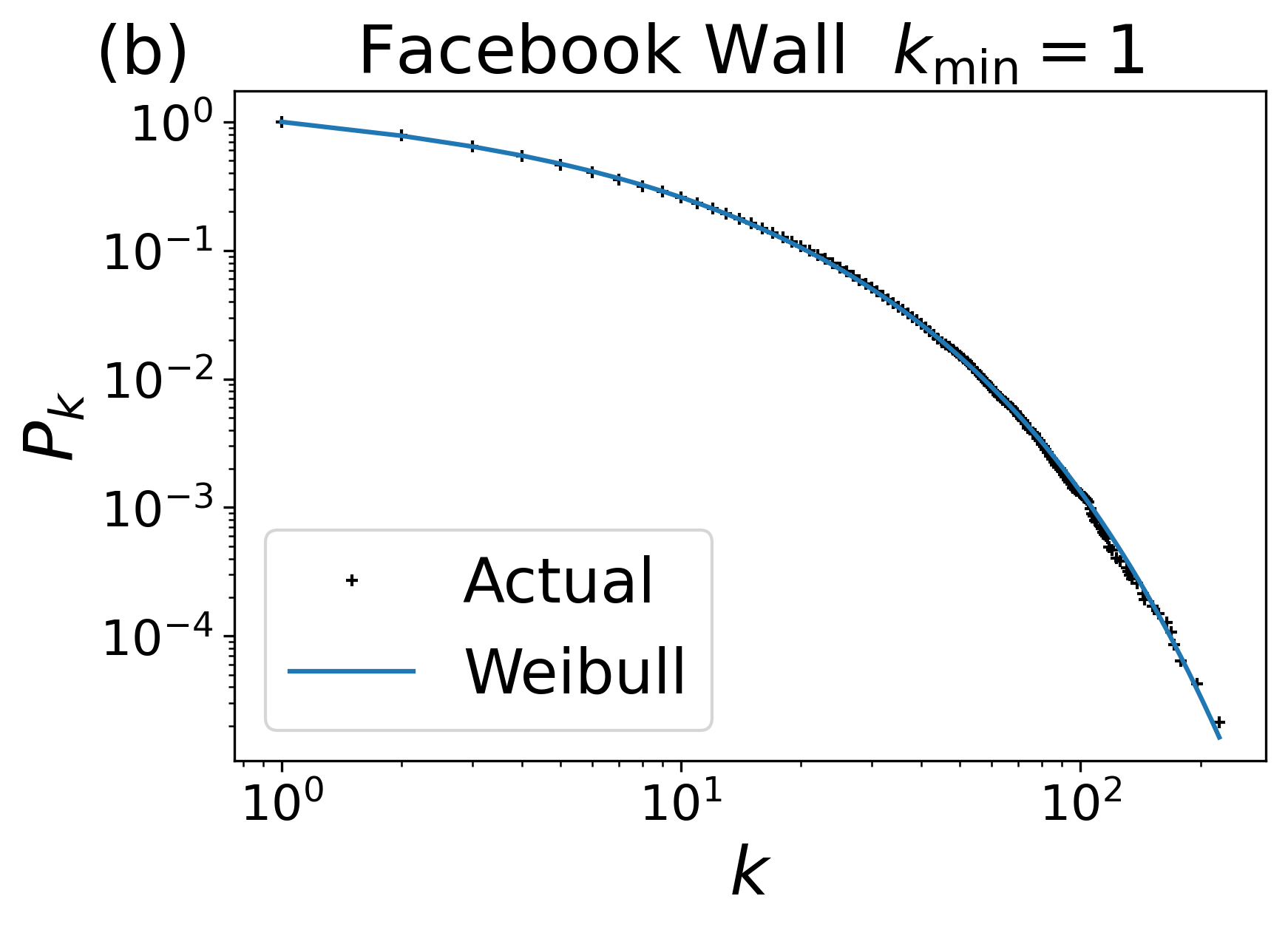}
\includegraphics[width=0.32\textwidth]{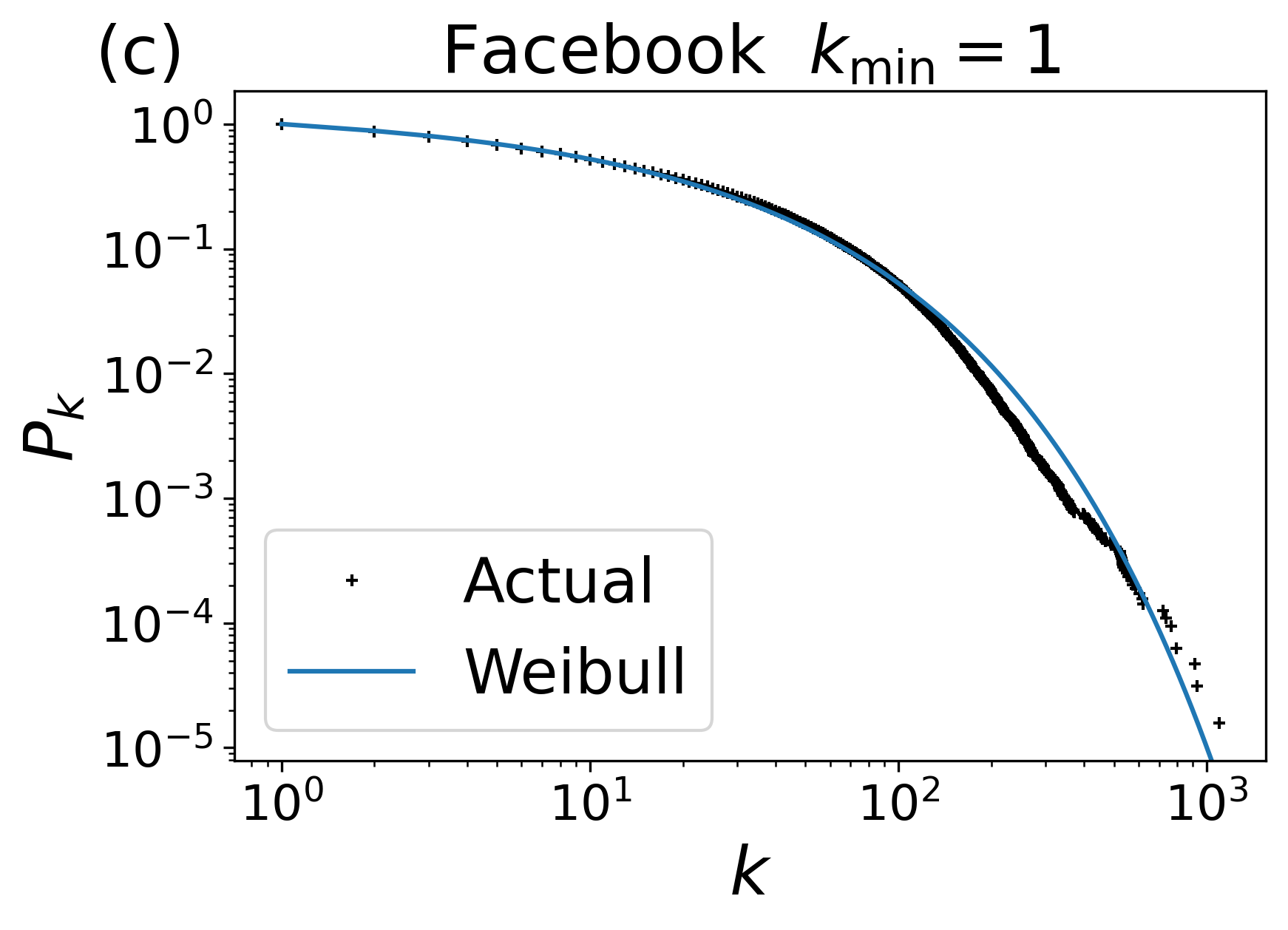}
\includegraphics[width=0.32\textwidth]{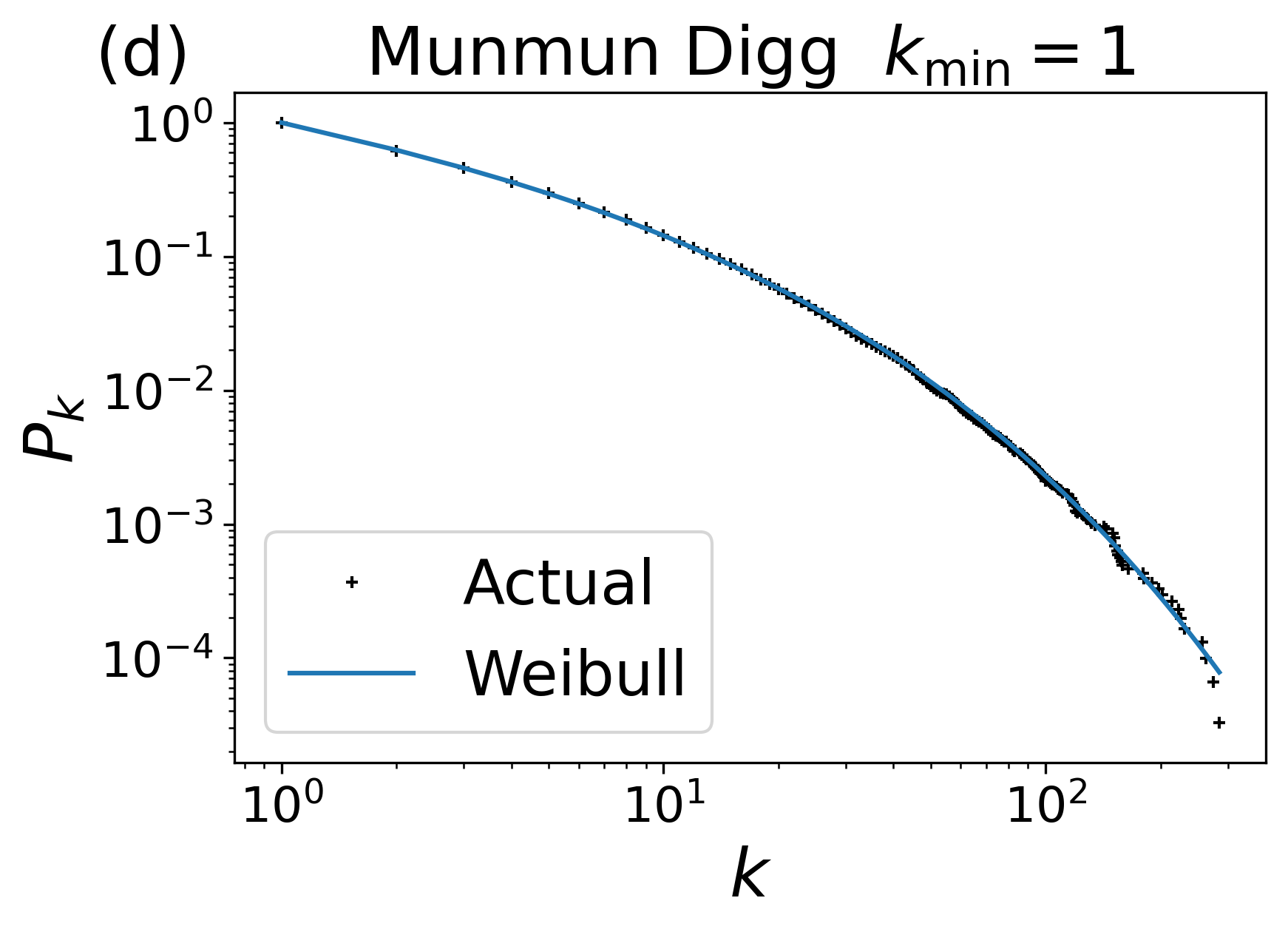}
\includegraphics[width=0.32\textwidth]{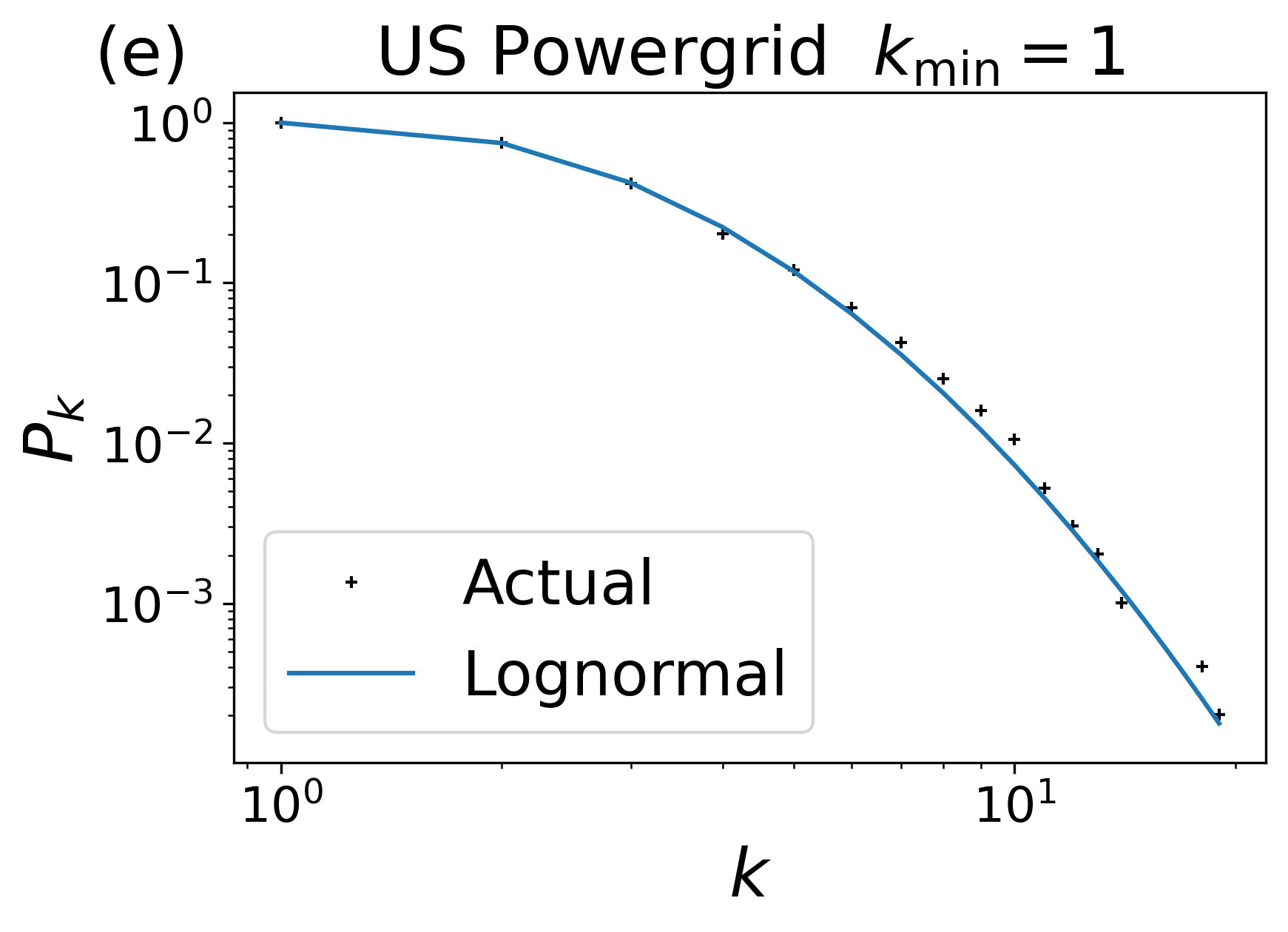}
\includegraphics[width=0.32\textwidth]{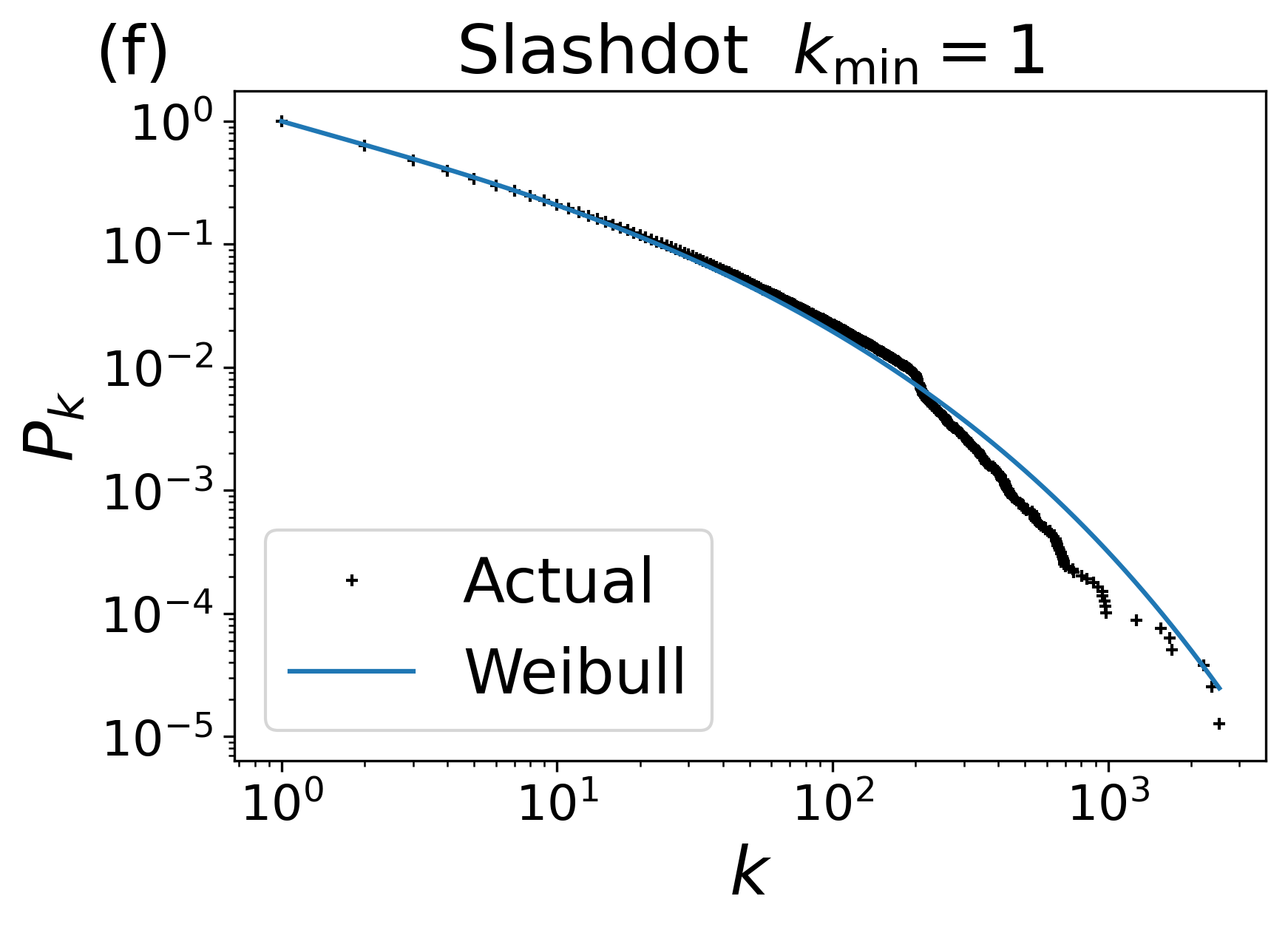}
\includegraphics[width=0.32\textwidth]{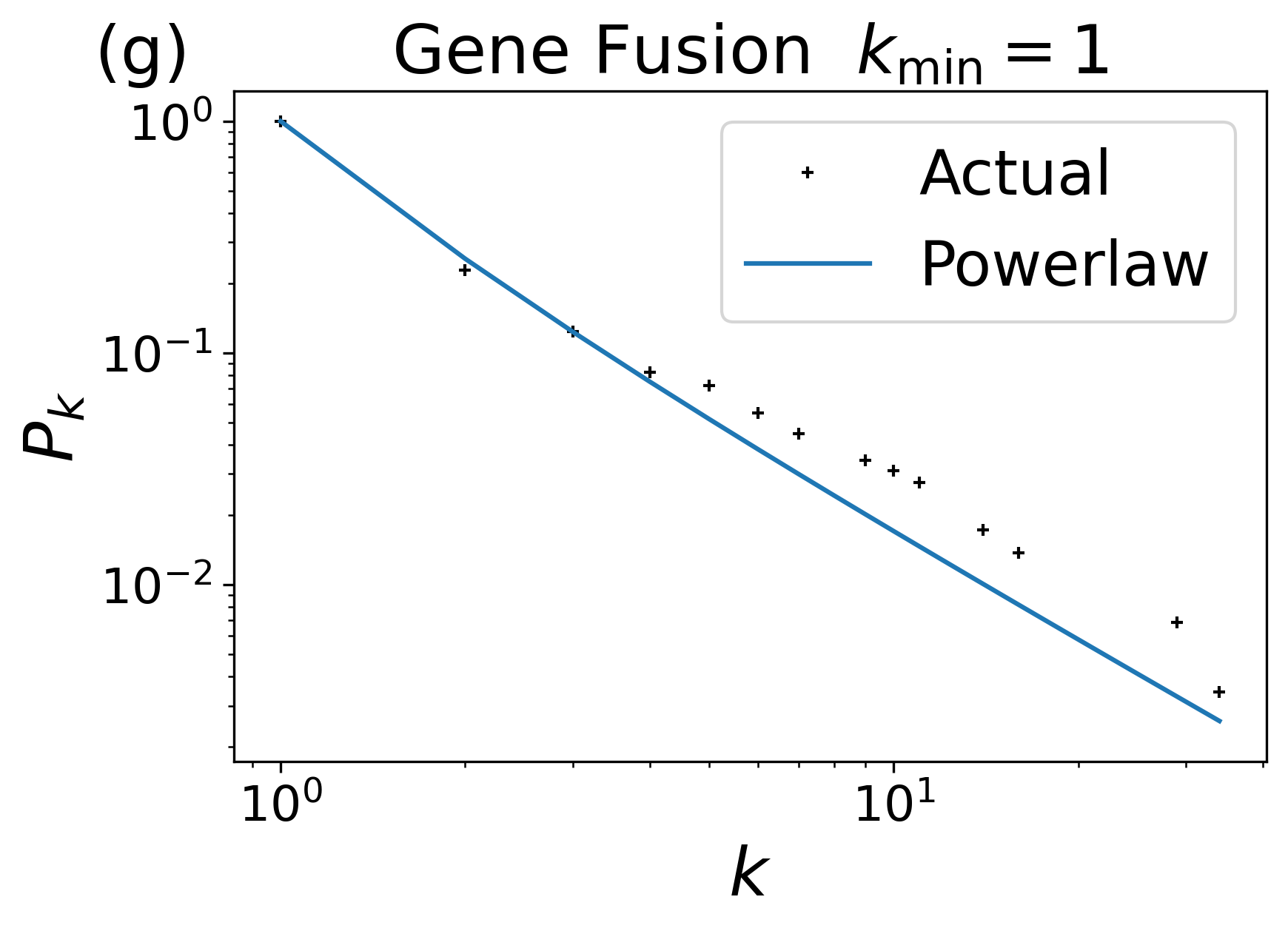}
\includegraphics[width=0.32\textwidth]{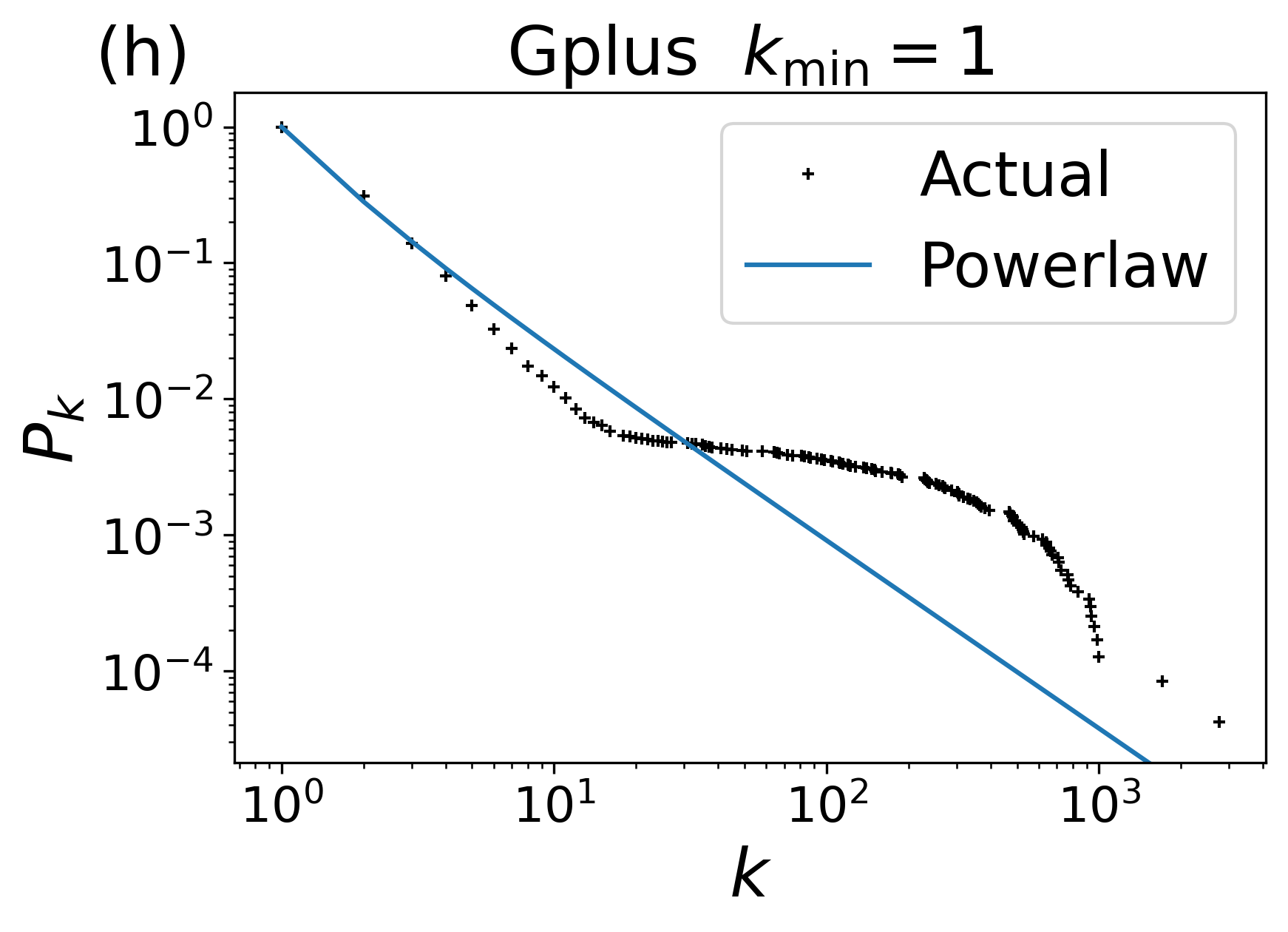}
\includegraphics[width=0.32\textwidth]{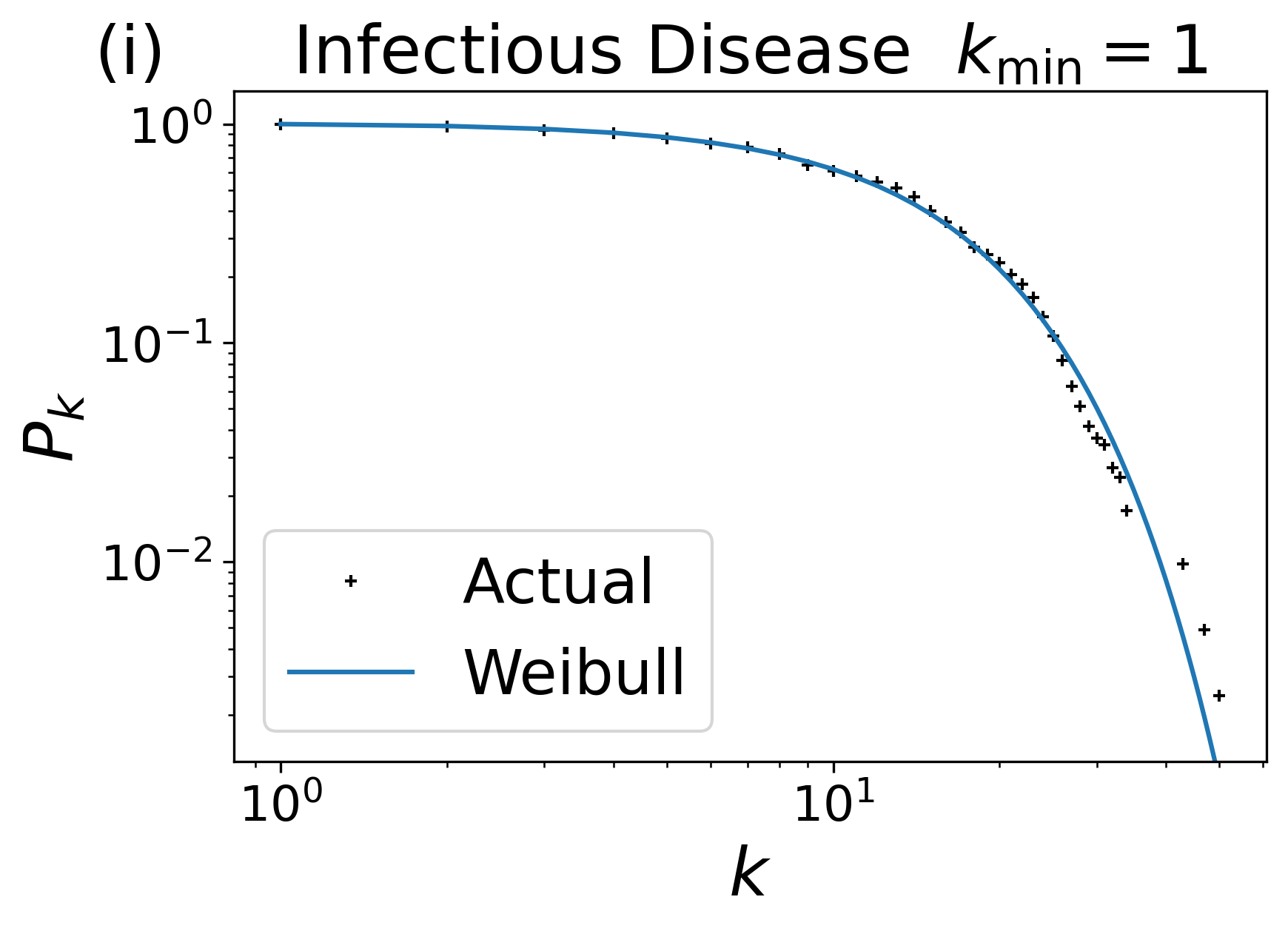}
\caption{Results for a selection of the networks whose degree distributions we fitted. As we can see in most cases, we manage to obtain excellent fits with low $k_{\rm min}$ values in all cases. Notable exceptions are Gene Fusion and Google+, which are discussed in turn in Section~\ref{section:results}.}
\label{fig:results}
\end{center}
\end{figure*}
We now look at some of the results obtained by using the above described methodology on a collection of network datasets, some of our own, one sexual contact network taken from Ref.~\cite{Itojapan}, and many taken from the Konect database. Konect stands for the Koblenz Network Collection~\cite{konect:all}; an open-access online database of network datasets from various areas of science. A full list of the datasets we applied the methodology to as well as a description of each dataset can be seen in Table~\ref{tab:networkdata} in Appendix~\ref{appendix:B}. A summary of some basic properties of these networks can be found in Table~\ref{tab:networksummary} in Appendix~\ref{appendix:B}.

For most networks that we looked at, we obtain good estimates for the parameters of the chosen distributions, as can be seen from the closely fitting curves to the datapoints. Note as well that in the majority of these cases we begin at $k_{\rm min}$ values of one and two, in line with our priority of minimizing $k_{\rm min}$. 

\begin{figure*}[t!]

\includegraphics[width=0.45\textwidth]{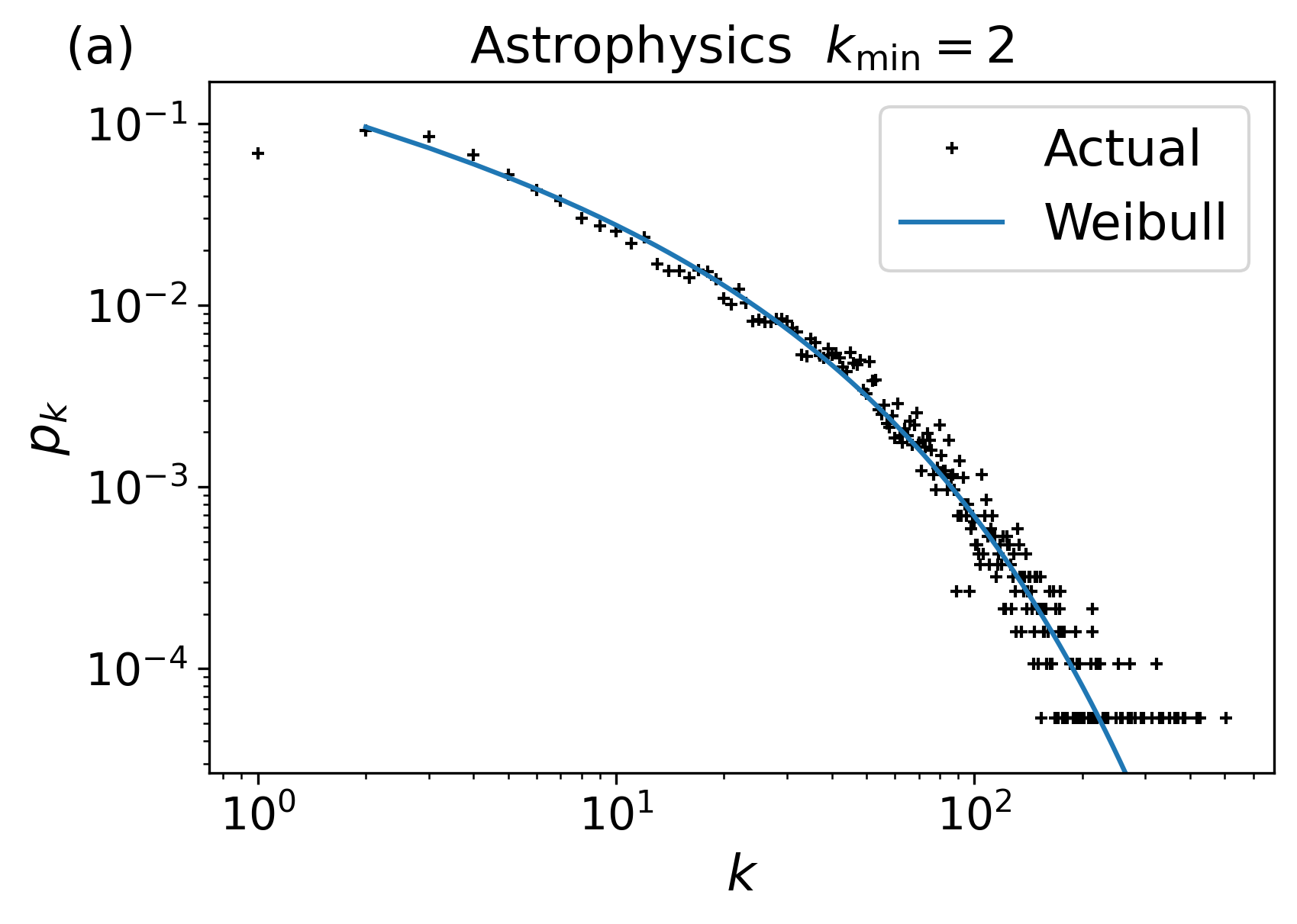}
\includegraphics[width=0.45\textwidth]{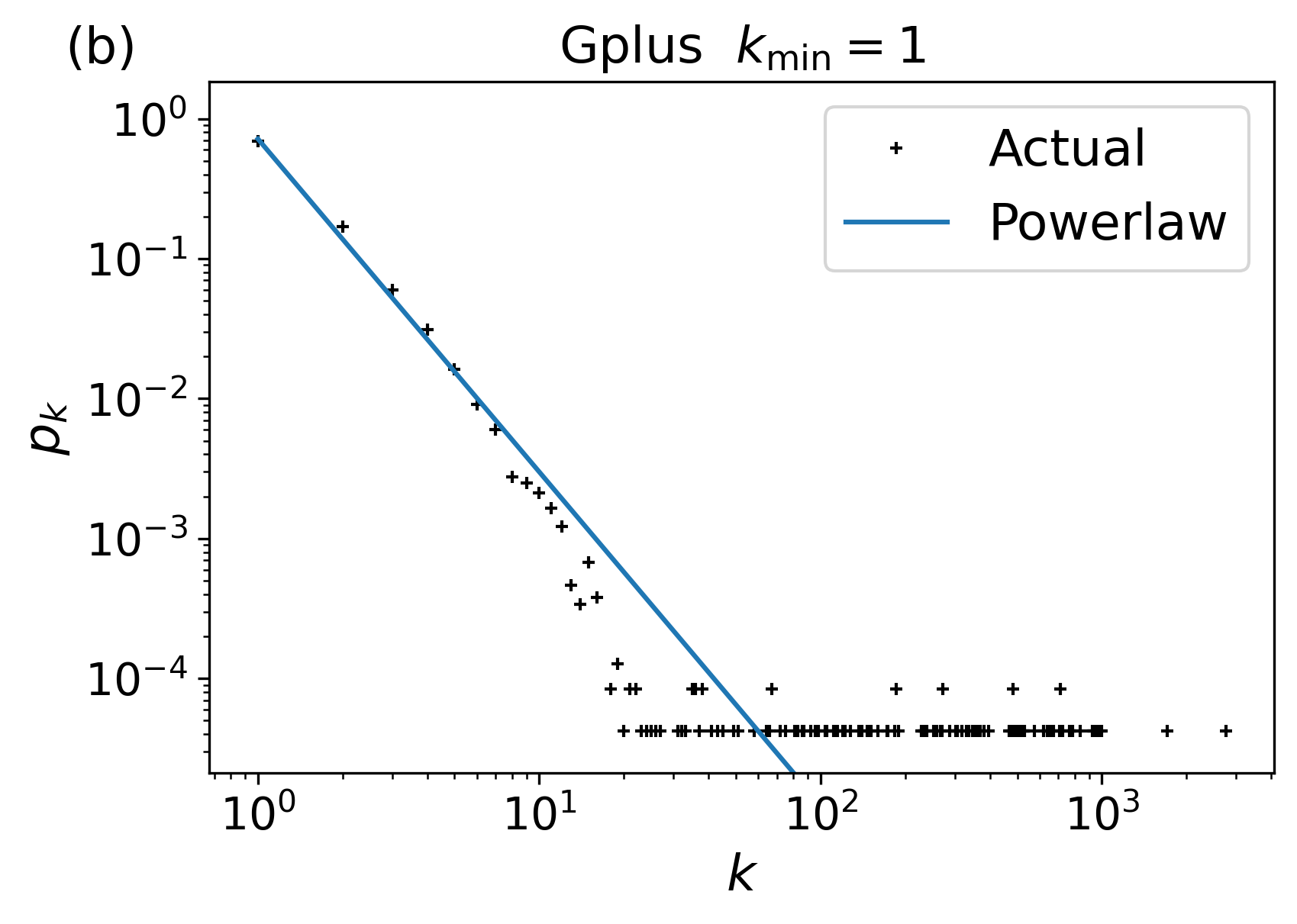}
\caption{Probability density functions for the Astrophysics dataset (left) and Google+ dataset (right). As we can see the Google+ data is far noisier in the tail, making the distribution more difficult to fit to.}
\label{fig:gpluspdf}

\end{figure*}

The results shown in Figure~\ref{fig:results} are representative of the results as a whole (results for all other datasets can be seen in Figures ~\ref{fig:all_ccdfs} and ~\ref{fig:all_pdfs}). In fact, based on its CCDF the Google+ (Figure~\ref{fig:results} (h)) dataset 
appears to be the worst fit of any dataset displayed.  One possible reason for this relatively poor fit is the very noisy tail of the distribution, which can be seen in the PDF of this distribution in Figure~\ref{fig:gpluspdf}. This noisy tail is caused by many hubs of different degrees, i.e., there are many points with different degrees but equal probabilities, hence the long horizontal line of points in Figure~\ref{fig:gpluspdf} (b). This is in contrast to the more `well-behaved' PDF of the astrophysics dataset in Figure~\ref{fig:gpluspdf} (a). Even in this case, looking at the PDF, we can see that there is not that much room for improvement in the fit. 

\begin{figure*}[t!]

\includegraphics[width=0.45\textwidth]{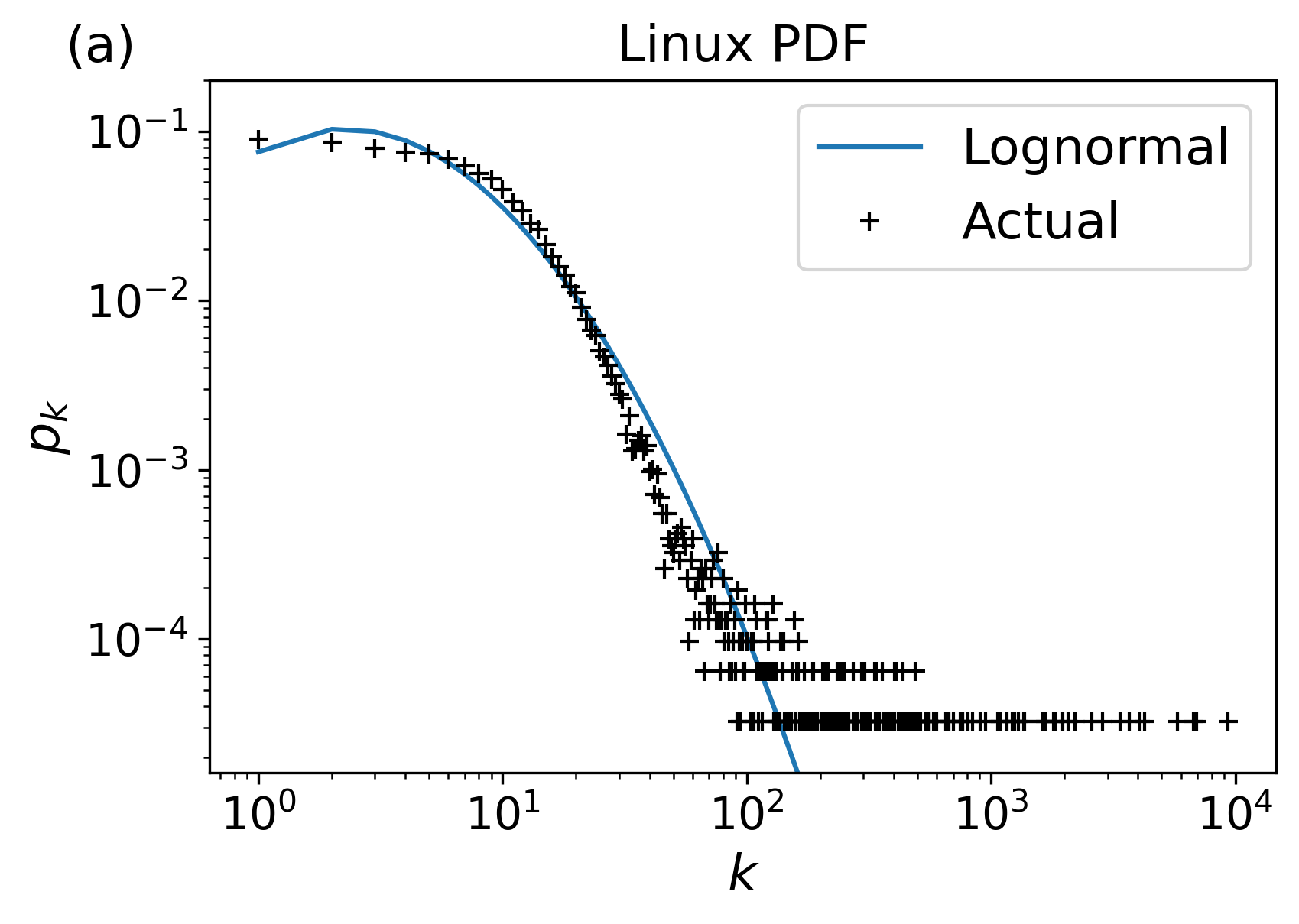}
\includegraphics[width=0.45\textwidth]{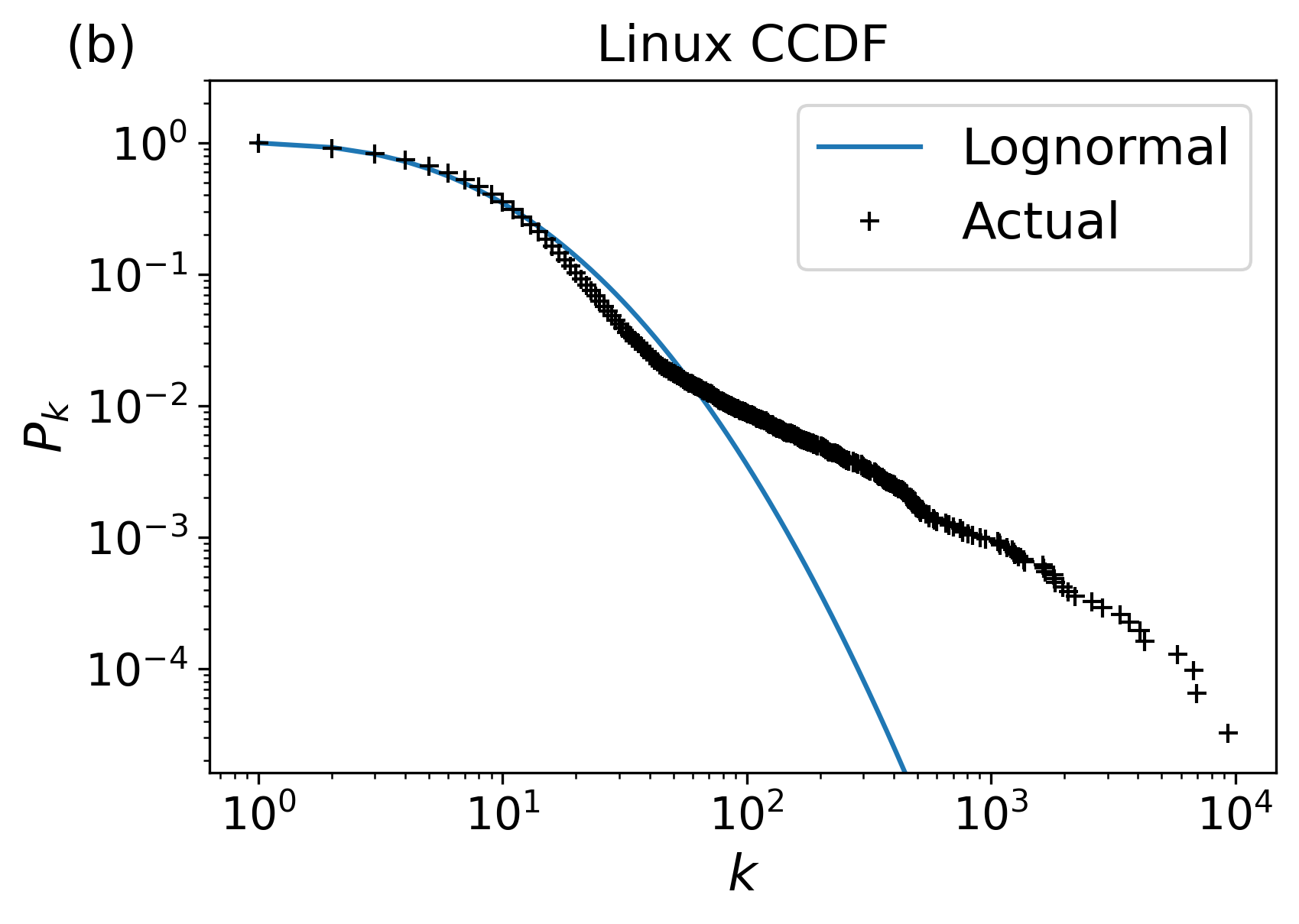}
\caption{PDF and CCDF of fits to the Linux dataset. As we can see, both the PDF and CCDF should be accounted for when assessing the fit.}
\label{fig:linux}

\end{figure*}

According to the CCDF in Figure~\ref{fig:linux} (b) for the Linux dataset we appear to obtain a bad fit here too. However, looking at the PDF (Figure~\ref{fig:linux} (a)) we see we actually have quite a good fit. In this case, the sharp drop off in CCDF values may have been caused by finite size effects, cumulative errors in individual probability values, or something else. The lesson here is that we cannot judge fit based solely on the CCDF.

A counter example to this is the Twitter dataset whose CCDF and PDF can be seen in Figures~\ref{fig:all_ccdfs} (v) and~\ref{fig:all_pdfs} (v) respectively. Here, we see a poor fit to both the PDF and CCDF, highlighting the need to look at both before determining if a distribution is a good fit or not.

In Figure~\ref{fig:results} (g) we have the Gene Fusion dataset. Overall this fits the data well however the dataset only contains approximately 250 nodes, hence we have large distances between the predicted and observed probabilities. As mentioned in step (iii) of the algorithm, in the case of such small datasets, the methodology is altered slightly so that only two votes for the same distribution are required for the distribution to be chosen, as we found that a larger value results in too large a portion of the data being excluded.

We can see that in cases such as the Facebook dataset in Figure~\ref{fig:results} (c), we obtain a good fit for the entire distribution. In the work of Clauset \textit{et. al}~\cite{Clauset2009}, it would be reasonable to choose a $k_{\rm min}$ value even as high as $k_{\rm min} \approx 110$. In doing so all of the data below this point would be ignored, which equates to over 99\% of the values, and with our methodology we have fit to the entire dataset. Other methodologies would similarly ignore the bulk of the data and fit a power-law to the tail, using one of the estimators shown in Table~\ref{tab:1} based on it appearing as a straight line when plotted on a log-log scale.

Some fits may be poor while prioritizing a low value of $k_{\rm min}$, but in this situation we would increase $k_{\rm min}$ to see if we can obtain a better fit. Examples of this can be seen in the Internet (CAIDA), Internet AS 2, and Internet Topology datasets, for which we get a very poor fit for low values of $k_{\rm min}$ but see improvement when we increase $k_{\rm min}$. This is in contrast to the Twitter dataset, for which we find that even larger values of $k_{\rm min}$ struggle to obtain a good fit. In this case, perhaps some less common distribution which we have not covered here would provide a better fit.

In the case of the RIP Ireland dataset, we see that the chosen distribution is a Weibull with parameters $\kappa = 183.14$, $\beta = 0.99$. Since $\beta$ is so close to one this is essentially an exponential distribution. However, since an exponential distribution is a special case of a Weibull distribution, and one can always achieve a higher likelihood by adding more parameters~\cite{Clauset2009}, the likelihood of an exponential will always be less than that of a Weibull. In this case, we rely on the information criteria to sufficiently penalize the distribution with more parameters. If using the Bayesian information criteria with $k_{\rm min} = 1$, we obtain an exponential fit for this distribution, though not for $k_{\rm min} = 2$ or greater. With the advantage of a lower $k_{\rm min}$ and a simpler distribution, an exponential distribution is likely the better choice here. Results for this daataset shown in Table~\ref{tab:networksummary} are for a Weibull distribution to illustrate our point.

In all other cases, we find that choosing BIC instead of AIC has no effect on what distribution is chosen.

\begin{figure*}[t!]
\includegraphics[width=0.45\textwidth]{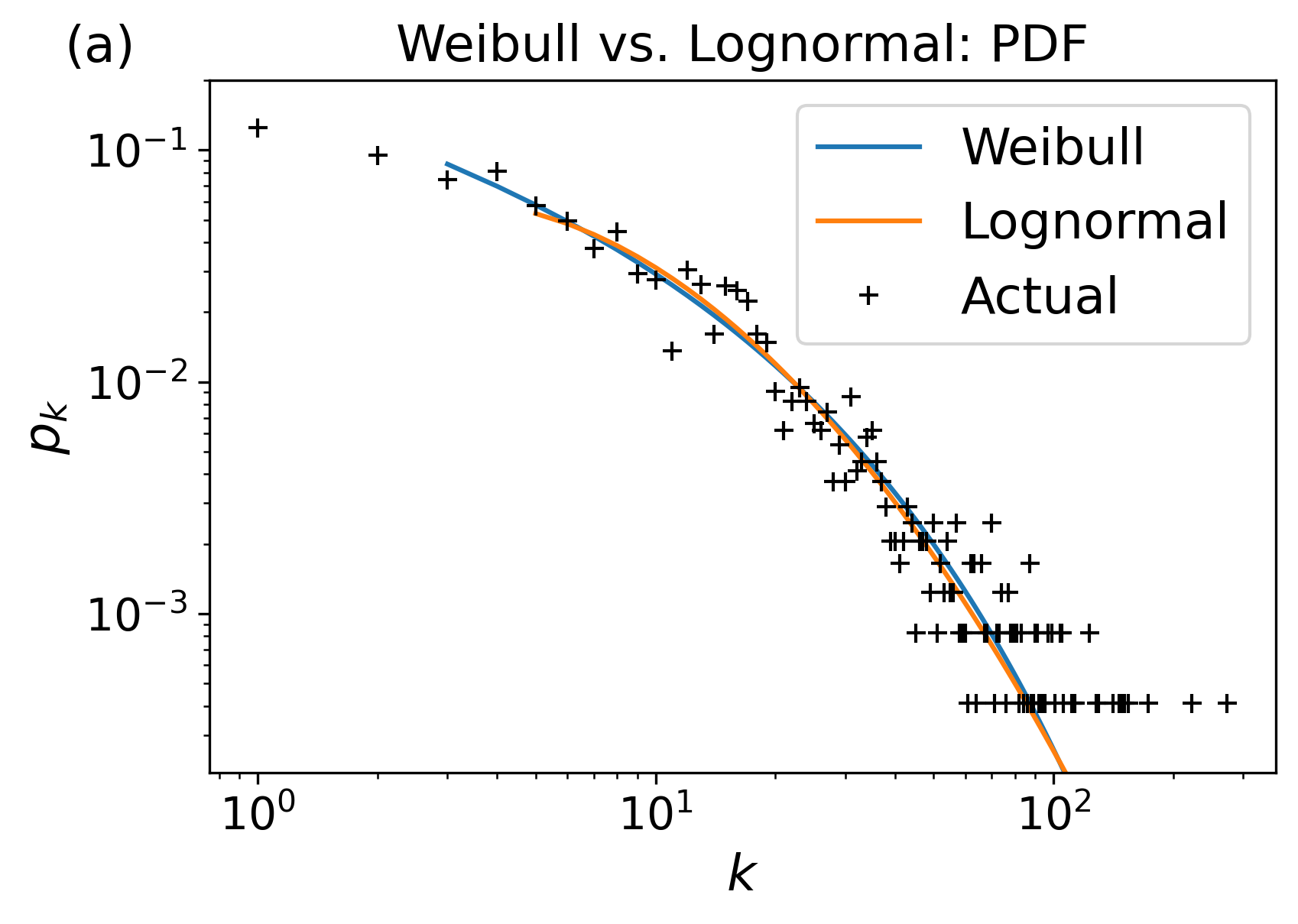}
\includegraphics[width=0.45\textwidth]{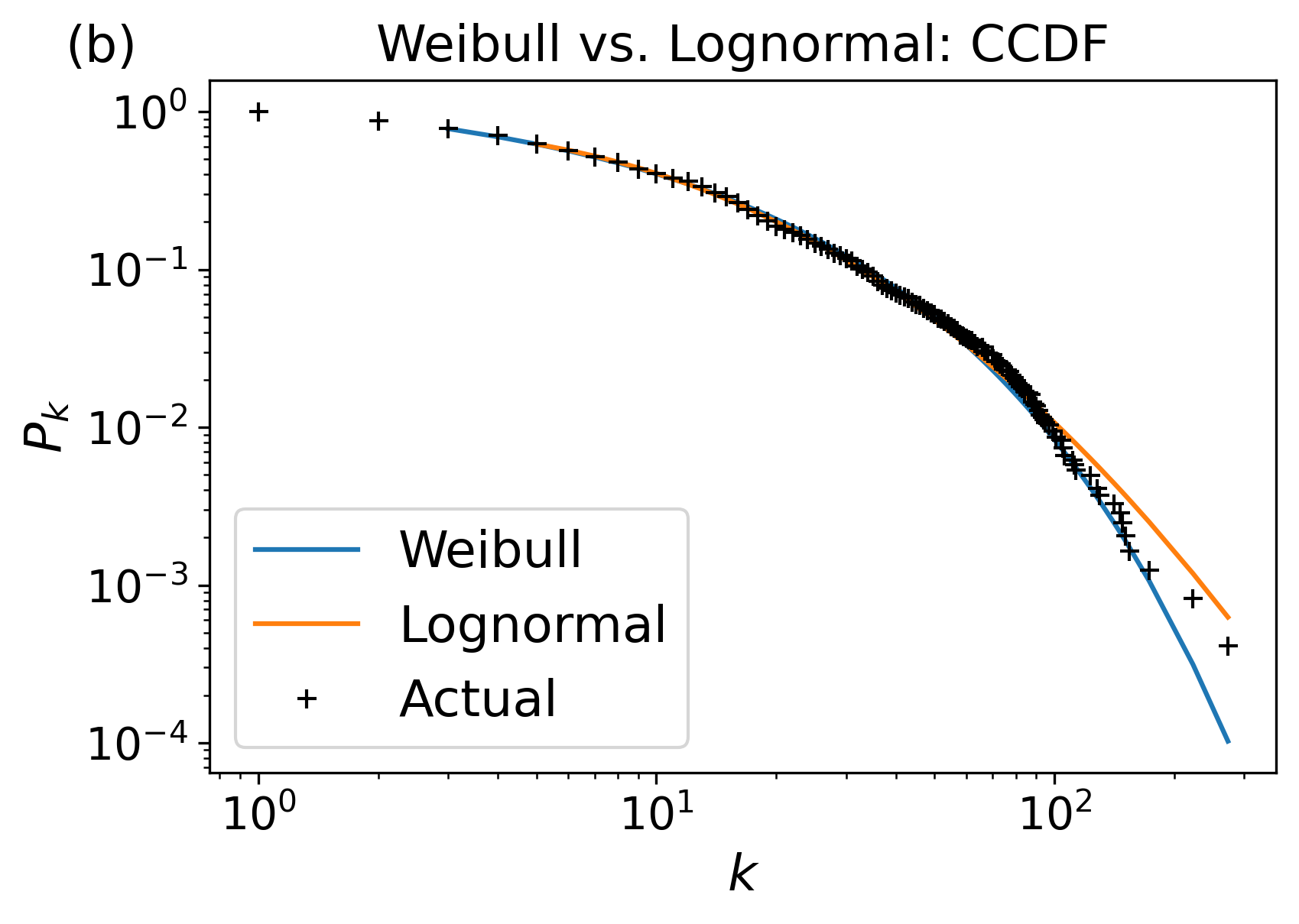}
\caption{Weibull distribution and a lognormal distribution fit to the Petster dataset at different $k_{\rm min}$ values as described in Section~\ref{section:results}.}
\label{fig:petsterwbvsln}
\end{figure*}
Lastly, we focus on the Petstser dataset, which is a more ambiguous example. Initially we fit using the steps outlined in the previous section, which tells us that the data follow a Weibull distribution for $k \ge 3$. This fit is shown in Figure~\ref{fig:petsterwbvsln}. 

Suppose we feel that based on the CCDF we could improve the fit to the tail of the distribution. We increase $k_{\rm min}$ to four and repeat the algorithm, which gives the lognormal again shown in Figure~\ref{fig:petsterwbvsln}. 

\begin{figure*}[t!]

\includegraphics[width=0.45\textwidth]{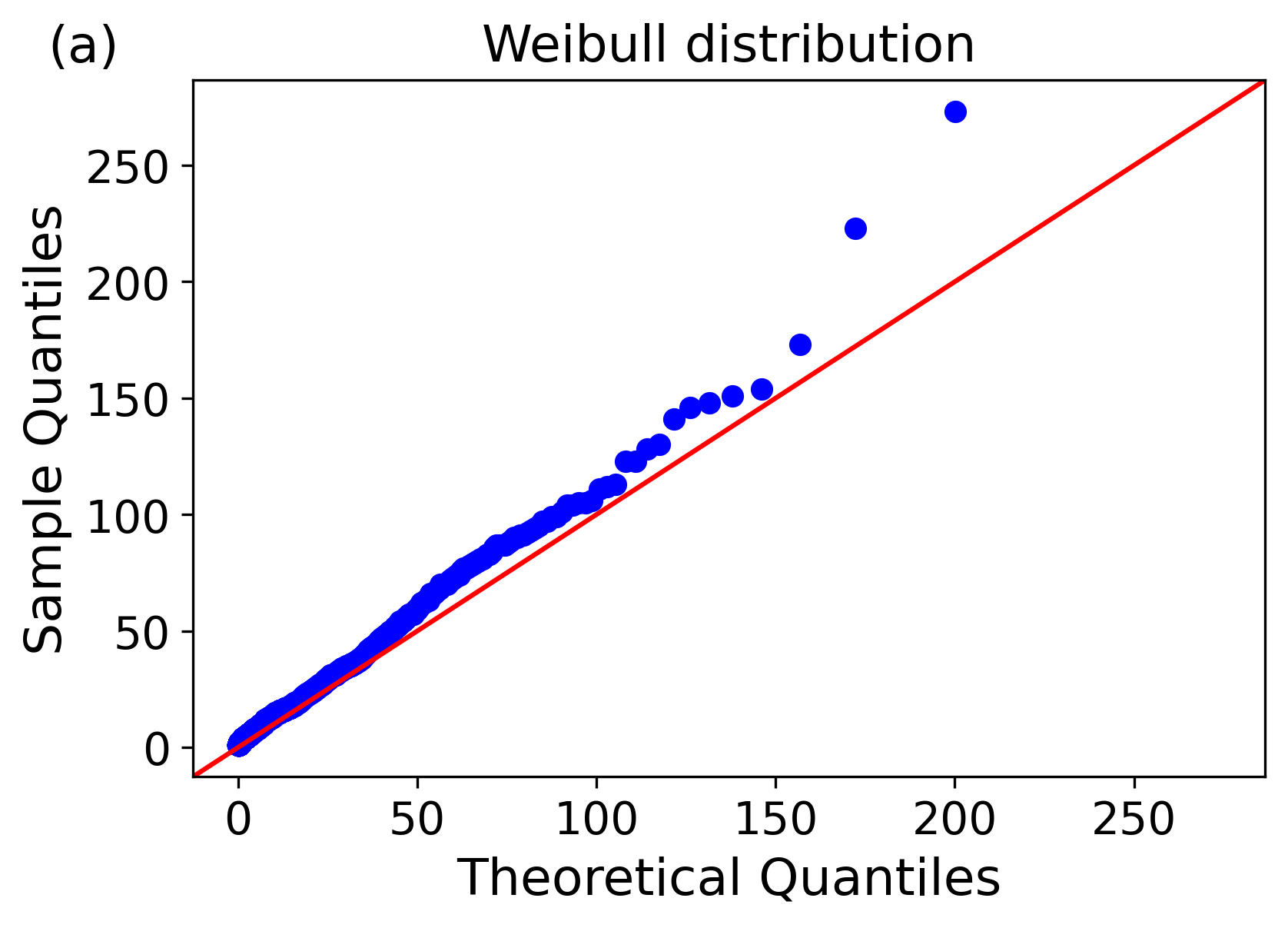}
\includegraphics[width=0.45\textwidth]{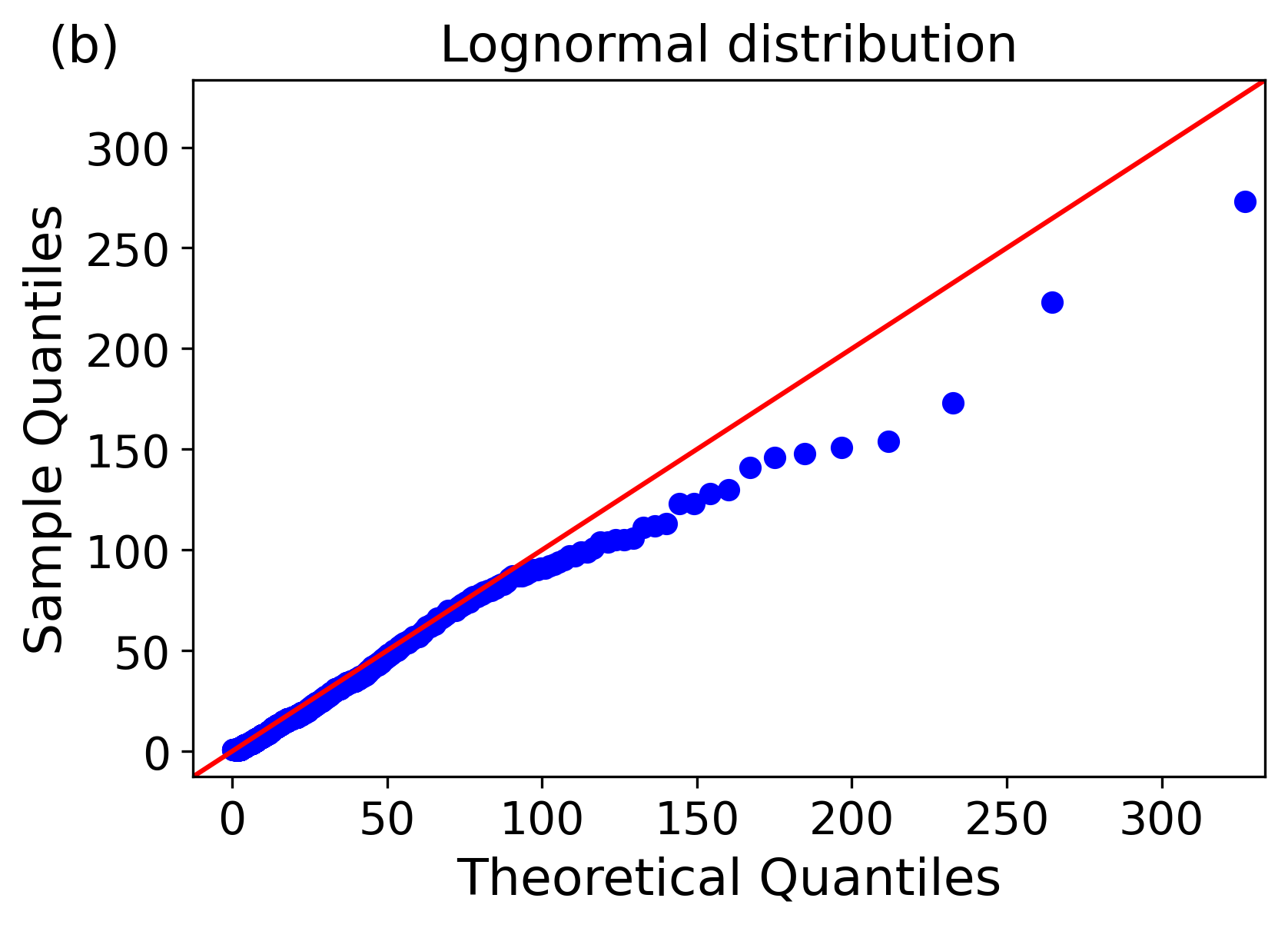}
\caption{Q-Q plots of a Weibull distribution (left) and a lognormal distribution (right) fit to the Petster dataset as described in Section~\ref{section:results}.}
\label{fig:qqplots}
\end{figure*}

This distribution arguably fits better in the tail of the distribution, so how do we choose between the two options? Looking at the PDF for both distributions we see there is very little appreciable difference, aside from the different $k_{\rm min}$ values. We then look at the Q-Q plots for each distribution, shown in Figure~\ref{fig:qqplots}. As we can see here the Weibull appears to be a better fit for the distribution, as the data follows the diagonal more closely (both begin to deviate at the tail which is to be expected when dealing with extreme values).

When the choice between two distributions is so close, we must look at the AIC weights. At $k_{\rm min} = 3$, the weights for the Weibull and lognormal distributions were 0.99 and 0.01 respectively. At $k_{\rm min} = 5$, these values were 0.3 and 0.7. Because of this, at $k_{\rm min} = 3$ we can rule out a lognormal distribution. However, at $k_{\rm min} = 5$, there is still some support for a Weibull distribution.

Combining all of this information, along with our priority of minimizing $k_{\rm min}$ we conclude that the Weibull distribution is the better fit for the data. In the case of more ambiguous fits, or wanting to improve the fit to a distribution we recommend this approach of looking at all available information.

All in all, we can see that the methodology described in this paper performs well in a variety of scenarios, and only does poorly in a few cases. This may be improved upon by testing other distributions, as here we have only tested a handful of commonly found distributions.

\section{Discussion}
\label{discussion}

For the majority of the time that the field of complex networks has been studied, questions around the degree distributions of networks have persisted to be left without satisfying answers. Here we have built upon the work of Clauset \textit{et al.}~\cite{Clauset2009}~to provide a methodology that introduces more rigour to the process of determining the degree distribution of a network dataset. We have three key differences to their method. 

Firstly, we suggest taking a low value of $k_{\rm{min}}$ to consider as many of the nodes as possible. Depending on the network's distribution, even a value of 3 or larger could end up fitting to less than 10\% of the network. This is only desirable when a network is poorly sampled and we are just interested in the tail. If we suspect the network is well sampled, then a high value of $k_{\rm{min}}$ will miss the majority of the network. These often-ignored low degree nodes are important in many contexts~\cite{tanaka2012dynamical}, playing crucial roles in processes such as percolation~\cite{newman2001random}, as well community detection in social networks~\cite{mehrabi2019community}. 

Secondly, we show that the parameter estimate obtained through numerical maximisation of Equation~\ref{eqn:pllnl} is  superior to the estimate given by Equation~\ref{eqn:clausetapprox}, especially for low values of $k_{\rm{min}}$. Taking the approximation for smaller networks with low values of $k_{\rm{min}}$ will give poor results and possibly rule out a power-law even if that is the true distribution. 

Thirdly, we use information criteria to choose the model rather than constructing a $p$-value from the Kolmogorov-Smirnov tests. These methods do not tell us about the goodness of fit but allow us to choose the most appropriate of the candidate models for this dataset.

As we can see from the results in Section~\ref{section:results}, using maximum likelihood estimation to determine the most appropriate fit from a set of candidate models performs well on a variety of network datasets with complex distributions. This work improves the standard of distribution fitting, with most if not all of the data being fit to in all cases, through minimisation of $k_{\rm min}$. 
This process is not entirely automated and we believe this should be the case. In some cases Q-Q plots may be necessary to interpret the results and one may need to check both the CCDF and PDF. These are noisy empirical datasets so caution must be taken and an automated process could easily miss these subtleties. 

After checking many different distributions, one may wonder why we are so interested in the specific distribution a complex network has. Networks have different underlying formation processes. For example, if we know the degree distribution follows a Poisson or binomial distribution, then we know this network if formed by purely a random process. If we know nodes prefer to connect to high degree nodes, we know it's a preferential attachment process. Hence, if we know the underlying distribution, we can come up with a theory that will give rise to that distribution.

This is one of the reasons why the power-law is so popular, not only is it instrumental in many statistical physics theories, and required for the epidemic threshold in disease transmission models, but it is the distribution underlying the preferential attachment model. It is also a single parameter distribution making it quite desirable for fitting with low statistical complexity. In our results, we find support for power-laws in many internet-based networks. However, this does not seem like a good mechanism for many of the social networks we analyse.

Power-laws suggest some nodes will have an extremely high degree, often approaching the system size, something not realistic for social systems. In the social networks here for example, they are better fitted by Weibull or log-normal distributions. In fact the RIP Ireland data is well fitted by a Weibull with $\beta \approx 1$ -- i.e. an exponential. With the exception of two outliers (people known in the media) no one has a lifetime degree even close to 2,000 (for a system size $>4$ million). 
Therefore, preferential attachment seems like a poor mechanism for social networks in particular with their fast decaying tails. 

Many other processes for social networks have been suggested even in the early years of complex networks, for example Ebel \textit{et al}.~\cite{ebel2002dynamics} propose a mechanism which will give rise to a stretched exponential. This is closer to the results that we observe. We aim to test this further going forward to identify an appropriate mechanism for social networks.


The foundation presented here performs well and can be extended to improve its performance further. In the future, we aim to extend the analysis both to more datasets and to test more parametric distributions. Another avenue for the extension of this work that we aim to explore is allowing for fitting of more than one distribution function to distributions in which we feel the tail behaves differently to the body of the distribution, i.e., splitting the distribution at some value $k^*$ and fitting two different distributions to the data above and below this value.  

We believe that this widely applicable and user-friendly approach provides much needed rigour in answering a fundamental question about the properties of a network.

\section*{Acknowledgements}
This publication has emanated from research conducted with the financial support of
Science Foundation Ireland under Grant number 18/CRT/6049. The authors would like to thank David O'Sullivan, Samuel Unicomb and James Gleeson for their suggestions and comments on the manuscript.

\bibliography{ref}
\appendix
\label{Appendix}
\section{Code \& Other Distributions}
\label{appendix:A}
\subsection{Code}
A documented package with which the results in this paper were produced is available at https://github.com/Shaneul/MLE.
\subsection{Truncated Power-Law Distribution}
Normalizing the truncated power-law from Equation~(\ref{eqn:truncpl}) yields
\begin{equation}
    p_k = (1-\Delta)\frac{{\rm e}^{k_{\rm min}/\kappa}}{Z(k_{\rm min})}k^{-\gamma}{\rm e}^{-k/\kappa}, \label{eqn:normalisedtruncpl}
\end{equation}
where $Z(x) = \Sigma_{m=0}^{\infty}(x + m)^{-\gamma}{\rm e}^{-m/\kappa}.$ The log likelihood then is straightforward to obtain from this and shown in Equation~(\ref{eqn:truncpllnl}).
\subsection{Exponential Distribution}
\label{section:appendix:exp}
Starting with Equation~(\ref{eqn:exponential}), we expand like so
$$
1 - \Delta = C\,  \sum_{k=k_{\rm min}}^{k_{\rm max}}  
{\rm e}^{-{{k}/{\kappa}}} = 
C\left({\rm e}^{{-k_{\rm min}}/{\kappa}} 
+ {\rm e}^{{-k_{\rm min} + 1}/{\kappa}} + ...\right).
$$
To normalize the distribution we multiply the summation here above and below by $1 - {\rm e}^{-1/\kappa}$ like so
$$
C\, \frac{1}{1 - {\rm e}^{-1/\kappa}}\,
\left(1 - {\rm e}^{-1/\kappa}\right)
\,\left({\rm e}^{{-k_{\rm min}}/{\kappa}} +
{\rm e}^{{-k_{\rm min} + 1}/{\kappa}} + ...\right)
$$
\begin{multline*}
= \frac{C}{1-{\rm e}^{-1/\kappa}}
 \bigl({\rm e}^{{-k_{\rm min}}/{\kappa}} 
+ {\rm e}^{{-k_{\rm min} + 1}/{\kappa}} + \\... 
 - {\rm e}^{{-\left(k_{\rm min} + 1\right)}
/{\kappa}} - 
{\rm e}^{-\left(k_{\rm max} + 1\right)/\kappa}\bigl).
\end{multline*}
All but the first and last terms here cancel, leaving us with
$$
\frac{C}{1- {\rm e}^{-1/\kappa}} 
\left( {\rm e}^{-k_{\rm min}/\kappa} - 
{\rm e}^{-\left(k_{\rm max}+1\right)/\kappa} \right),
$$
and so
$$
C = (1 - \Delta)\,
\frac{1 - {\rm e}^{-1/\kappa}}{{\rm e}^{{-k_{\rm min}}/\kappa} 
- {\rm e}^{-\left(k_{\rm max} + 1\right)/\kappa}}.
$$
If $k_{\rm max} \rightarrow \infty$ then we get
$$
C = (1 - \Delta)\frac{1 - {\rm e}^{-1/\kappa}}{{\rm e}^{-k_{\rm min}/\kappa}},
$$
and this leads to the normalized exponential distribution 
\begin{equation}
p(k) = (1 - \Delta)\left(\frac{1 - {\rm e}^{-1/\kappa}}{{\rm e}^{-k_{\rm min}/\kappa}}\right) {\rm e}^{-k/\kappa}, \label{eqn:normalisedexp}
\end{equation}
from which the log likelihood Equation~(\ref{eqn:exponentiallnl}) is obtained.
\subsection{Weibull Distribution}
Starting with Equation~(\ref{eqn:weibull}), we first define the constant to normalize the distribution.
\[(1 - \Delta) = C\sum_{i = k_{\rm min}}^{\infty} ( k/\kappa)^{\beta - 1} e^{-(k/\kappa)^\beta}.\]
This gives
\[C = \frac{(1 - \Delta)}{\sum_{i = k_{\rm min}}^{\infty} (k/\kappa)^{\beta - 1} e^{-(k/\kappa)^\beta}},\]
and with \(C\) defined as above, we have
\[p(k) = C(k/\kappa)^{\beta - 1}{\rm e}^{-(\lambda k)^\beta}.\]
The likelihood then is given by
$$
{\cal{L}} = \prod^{N} \frac{(1 - \Delta)}{\sum_{m = k_{\rm min}}^{\infty} (m/\kappa)^{\beta - 1} {\rm e}^{-(m/\kappa)^\beta}} (k/\kappa)^{\beta - 1} {\rm e}^{-(k/\kappa)^\beta},
$$
from which the log likelihood Equation~(\ref{eqn:weibulllnl}) is obtained
\subsection{Poisson Distribution}
Starting with Equation~(\ref{eqn:poisson}), we normalize like so
\[1 - \Delta = C\sum_{k=k_{\rm min}}^{\infty} \frac{\lambda^k}{k!} {\rm e}^{-\lambda}.\]
Adding and subtracting the same term from the above equation leads to
\begin{equation*}    
\begin{split}
1 - \Delta & = C\left[\sum_{k = k_{\rm min}}^{\infty} \frac{\lambda^k}{k!} {\rm e}^{-\lambda} + \sum_{k=0}^{k_{\rm min} - 1} \frac{\lambda^k}{k!} {\rm e}^{-\lambda} - \sum_{k=0}^{k_{\rm min} - 1} \frac{\lambda^k}{k!} {\rm e}^{-\lambda}\right] \\
& = C{\rm e}^{-\lambda}\left(\sum_{k=0}^{\infty}\frac{\lambda^k}{k!} - \sum_{k=0}^{k_{\rm min} - 1}\frac{\lambda^k}{k!}\right)\\
& = C{\rm e}^{-\lambda}\left({\rm e}^{\lambda} - \sum_{k=0}^{k_{\rm min} - 1}\frac{\lambda^k}{k!}\right)\\
& = C\left(1 - {\rm e}^{-\lambda}\sum_{k=0}^{k_{\rm min} - 1}\frac{\lambda^k}{k!}\right).\\
\end{split}
\end{equation*}
This gives
\begin{equation*}
C = (1 - \Delta)\left(1 - {\rm e}^{-\lambda}\sum_{k=0}^{k_{\rm min} - 1}\frac{\lambda^k}{k!}\right)^{-1}.
\end{equation*}
Equation~\ref{eqn:poisson} then becomes
$$
p(k) = (1 - \Delta)\left(1 - {\rm e}^{-\lambda}\sum_{k=0}^{k_{\rm min}-1}\frac{\lambda^k}{k!}{\rm e}^{-\lambda}\right)^{-1}\frac{\lambda^k}{k!}{\rm e}^{-\lambda},
$$
and our log likelihood is as shown in Equation~(\ref{eqn:poissonlnl}).
\subsection{Lognormal Distribution}
Starting with Equation~(\ref{eqn:logn}) we normalize in the usual way
$$
1-\Delta = C\sum_{k=k_{\rm min}}^{\infty}\frac{1}{k_i}{\rm e}^{-\frac{(\ln{k_i} - \mu)^2}{2\sigma^2}}.
$$
This gives 
$$
C = (1-\Delta)\left(\sum_{k=k_{\rm min}}^{\infty}\frac{1}{k}{\rm e}^{-\frac{(\ln{k} - \mu)^2}{2\sigma^2}}\right)^{-1}.
$$
Equation~(\ref{eqn:logn}) then becomes
$$
p(k) = (1-\Delta)\left(\sum_{k=k_{\rm min}}^{\infty}\frac{1}{k}{\rm e}^{-\frac{(\ln{k} - \mu)^2}{2\sigma^2}}\right)^{-1} \frac{1}{k}{\rm e}^{-\frac{(\ln{k} - \mu)^2}{2\sigma^2}},
$$
and the log likelihood Equation~(\ref{eqn:lognlnl}) is obtained from this.
\section{Additional Tables \& Figures}
\label{appendix:B}
Below are a collection of supporting tables and figures referenced throughout the paper.

\begin{table*}[]
\renewcommand{\arraystretch}{1.2}

\caption{ \small Values for each power-law exponent estimate discussed in section~\ref{sec:comparison} obtained on 1,000 datasets each of size $N=100$ with true parameter value $\gamma = 2.5$. Values shown are means with standard deviations in brackets. $\hat{\gamma}_c$ is the continuous estimator for the exponent from eq.~(\ref{eqn:gammahat}), $\gamma_{\rm pdf}$ 
and $\gamma_{\rm ccdf}$ are the values obtained by least squares regression of the pdf and ccdf respectively. $\gamma_{\rm d}$ is the value obtained by the discrete approximation eq.~\ref{eqn:clausetapprox}, $\gamma_{\langle k \rangle}$ is the value obtained by the method of moments and $\gamma_{\rm mle}$ is the value obtained by maximum likelihood estimation. Best estimates for each value of $k_{\rm min}$ are highlighted in bold.}
\label{tab:comparisontablen100}
\begin{tabular}{|l|x{2.2cm}|x{2.2cm}|x{2.2cm}|x{2.2cm}|x{2.2cm}|x{2.2cm}|}
    \hline
$k_{\rm min}$ & $\gamma_{\rm c}$ & $\gamma_{\rm pdf}$ & $\gamma_{\rm ccdf}$ & $\gamma_{\rm d}$ & $\gamma_{\langle k \rangle}$ & $\gamma_{\rm mle}$ \\
    \hline
1                                   & 4.49\,\,(0.19)                                  & 2.76\,\,(0.23)                              & 2.72\,\,(0.21)                               & 2.02\,\,(0.02)                                  & 3.31\,\,(0.13)                                        & \textbf{2.51\,\,(0.04)}                            \\ \hline
2                                   & 3.23\,\,(0.21)                                  & 2.24\,\,(0.28)                              & 2.37\,\,(0.24)                               & 2.35\,\,(0.08)                                  & 2.27\,\,(0.06)                                        & \textbf{2.48\,\,(0.10)}                             \\ \hline
3                                   & 2.98\,\,(0.20)                                   & 2.20\,\,(0.24)                               & 2.46\,\,(0.22)                               & 2.45\,\,(0.11)                                  & 2.17\,\,(0.02)                                        & \textbf{2.51\,\,(0.12)}                            \\ \hline
4                                   & 2.90\,\,(0.27)                                   & 2.09\,\,(0.37)                              & 2.48\,\,(0.36)                               & \textbf{2.51\,\,(0.17)}                                  & 2.12\,\,(0.02)                                        & 2.55\,\,(0.18)                            \\ \hline
5                                   & \textbf{2.58\,\,(0.20)}                                   & 1.71\,\,(0.17)                              & 2.20\,\,(0.21)                                & 2.36\,\,(0.14)                                  & 2.08\,\,(0.02)                                        & 2.37\,\,(0.15)                            \\ \hline
6                                   & 2.77\,\,(0.25)                                  & 1.75\,\,(0.26)                              & 2.40\,\,(0.33)                                & \textbf{2.53\,\,(0.19)}                                  & 2.08\,\,(0.02)                                        & 2.54\,\,(0.19)                            \\ \hline
7                                   & 2.75\,\,(0.43)                                  & 1.59\,\,(0.29)                              & 2.29\,\,(0.44)                               & \textbf{2.54\,\,(0.34)}                                  & 2.06\,\,(0.02)                                        & 2.55\,\,(0.34)                            \\ \hline
8                                   & 2.63\,\,(0.43)                                  & 1.45\,\,(0.22)                              & 2.27\,\,(0.32)                               & 2.46\,\,(0.35)                                  & 2.05\,\,(0.02)                                        & \textbf{2.47\,\,(0.36)}                            \\ \hline
9                                   & 2.91\,\,(0.58)                                  & 1.44\,\,(0.24)                              & \textbf{2.47\,\,(0.34)}                               & 2.71\,\,(0.46)                                  & 2.06\,\,(0.01)                                        & 2.72\,\,(0.47)                            \\ \hline
10                                  & 2.59\,\,(0.49)                                  & 1.27\,\,(0.18)                              & 2.33\,\,(0.61)                               & \textbf{2.46\,\,(0.41)}                                  & 2.04\,\,(0.02)                                        & 2.46\,\,(0.42)                            \\ \hline

\end{tabular}
\end{table*}

\begin{table*}[]
\renewcommand{\arraystretch}{1.2}

\caption{ \small Values for each power-law exponent estimate discussed in section~\ref{sec:comparison} obtained on 1,000 datasets each of size $N=10,000$ with true parameter value $\gamma = 2.5$. Values shown are means with standard deviations in brackets. $\hat{\gamma}_c$ is the continuous estimator for the exponent from eq.~(\ref{eqn:gammahat}), $\gamma_{\rm pdf}$ 
and $\gamma_{\rm ccdf}$ are the values obtained by least squares regression of the pdf and ccdf respectively. $\gamma_{\rm d}$ is the value obtained by the discrete approximation eq.~\ref{eqn:clausetapprox}, $\gamma_{\langle k \rangle}$ is the value obtained by the method of moments and $\gamma_{\rm mle}$ is the value obtained by maximum likelihood estimation. Best estimates for each value of $k_{\rm min}$ are highlighted in bold.}
\label{tab:comparisontablen10000}
\begin{tabular}{|l|x{2.2cm}|x{2.2cm}|x{2.2cm}|x{2.2cm}|x{2.2cm}|x{2.2cm}|}
    \hline
$k_{\rm min}$ & $\gamma_{\rm c}$ & $\gamma_{\rm pdf}$ & $\gamma_{\rm ccdf}$ & $\gamma_{\rm d}$ & $\gamma_{\langle k \rangle}$ & $\gamma_{\rm mle}$ \\
    \hline
1                                   & 4.46\,\,(0.07)                                  & 2.60\,\,(0.19)                               & 2.51\,\,(0.14)                               & 2.02\,\,(0.01)                                  & 3.08\,\,(0.13)                                        & \textbf{2.50\,\,(0.02)}                             \\ \hline
2                                   & 3.27\,\,(0.06)                                  & 2.48\,\,(0.20)                               & 2.48\,\,(0.16)                               & 2.37\,\,(0.02)                                  & 2.27\,\,(0.03)                                        & \textbf{2.50\,\,(0.03)}                             \\ \hline
3                                   & 2.96\,\,(0.07)                                  & 2.39\,\,(0.19)                              & 2.47\,\,(0.16)                               & 2.45\,\,(0.04)                                  & 2.16\,\,(0.02)                                        & \textbf{2.50\,\,(0.04)}                             \\ \hline
4                                   & 2.83\,\,(0.08)                                  & 2.31\,\,(0.20)                               & 2.46\,\,(0.17)                               & 2.47\,\,(0.05)                                  & 2.11\,\,(0.01)                                        & \textbf{2.50\,\,(0.05)}                             \\ \hline
5                                   & 2.76\,\,(0.09)                                  & 2.22\,\,(0.20)                               & 2.44\,\,(0.18)                               & 2.49\,\,(0.07)                                  & 2.08\,\,(0.01)                                        & \textbf{2.50\,\,(0.07)}                             \\ \hline
6                                   & 2.72\,\,(0.10)                                   & 2.17\,\,(0.20)                               & 2.45\,\,(0.19)                               & 2.49\,\,(0.07)                                  & 2.07\,\,(0.01)                                        & \textbf{2.50\,\,(0.08)}                             \\ \hline
7                                   & 2.68\,\,(0.11)                                  & 2.09\,\,(0.20)                               & 2.43\,\,(0.20)                                & \textbf{2.50\,\,(0.09)}                                   & 2.06\,\,(0.01)                                        & \textbf{2.50\,\,(0.09)}                             \\ \hline
8                                   & 2.66\,\,(0.12)                                  & 2.02\,\,(0.19)                              & 2.43\,\,(0.20)                                & \textbf{2.50\,\,(0.10)}                                    & 2.05\,\,(0.01)                                        & 2.51\,\,(0.10)                             \\ \hline
9                                   & 2.65\,\,(0.13)                                  & 1.95\,\,(0.19)                              & 2.42\,\,(0.22)                               & \textbf{2.50\,\,(0.11)}                                   & 2.04\,\,(0.01)                                        & 2.51\,\,(0.11)                            \\ \hline
10                                  & 2.63\,\,(0.14)                                  & 1.91\,\,(0.19)                              & 2.43\,\,(0.22)                               & \textbf{2.51\,\,(0.12)}                                  & 2.04\,\,(0.01)                                        & \textbf{2.51\,\,(0.12)}                            \\ \hline
\end{tabular}
\end{table*}

\begin{table*}[h]
\renewcommand{\arraystretch}{1.2}

\caption{ 
\small List of network datasets and a brief explanation of their sources. All datasets can be found on the Konect website~\cite{konect:all}
}

\centering
\begin{tabular}{|l|l|c|}
\hline
\textbf{Network} & \textbf{Description} & \textbf{Source} \\ \hline \hline
Gene Fusion                      & \begin{tabular}[c]{@{}l@{}}Nodes represent genes, edges represent that two have fused in the \\ emergence of cancer.\end{tabular}   &    \cite{dataset:genefusion}         \\ \hline
Laxdaela & \begin{tabular}[c]{@{}l@{}}Character interaction network from Laxdaella \\ - one of the Icelandic sagas.\end{tabular} & \cite{mac2013icelanders} \\ \hline
Infectious Disease               & Nodes represent people, edges represent face-to-face contact. &  \cite{dataset:Isellainfectious}                                                                            \\ \hline
Network Science                  & Collaboration network of network scientists.  &      \cite{dataset:Newmannetsci}                                                                                             \\ \hline
ASoIaF & Charcter interaction network from the fantasy series A Song of Ice and Fire. & \cite{gessey2020asoiaf} \\ \hline
Yeast                   & Nodes represent proteins, edges represent interactions.                                    &   \cite{Jeongprotein}                                                \\ \hline
Petster                          & Friendship network between users of the Petster website.         &                                                      \cite{konect:petser}                        \\ \hline
Moreno Health                    & Friendship network at Moreno high school.                       &                                                \cite{dataset:Moodymoreno}                                 \\ \hline
Human Proteins                & Nodes represent proteins, edges represent interactions. &                                             \cite{dataset:humanprotein1vidal}                                             \\ \hline
US powergrid                     & \begin{tabular}[c]{@{}l@{}}Nodes represent substations, transformators, or generators,\\ edges represent connections between them.\end{tabular} &\cite{wattsstrogatz}\\ \hline
Sexual Contact & Sexual contact network in Japan. & \cite{Itojapan} \\ \hline
Java Classes          & \begin{tabular}[c]{@{}l@{}}Nodes represent Java classes, edges represent dependency between\\ two classes.\end{tabular}            &     \cite{dataset:java}        \\ \hline
Reactome                 & Nodes represent proteins, edges represent interactions.  &                            \cite{joshi2005reactome}                                                        \\ \hline
Internet of AS 1 & \begin{tabular}[c]{@{}l@{}}Nodes represent autonomous systems, edges represent connections \\ between them.\end{tabular}        & \cite{konect:internetauto1}           \\ \hline
PGP                              & \begin{tabular}[c]{@{}l@{}}Key-sharing network from the PGP (Pretty Good Privacy) web of \\ trust.\end{tabular}     & \cite{dataset:pgp}                            \\ \hline

RIP Ireland             & \begin{tabular}[c]{@{}l@{}}Online book of condolences for deaths in Ireland. Number of signatures \\ corresponds to node degree.\end{tabular} &  \cite{dempsey2020ripie} \\ \hline
Astrophysics                     & Collaboration network of astrophysicists.      &  \cite{Newmanscientific}                                                                                                \\ \hline
Internet AS 2 & \begin{tabular}[c]{@{}l@{}}Nodes represent autonomous systems, edges represent \\ communication between them.\end{tabular}                    & \cite{dataset:internetauto2routeviews}  \\ \hline
Twitter                          & Twitter user network.                           &     \cite{dataset:twitter}                                                                                            \\ \hline

Gplus                            & Google+ user network.                          &                                                      \cite{dataset:twitter}                                          \\ \hline
Internet (CAIDA)                 & \begin{tabular}[c]{@{}l@{}}Nodes represent autonomous systems from the CAIDA project and\\ edges represent communication.\end{tabular}     & \cite{dataset:internetauto2routeviews}     \\ \hline
Munmun Digg                      & Reply network on the news site Digg.                   &     \cite{dataset:munmun}                                                                                     \\ \hline
Linux                            & \begin{tabular}[c]{@{}l@{}}Nodes represent files, edges represent that they contain each \\ other.\end{tabular}  &  \cite{konect:linux}                             \\ \hline
Internet Topology                & \begin{tabular}[c]{@{}l@{}}Nodes represent autonomous systems, edges represent connections\\ between them.\end{tabular}          &        \cite{dataset:internettopology}        \\ \hline
Facebook                         & Users of Facebook in the New Orleans region.           &      \cite{dataset:fbwall}                                                                                   \\ \hline

Facebook wall                    & Facebook wall post network.                               &                                           \cite{dataset:fbwall}                                            \\ \hline
Slashdot                         & User network of technology news site Slashdot.      &                                     \cite{dataset:slashdot}                                                        \\ \hline
Enron                            &        \begin{tabular}[c]{@{}l@{}}Network of emails sent at Enron. Nodes represent employees, \\ edges represent emails sent between them. \end{tabular}                 &   \cite{enronsource}                                                                                    \\ \hline

\hline
\end{tabular}

\label{tab:networkdata}
\end{table*}

\begin{table*}[]
\renewcommand{\arraystretch}{1.2}
\caption{\small Quantities for each of our studied networks. $N$ is number of nodes, $L$ is number of edges, $\langle k \rangle$ is the mean degree and $k_{\rm max}$ is the maximum degree.}
\label{tab:networksummary}

\begin{tabular}{|L{4cm}|R{2.5cm}|R{2.5cm}|R{2.5cm}|R{2.5cm}|}
\hline
Name               & \multicolumn{1}{c|}{$N$} & \multicolumn{1}{c|}{$L$} & \multicolumn{1}{c|}{$\langle k \rangle$} & \multicolumn{1}{c|}{$k_{\rm max}$} \\ \hline \hline
Gene Fusion        & 291                    & 279                    & 1.92                       & 34                        \\ \hline
Laxdaela           & 338                    & 894                    & 5.29                       & 45                        \\ \hline
Infectious Disease & 410                    & 2,765                  & 13.49                      & 50                        \\ \hline
Network Science    & 1,461                  & 2,742                  & 3.75                       & 34                        \\ \hline
ASoIaF             & 1,786                  & 14,478                 & 16.21                      & 215                       \\ \hline
Yeast              & 1,870                  & 2,277                  & 2.44                       & 56                        \\ \hline
Petster            & 2,426                  & 16,631                 & 13.71                      & 273                       \\ \hline
Moreno Health      & 2,539                  & 10,455                 & 8.24                       & 27                        \\ \hline
Human Proteins     & 3,133                  & 6,726                  & 4.29                       & 129                       \\ \hline
US Powergrid       & 4,941                  & 6,594                  & 2.67                       & 19                        \\ \hline
Sexual Contact     & 4,999                  & 23,979                 & 9.59                       & 985                       \\ \hline
Java Classes       & 6,120                  & 50,290                 & 16.43                      & 5,655                     \\ \hline
Reactome           & 6,327                  & 147,547                & 46.64                      & 855                       \\ \hline
Internet AS 1      & 6,474                  & 13,895                 & 4.29                       & 1,460                     \\ \hline
PGP                & 10,680                 & 24,316                 & 4.55                       & 205                       \\ \hline
RIP Ireland        & 15,310                 & 1,410,272              & 184.23                     & 4,235                     \\ \hline
Astrophysics       & 18,771                 & 198,050                & 21.10                      & 504                       \\ \hline
Internet AS 2      & 22,963                 & 48,436                 & 4.22                       & 2,390                     \\ \hline
Twitter            & 23,370                 & 32,831                 & 2.81                       & 238                       \\ \hline
Gplus              & 23,628                 & 39,194                 & 3.32                       & 2,761                     \\ \hline
Internet (CAIDA)   & 26,475                 & 53,381                 & 4.03                       & 2,628                     \\ \hline
Munmun Digg        & 30,398                 & 86,312                 & 5.68                       & 285                       \\ \hline
Linux              & 30,837                 & 213,747                & 13.86                      & 9,340                     \\ \hline
Internet Topology  & 34,761                 & 107,720                & 6.20                       & 2,760                     \\ \hline
Facebook Wall      & 46,952                 & 193,494                & 8.24                       & 223                       \\ \hline
Facebook           & 63,731                 & 817,035                & 25.64                      & 1,098                     \\ \hline
Slashdot           & 79,116                 & 467,731                & 11.82                      & 2,534                     \\ \hline
Enron              & 87,273                 & 299,220                & 6.86                       & 1,728                     \\ \hline
\end{tabular}
\end{table*}

\begin{table*}[t]
\renewcommand{\arraystretch}{1.2}
\caption{Network datasets and their corresponding distributions, all figures rounded to two decimal places. $x_1$ is the first parameter for a distribution, $\bar{x}_1$ is the mean parameter obtained from fitting this distribution to bootstrapped samples. $\sigma_{x_1}$ is the standard deviation for this parameter. $P_{2.5}(x_1)$ and $P_{97.5}(x_1)$ are the 2.5th and 97.5th percentiles of the parameters respectively. These values are then repeated for $x_2$ in the case of two-parameter distributions.}
\label{tab:results}
\begin{tabular}{|l|l|R{0.7cm}|R{1cm}|R{1cm}|R{1cm}|R{1.2cm}|R{1.2cm}|R{1cm}|R{1cm}|R{1cm}|R{1.2cm}|R{1.2cm}|}
\hline
Network                          & Distribution & $k_{\rm min}$ & $x_1$ & $\bar{x}_1$ & $\sigma_{x_1}$ & $P_{2.5}(x_1)$ & $P_{97.5}(x_1)$ & $x_2$ & $\bar{x}_2$ & $\sigma_{x_2}$ & $P_{2.5}(x_2)$ & $P_{97.5}(x_2)$ \\ \hline \hline
Gene Fusion                      & Powerlaw     & 1     & 2.50   & 2.51  & 0.11 & 2.32  & 2.72   &        &      &      &       &        \\ \hline
Laxdaela                         & Lognormal    & 1     & 1.27  & 1.28  & 0.05 & 1.17   & 1.37   & 0.90 & 0.92 & 0.05 & 0.83  & 1.02   \\ \hline
Infectious Disease               & Weibull      & 1     & 14.99  & 15.04 & 0.50  & 14.08 & 16.07  & 1.62   & 1.60  & 0.08 & 1.45  & 1.76   \\ \hline
Network Science                          & Lognormal      & 1     & 1.02   & 1.02  & 0.02 & 0.98  & 1.06   & 0.77   & 0.78 & 0.02 & 0.74  & 0.82   \\ \hline
ASoIaF                           & Weibull      & 1     & 12.68 & 12.69 & 0.49 & 11.73  & 13.62  & 0.75 & 0.75 & 0.02 & 0.71  & 0.80   \\ \hline
Yeast Proteins             & Lognormal      & 2     & 0.01  & 0.24 & 0.85  & 0.00 & 4.17  & 1.11   & 1.06  & 0.21 & 0.01  & 1.17   \\ \hline
Petster                          & Weibull      & 3     & 7.40   & 7.31  & 0.65 & 6.06  & 8.69   & 0.62   & 0.62 & 0.03 & 0.57  & 0.67   \\ \hline
Moreno Health                    & Lognormal    & 2     & 0.0082 & 0.26  & 0.92 & 0.00     & 4.31   & 1.11   & 1.05 & 0.23 & 0.01  & 1.17   \\ \hline
Human Proteins                   & Lognormal    & 1     & 0.64   & 0.66  & 0.52 & 0.21  & 0.72   & 1.16   & 1.17 & 0.12 & 1.11  & 1.41   \\ \hline
US Powergrid                     & Lognormal    & 1     & 0.82   & 0.82  & 0.01 & 0.80   & 0.84   & 0.59   & 0.59 & 0.01 & 0.58  & 0.60    \\ \hline
Sexual Contact        & Lognormal    & 1     & 1.31 & 1.31  & 0.28 & 1.24   & 1.36   & 1.30 & 1.30  & 0.07 & 1.25  & 1.37   \\ \hline
Java Classes          & Lognormal    & 1     & 2.05   & 2.05  & 0.01 & 2.04  & 2.07   & 0.84   & 0.84 & 0.01 & 0.81  & 0.87   \\ \hline
Reactome                         & Weibull      & 1     & 25.60  & 25.40  & 1.79 & 23.88 & 27.14  & 0.56   & 0.55 & 0.11 & 0.52  & 0.56   \\ \hline
Internet AS 1 & Powerlaw     & 2     & 2.30   & 2.30   & 0.02 & 2.26  & 2.34   &        &      &      &       &        \\ \hline
PGP                              & Weibull      & 2     & 0.17   & 0.14  & 0.05 & 0.07  & 0.25   & 0.33   & 0.32 & 0.07 & 0.29  & 0.36   \\ \hline
RIP Ireland                      & Weibull      & 1     & 183.14 & 183.20 & 1.60  & 180.16 & 186.45 & 0.99 & 0.99 & 0.01 & 0.97  & 1.00      \\ \hline
Astrophysics                     & Weibull      & 2     & 10.87  & 10.82 & 0.26 & 10.30  & 11.31  & 0.59   & 0.59 & 0.01 & 0.57  & 0.60    \\ \hline
Internet AS 2 & Powerlaw     & 5     & 2.71   & 2.71  & 0.02 & 2.68  & 2.75   &        &      &      &       &        \\ \hline
Twitter                          & Powerlaw     & 1     & 2.47   & 2.47  & 0.01 & 2.44  & 2.49   &        &      &      &       &        \\ \hline
Gplus                            & Powerlaw     & 1     & 2.38   & 2.38  & 0.01 & 2.36  & 2.40    &        &      &      &       &        \\ \hline
Internet (CAIDA)                 & Powerlaw     & 5     & 2.70   & 2.70   & 0.02 & 2.67  & 2.73   &        &      &      &       &        \\ \hline
Munmun Digg                      & Weibull      & 1     & 0.48   & 0.47  & 0.03 & 0.41  & 0.54   & 0.37   & 0.37 & 0.01 & 0.36  & 0.38   \\ \hline
Linux                            & Lognormal    & 1     & 1.85   & 1.85  & 0.01 & 1.84  & 1.86   & 1.02   & 1.02 & 0.01 & 1.01  & 1.03   \\ \hline
Internet topology                & Powerlaw     & 5     & 2.39   & 2.40   & 0.01 & 2.37  & 2.42   &        &      &      &       &        \\ \hline
Facebook wall                    & Weibull      & 1     & 4.29   & 4.29  & 0.06 & 4.18  & 4.40    & 0.61   & 0.61 & 0.00    & 0.61  & 0.62   \\ \hline
Facebook                         & Weibull      & 1     & 14.00  & 14.01 & 0.15 & 13.72 & 14.30   & 0.58   & 0.58 & 0.00    & 0.57  & 0.58   \\ \hline
Slashdot                         & Weibull      & 1     & 0.05   & 2.34  & 4.67 & 0.05  & 12.02  & 0.23   & 1.54 & 1.66 & 0.18  & 4.00      \\ \hline
Enron                            & Weibull      & 3     & 0.05   & 0.05  & 0.00    & 0.05  & 0.05   & 0.22   & 0.26 & 0.39 & 0.22  & 0.22   \\ \hline
\end{tabular}
\end{table*}
\begin{figure*}[t!]
\begin{center}
\includegraphics[width=0.24\textwidth]{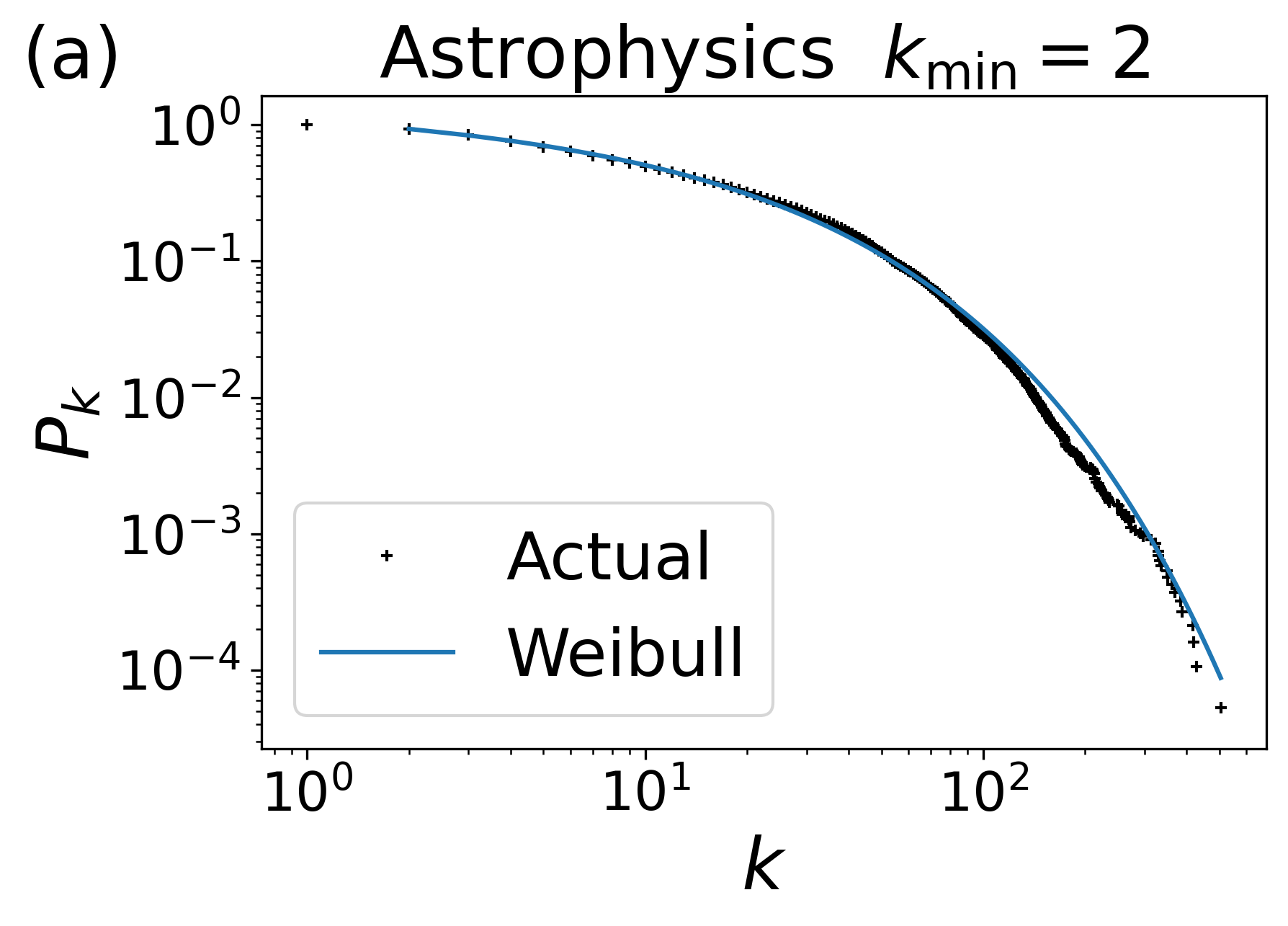}
\includegraphics[width=0.24\textwidth]{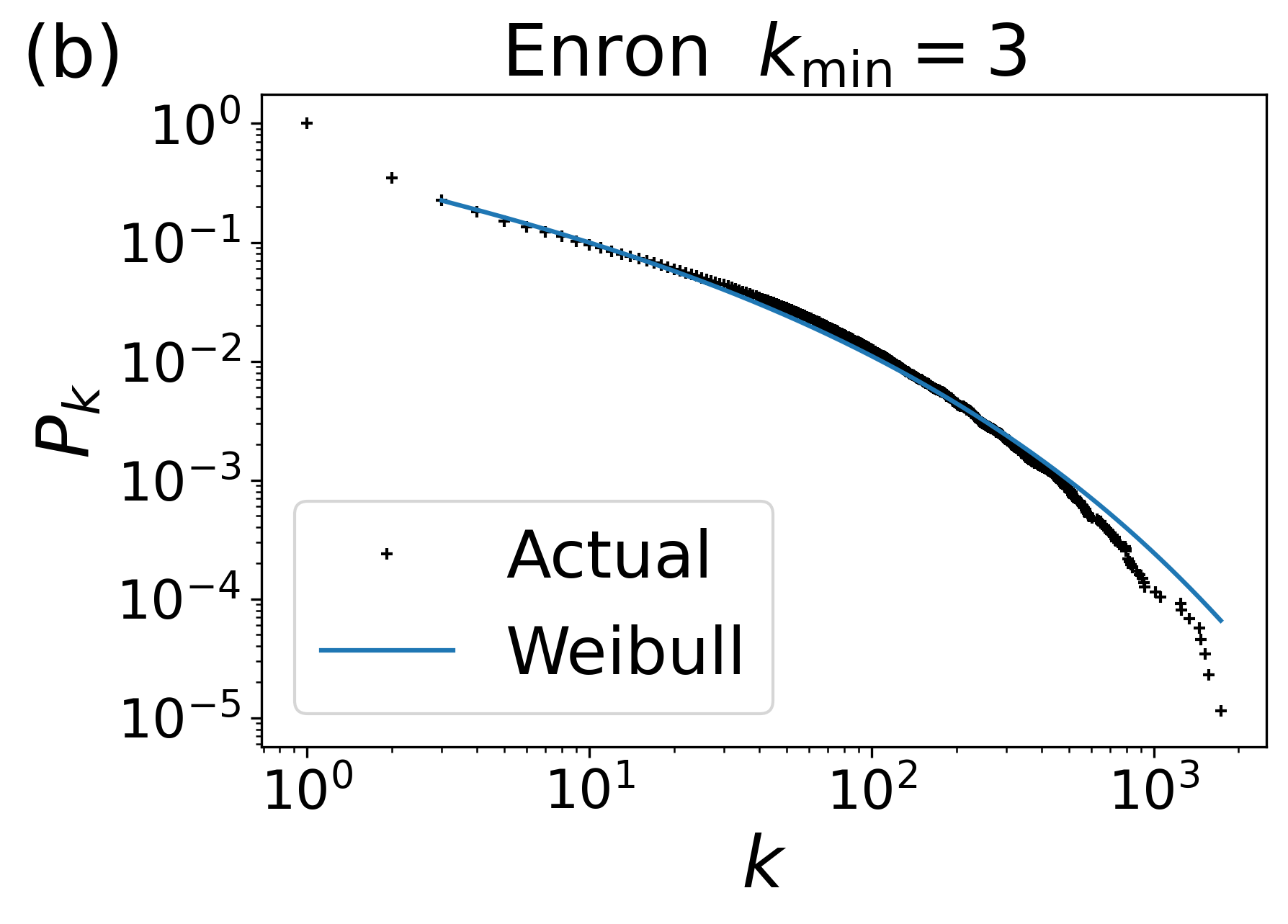}
\includegraphics[width=0.24\textwidth]{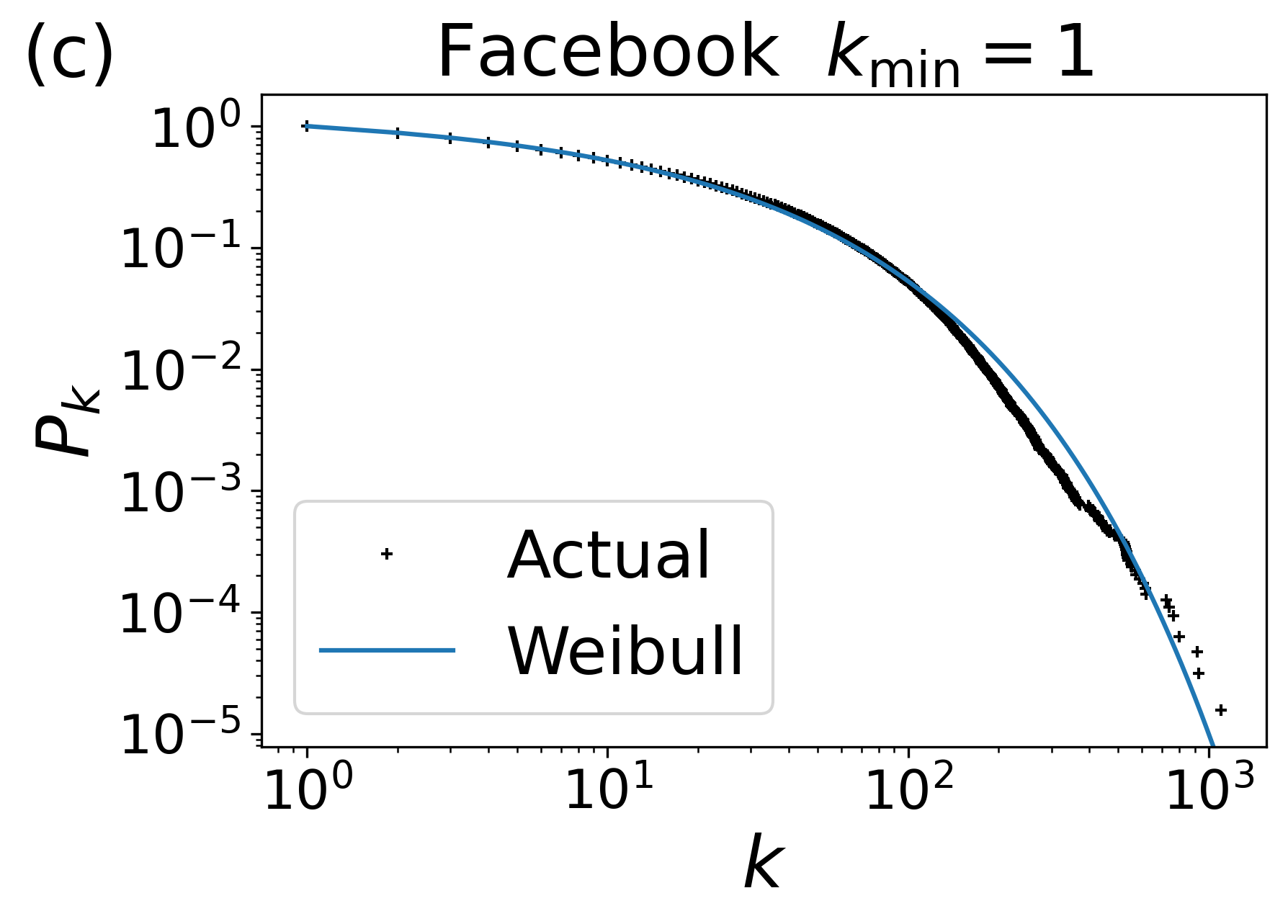}
\includegraphics[width=0.24\textwidth]{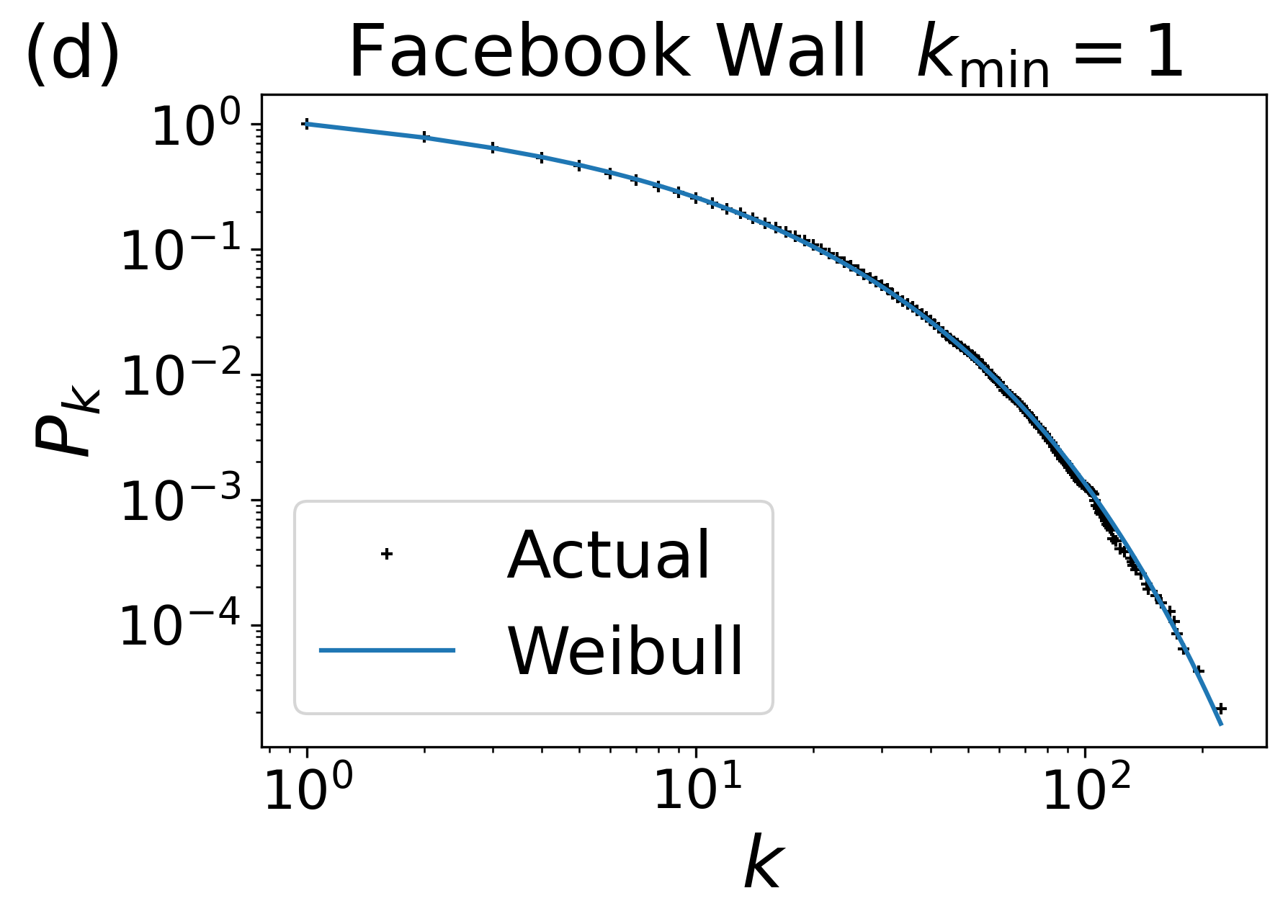}
\includegraphics[width=0.24\textwidth]{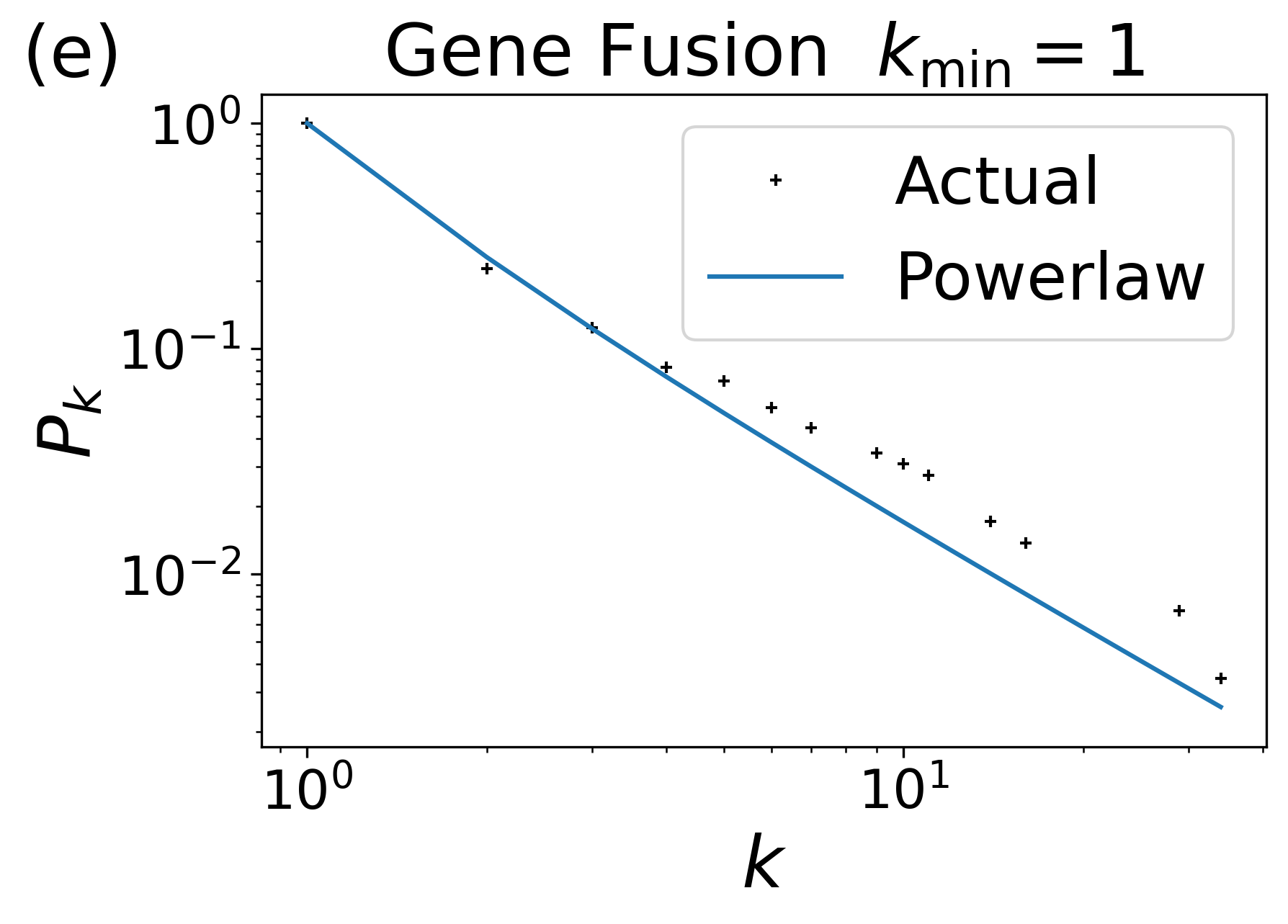}
\includegraphics[width=0.24\textwidth]{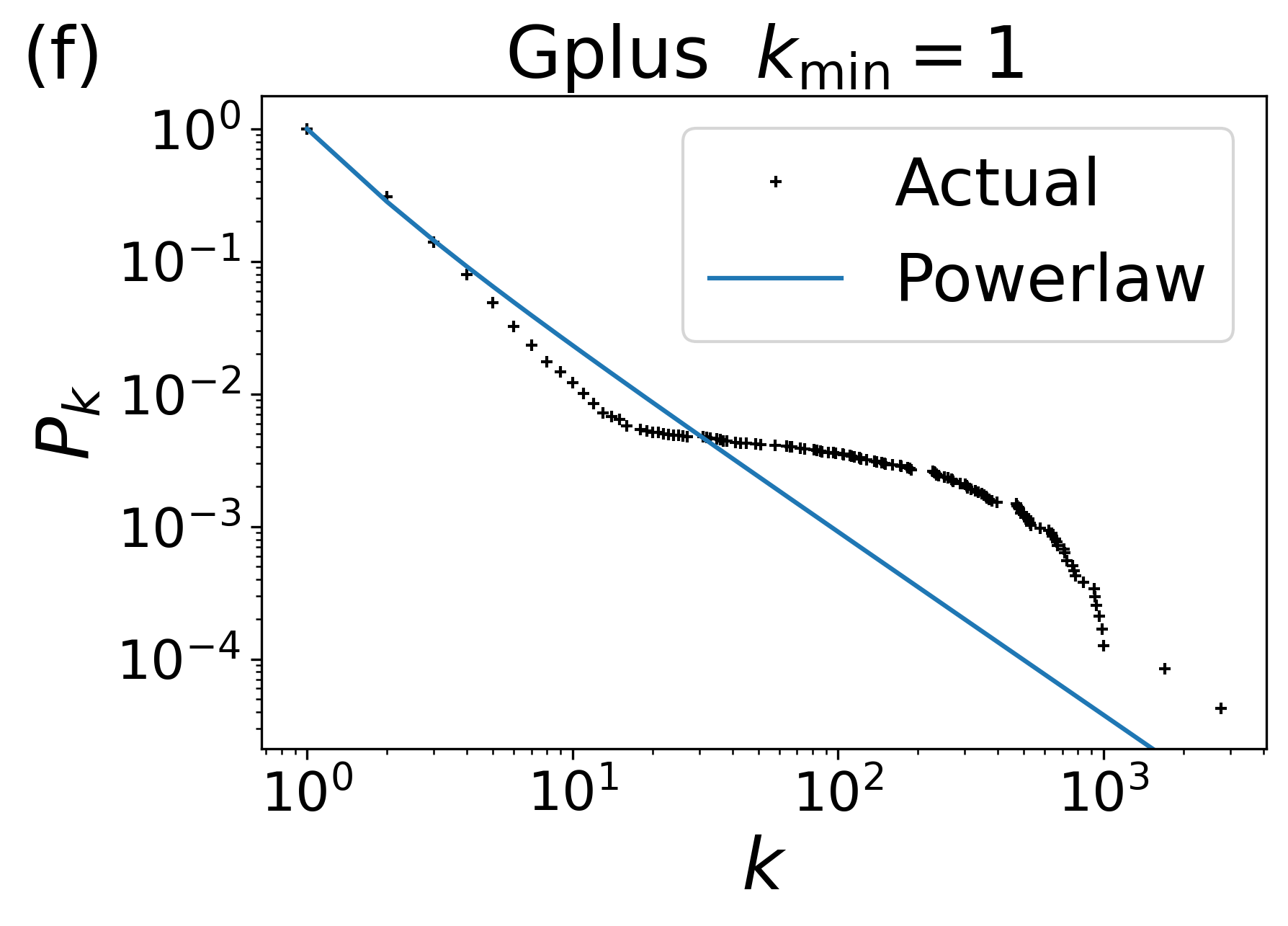}
\includegraphics[width=0.24\textwidth]{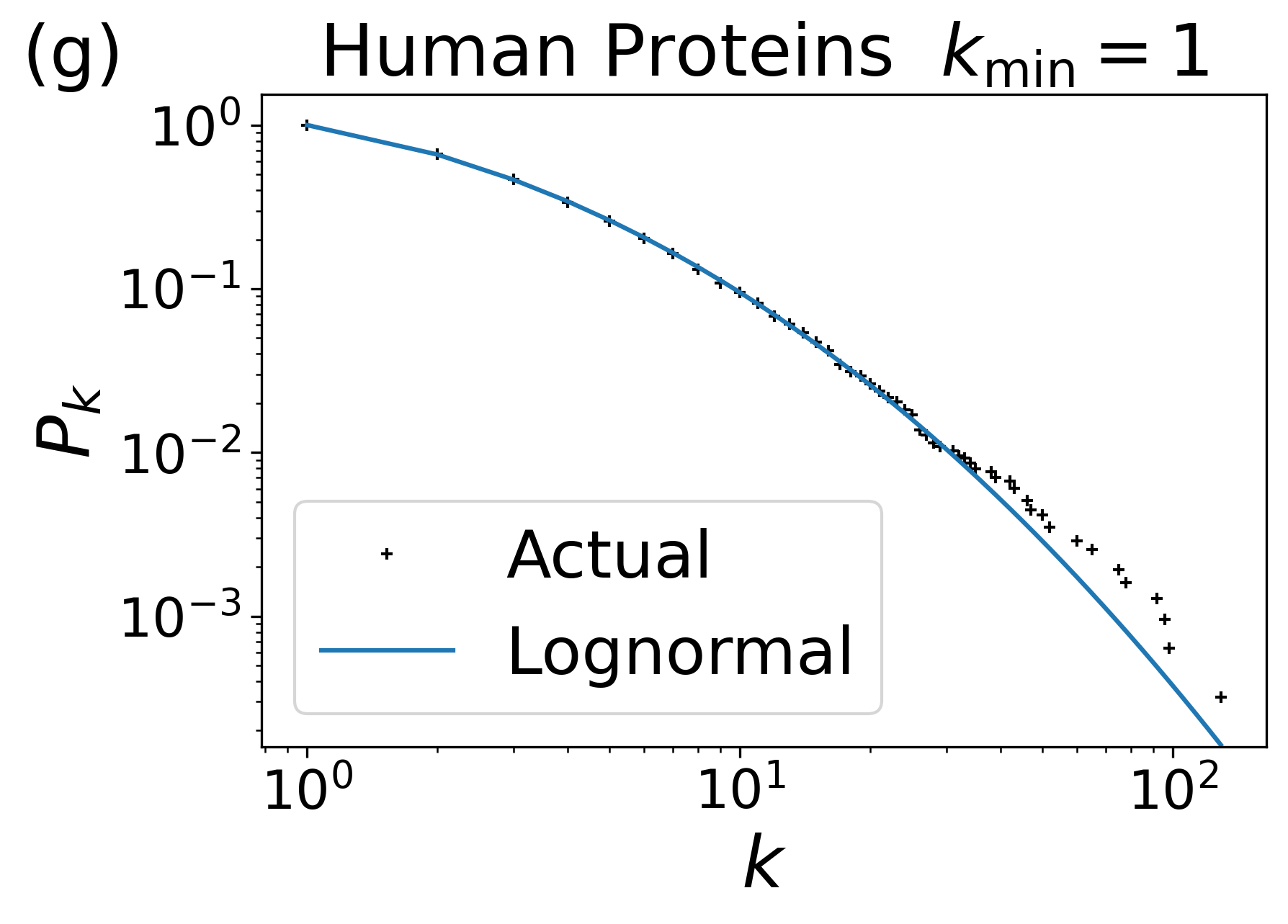}
\includegraphics[width=0.24\textwidth]{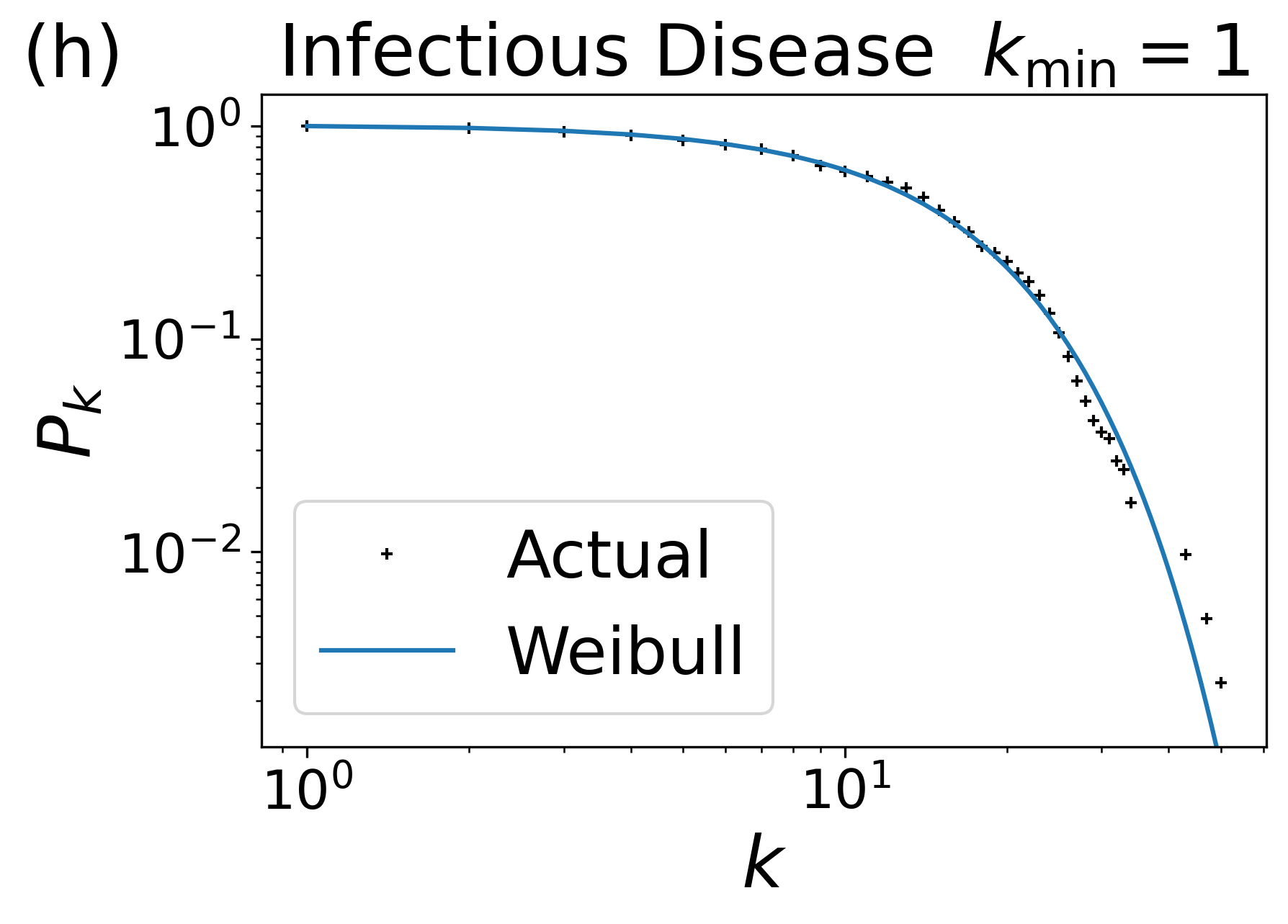}
\includegraphics[width=0.24\textwidth]{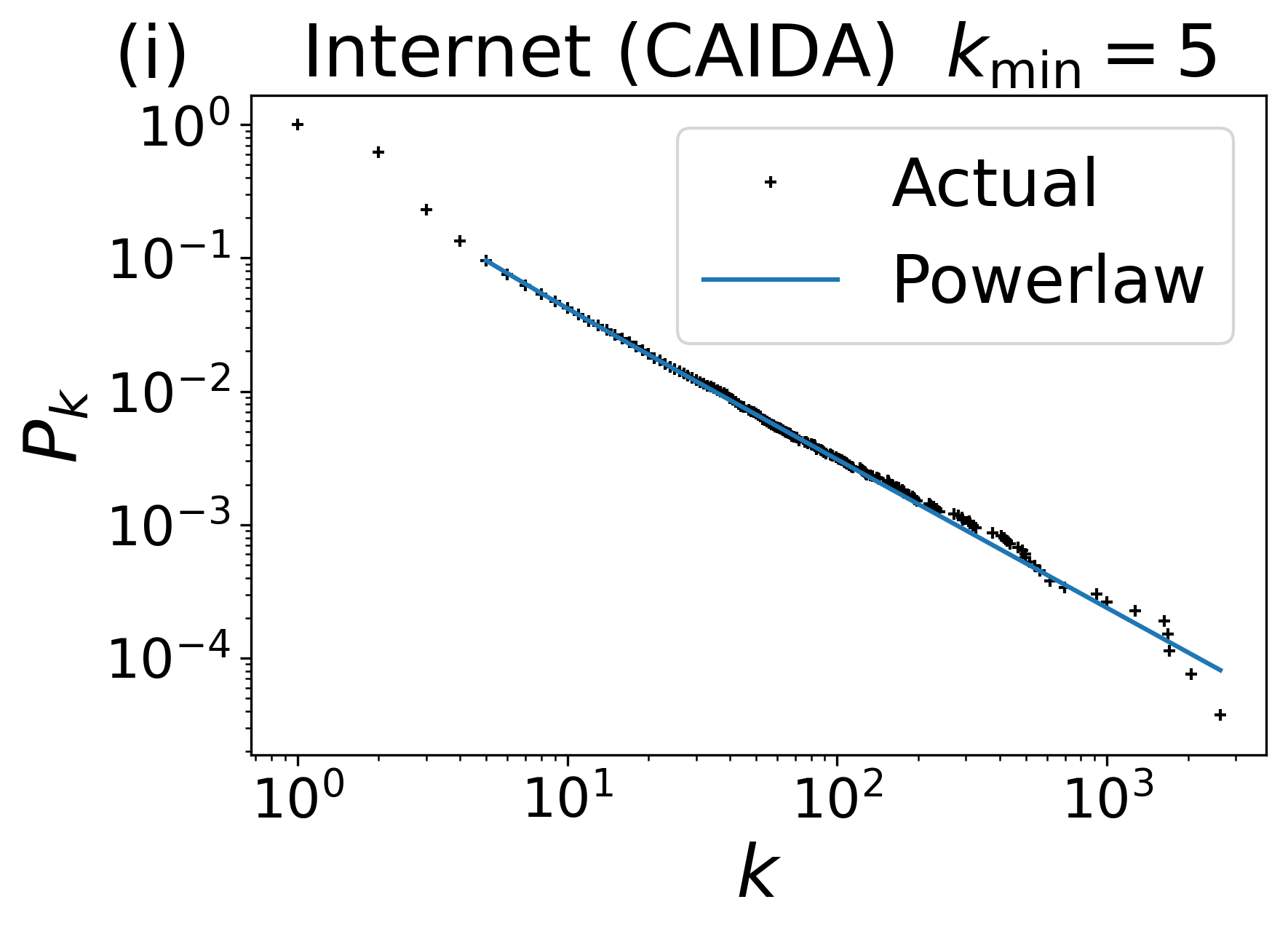}
\includegraphics[width=0.24\textwidth]{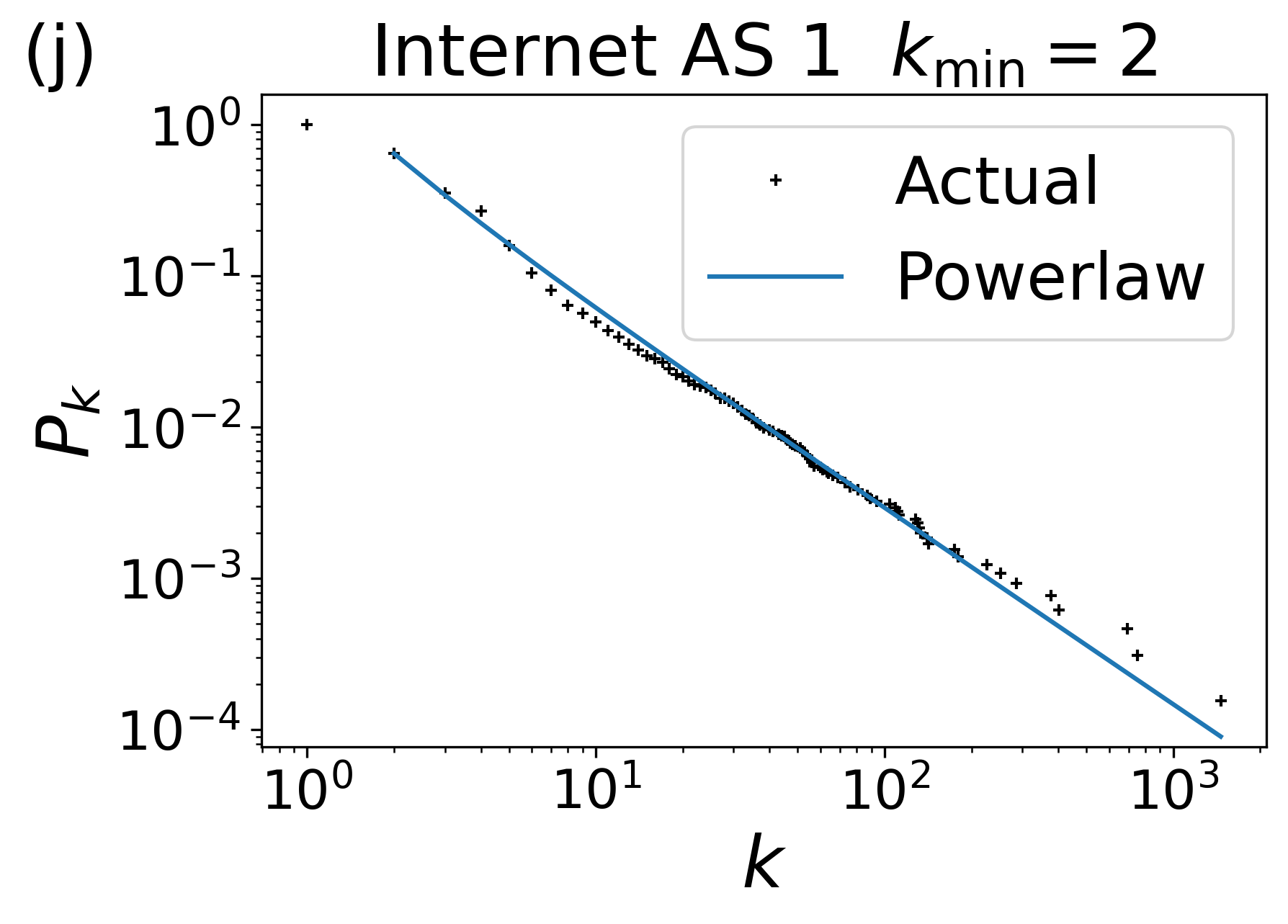}
\includegraphics[width=0.24\textwidth]{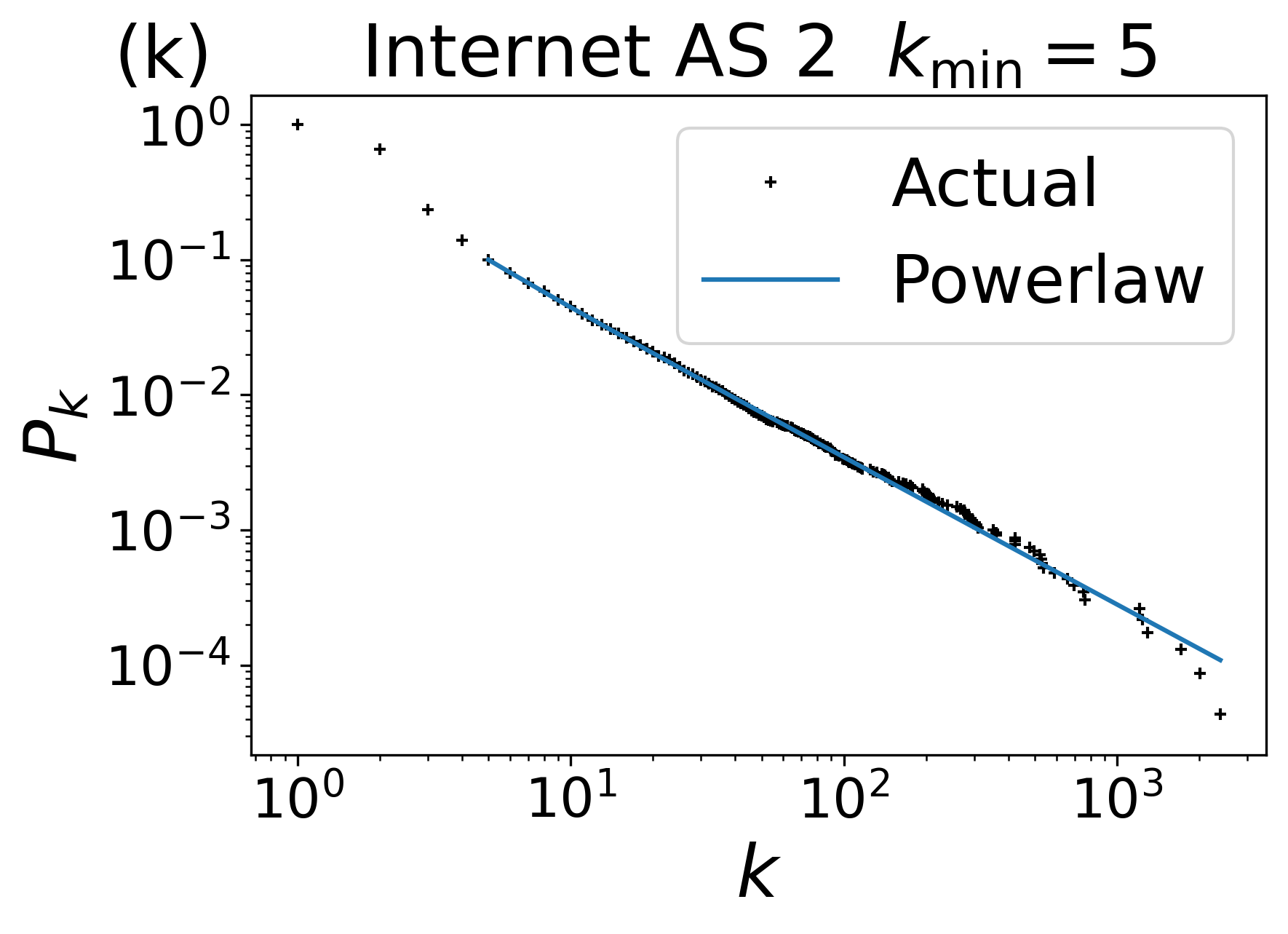}
\includegraphics[width=0.24\textwidth]{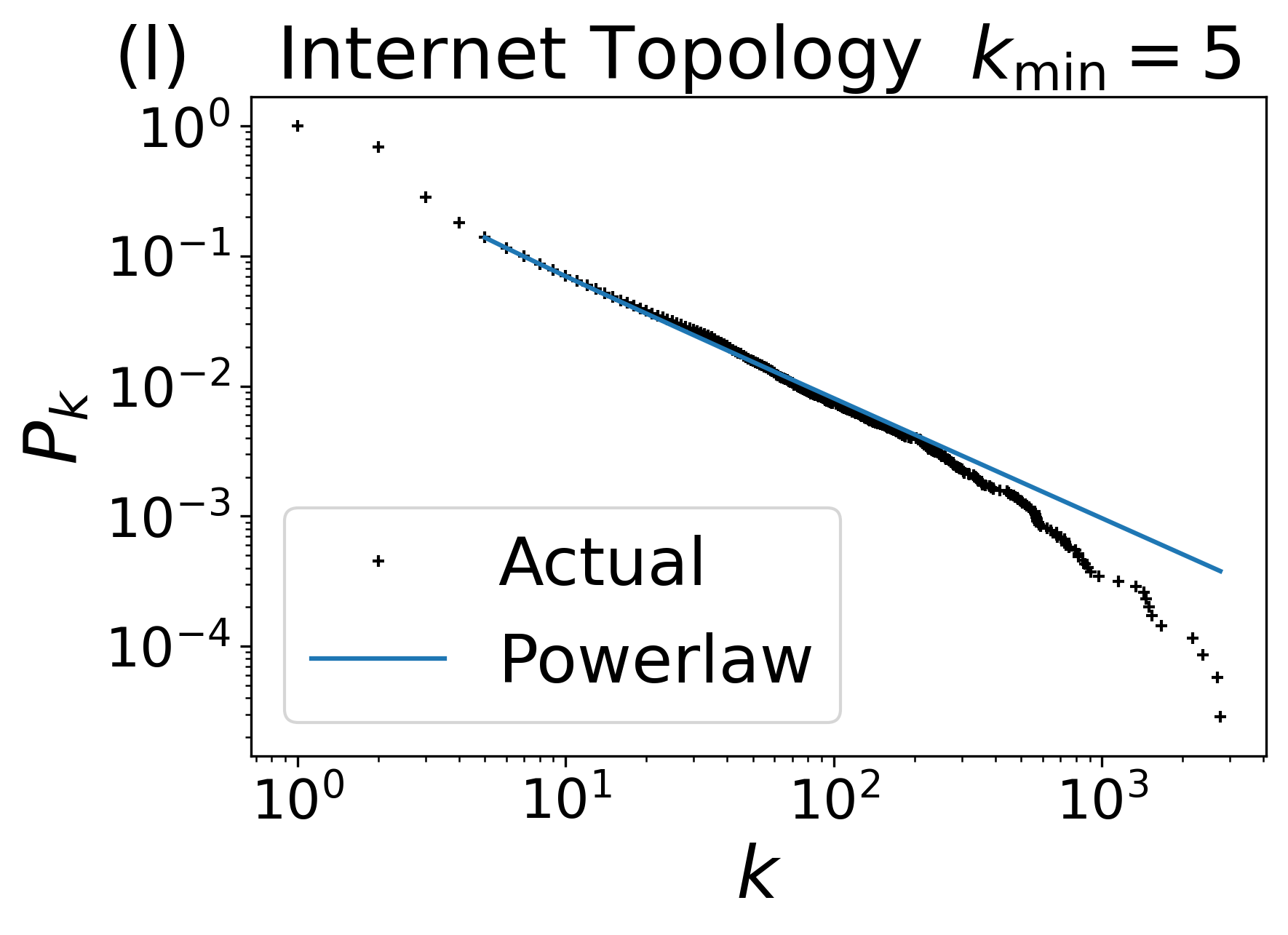}
\includegraphics[width=0.24\textwidth]{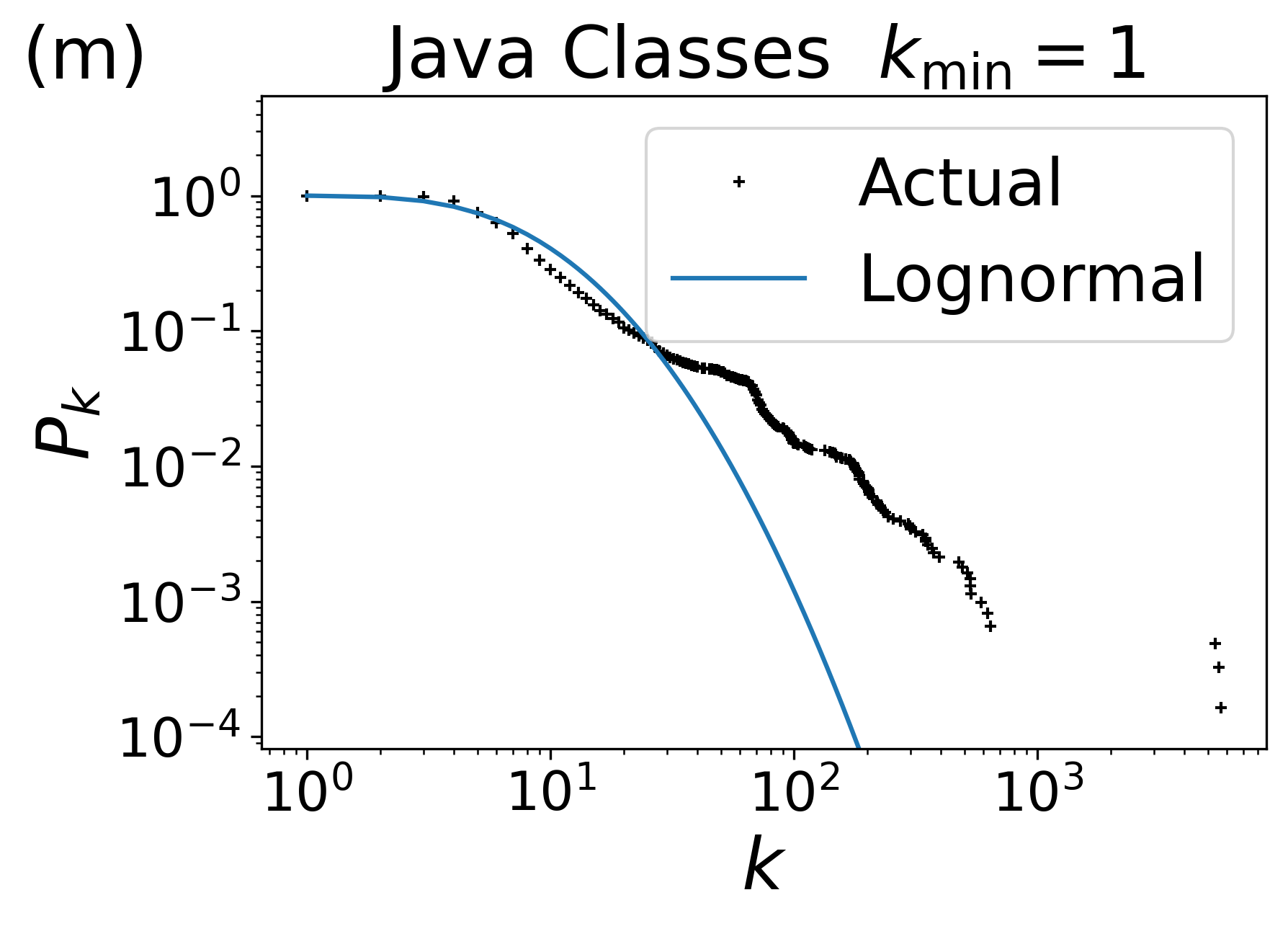}
\includegraphics[width=0.24\textwidth]{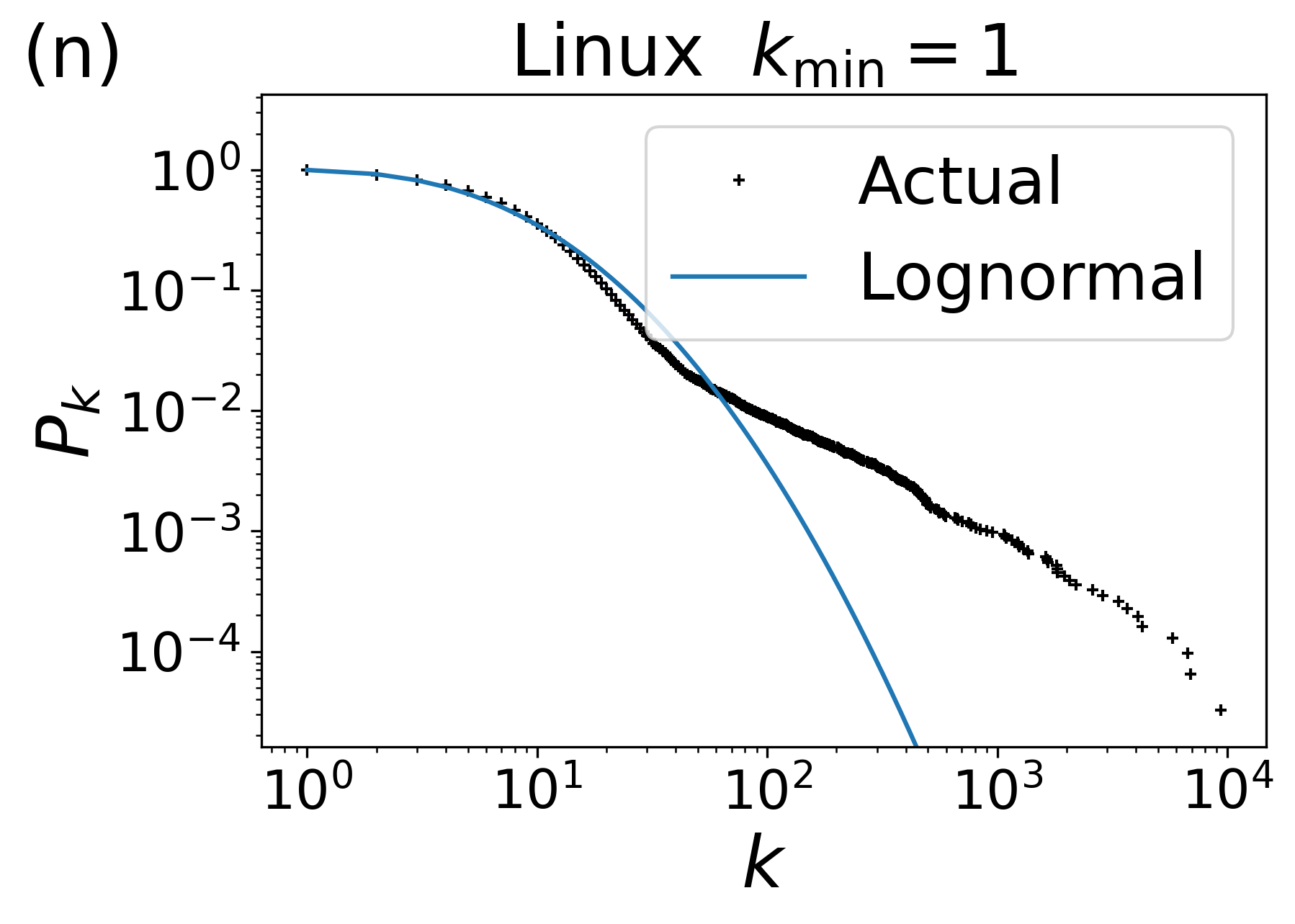}
\includegraphics[width=0.24\textwidth]{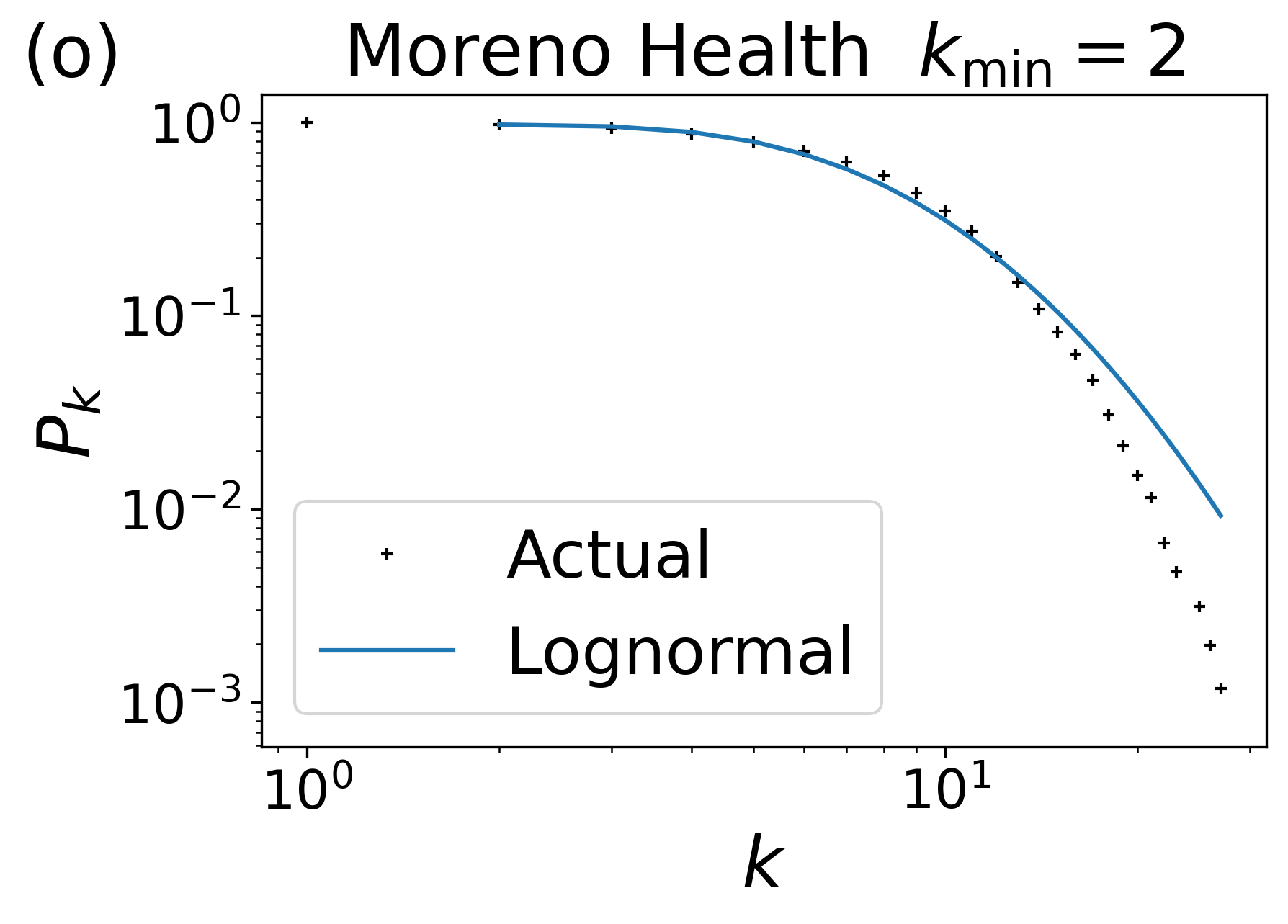}
\includegraphics[width=0.24\textwidth]{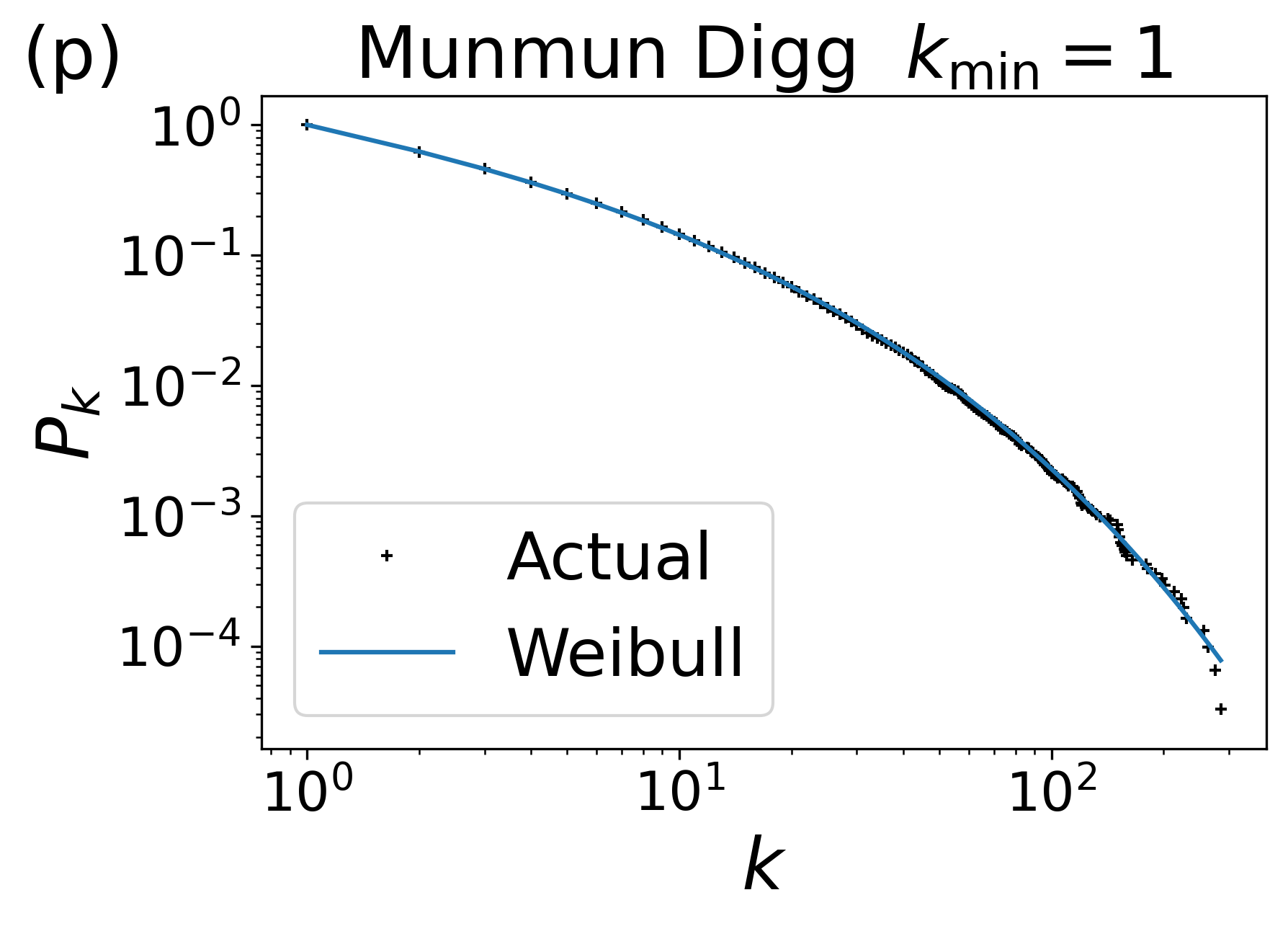}
\includegraphics[width=0.24\textwidth]{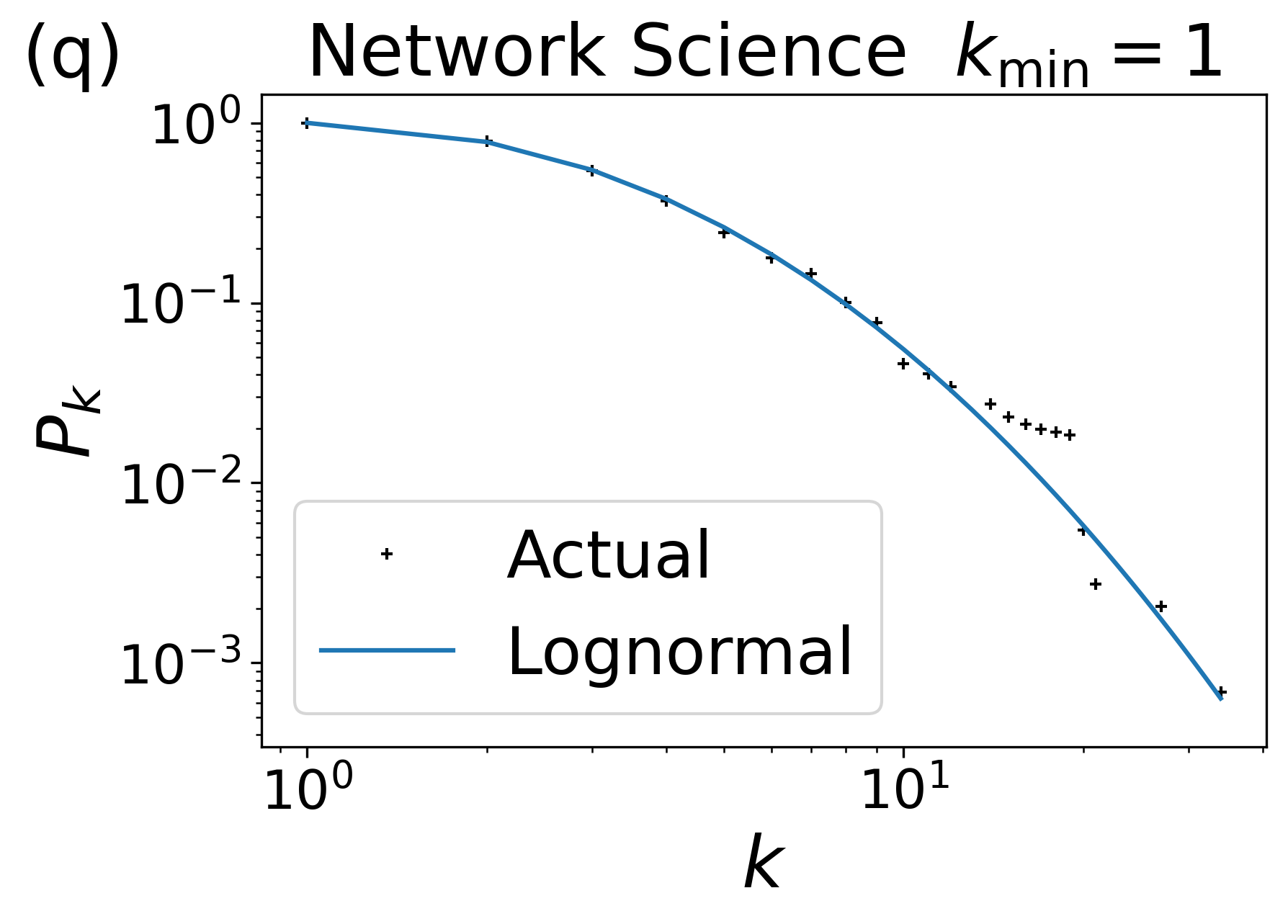}
\includegraphics[width=0.24\textwidth]{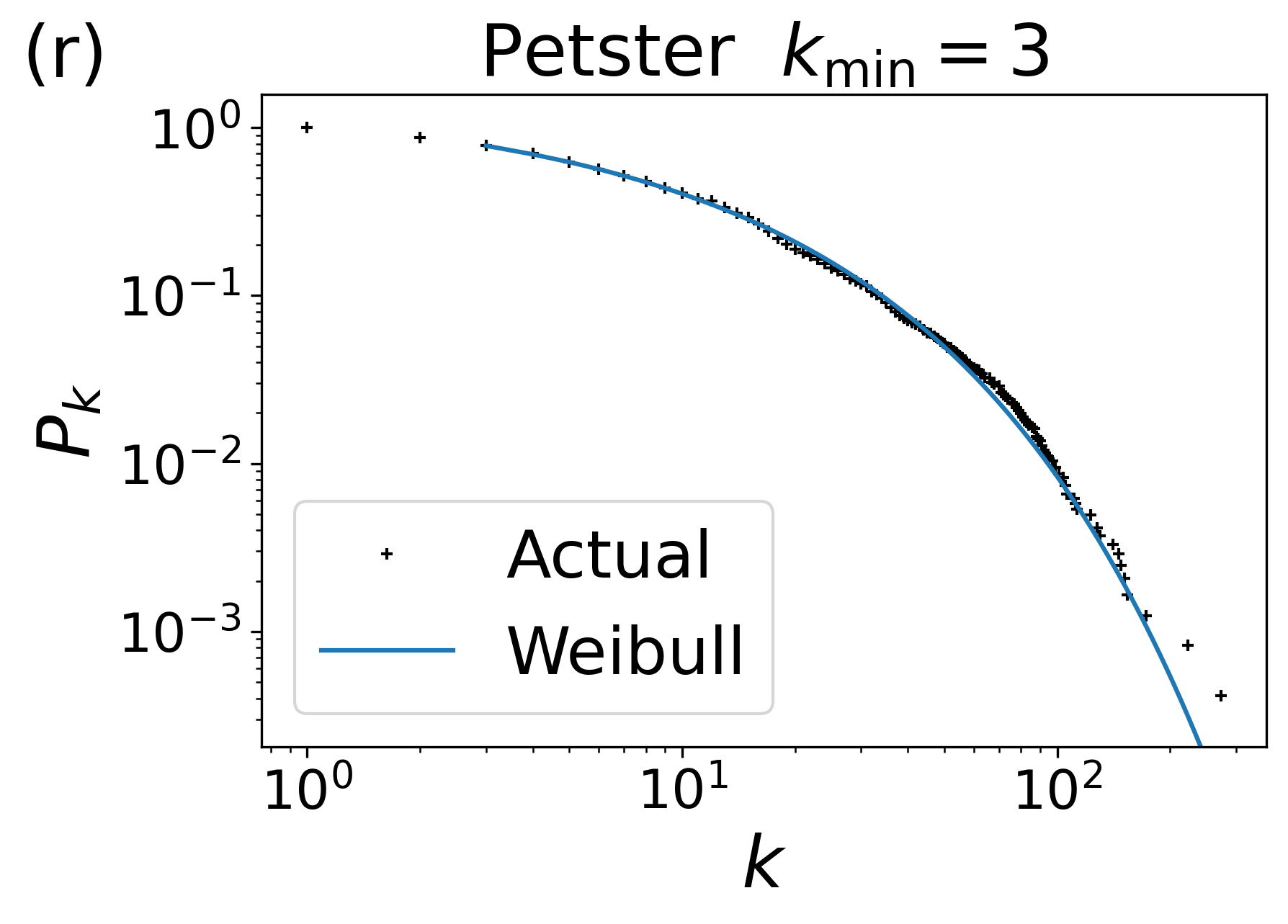}
\includegraphics[width=0.24\textwidth]{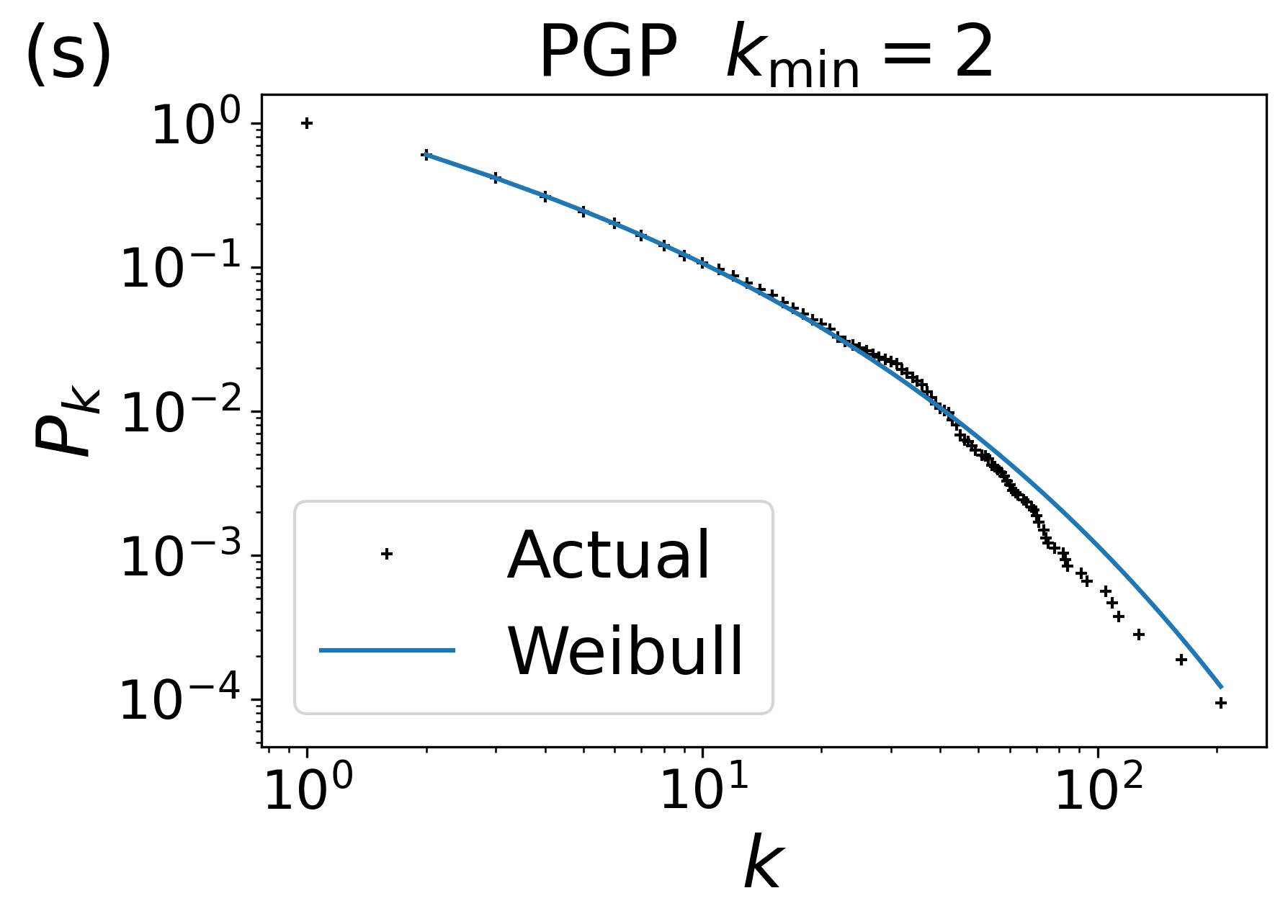}
\includegraphics[width=0.24\textwidth]{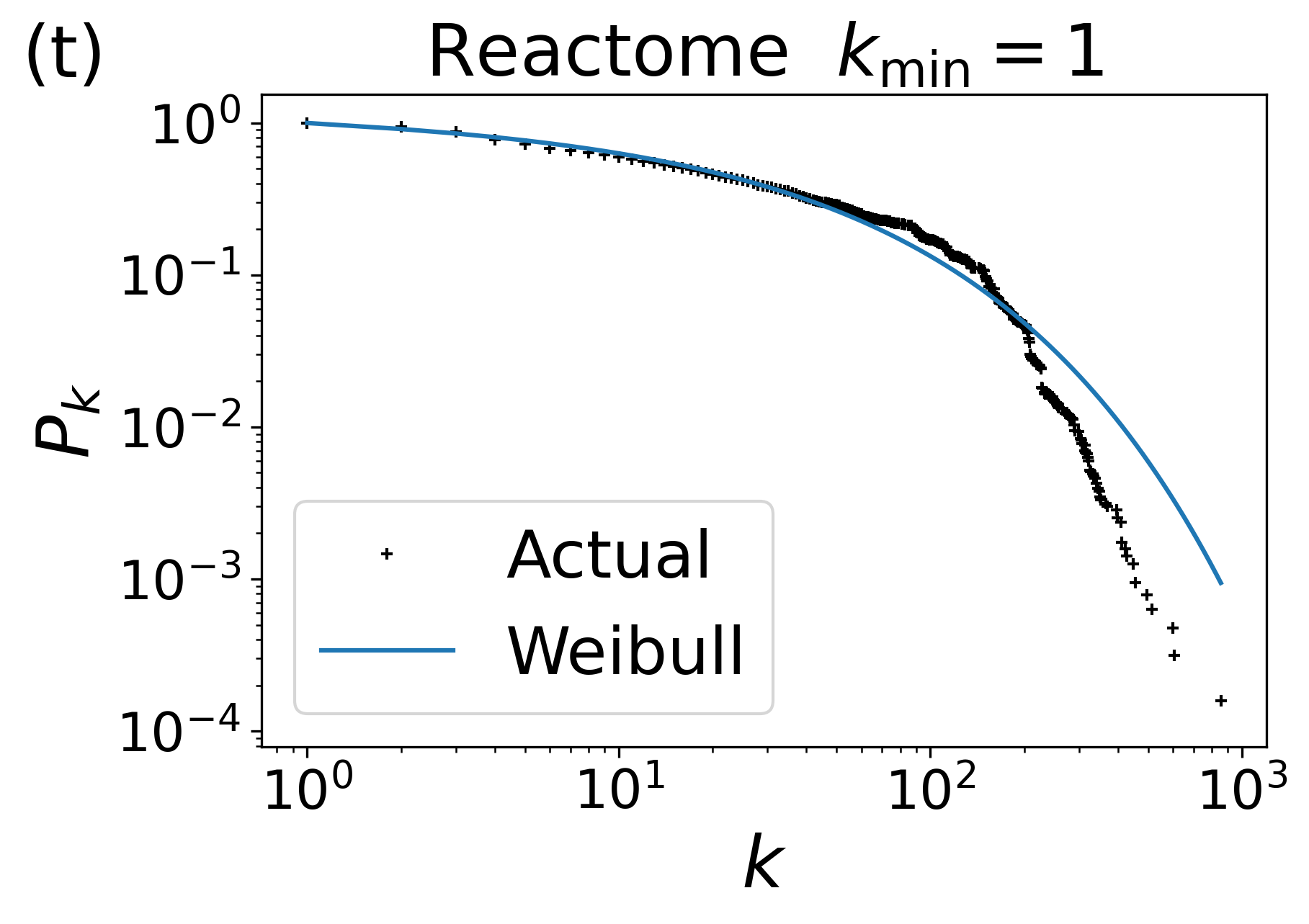}
\includegraphics[width=0.24\textwidth]{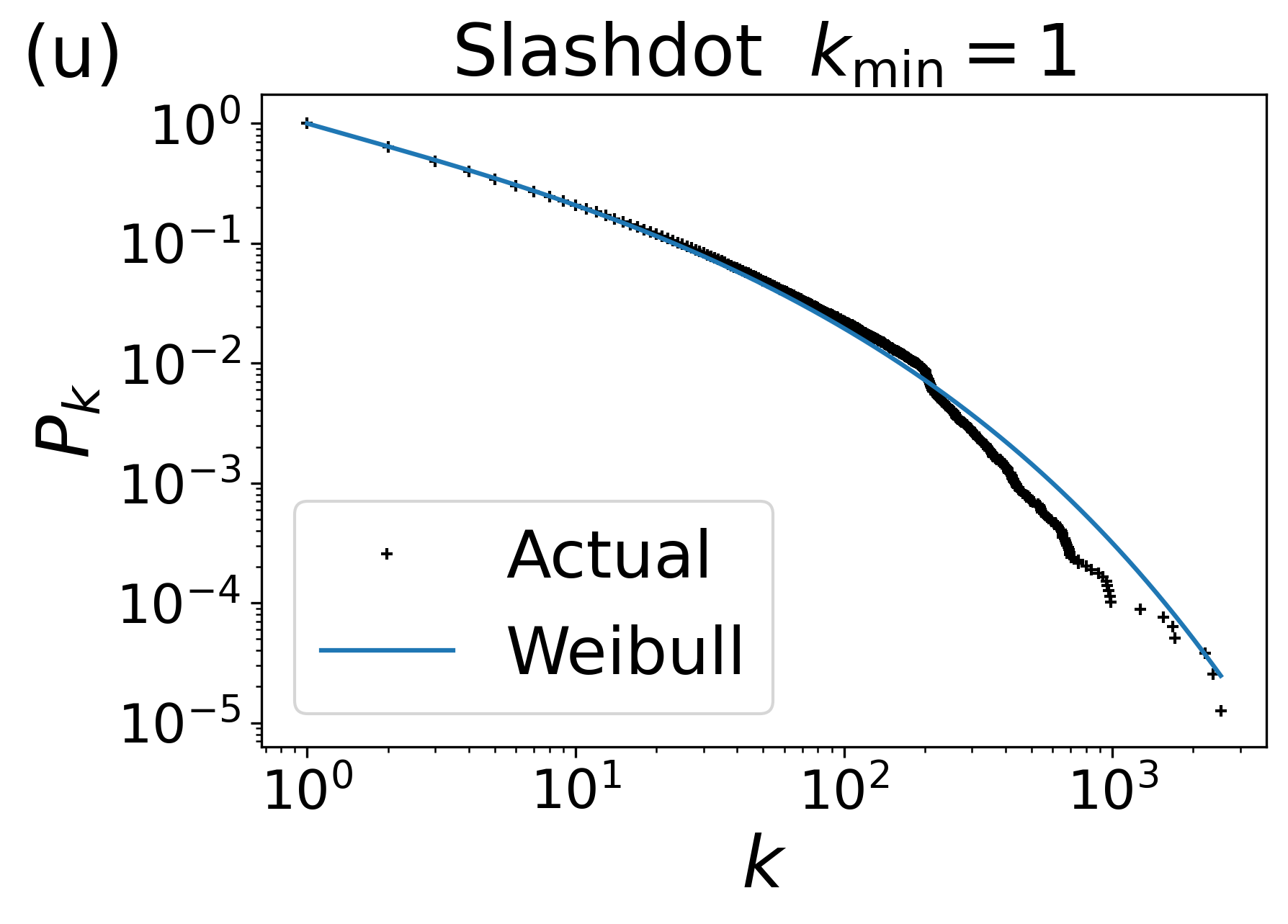}
\includegraphics[width=0.24\textwidth]{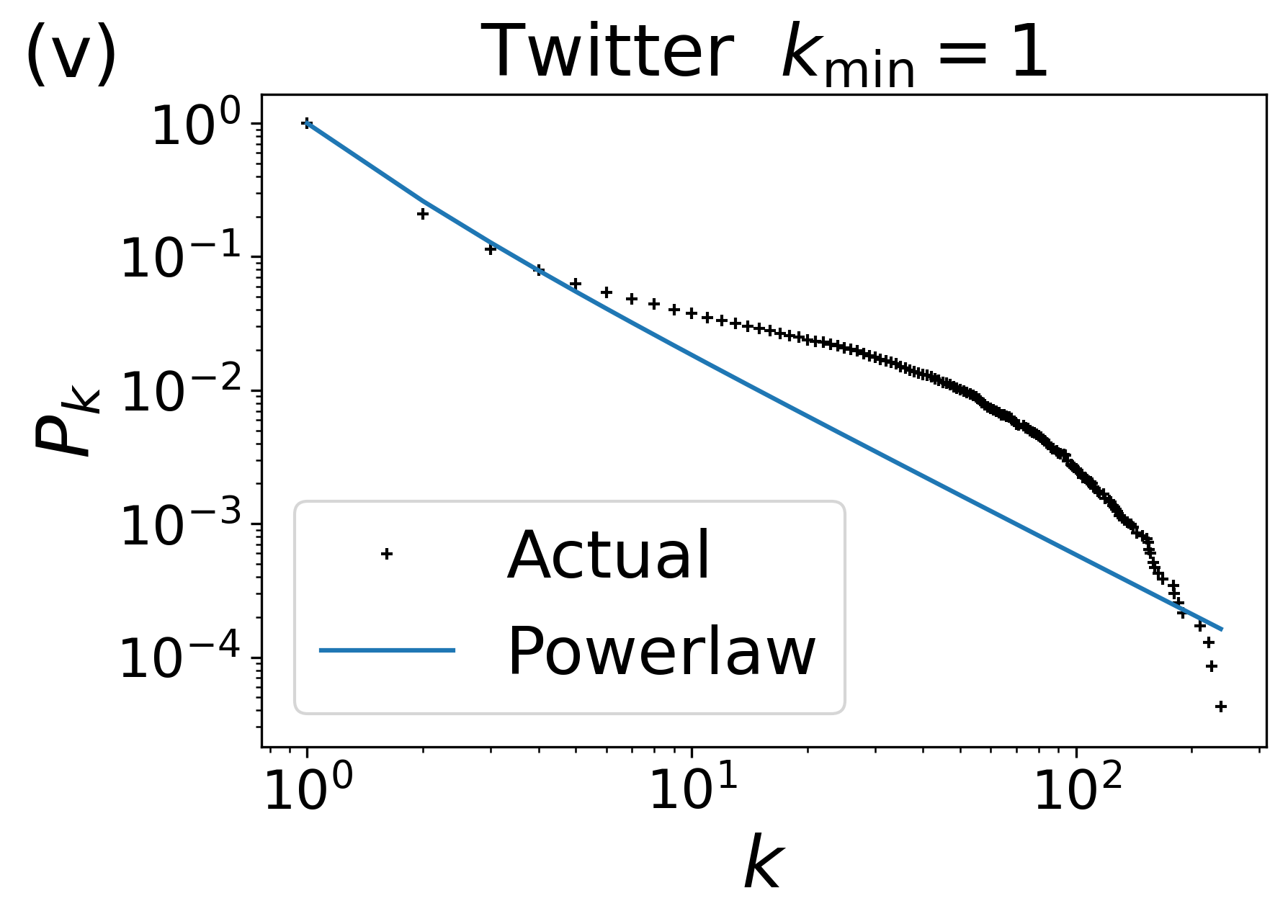}
\includegraphics[width=0.24\textwidth]{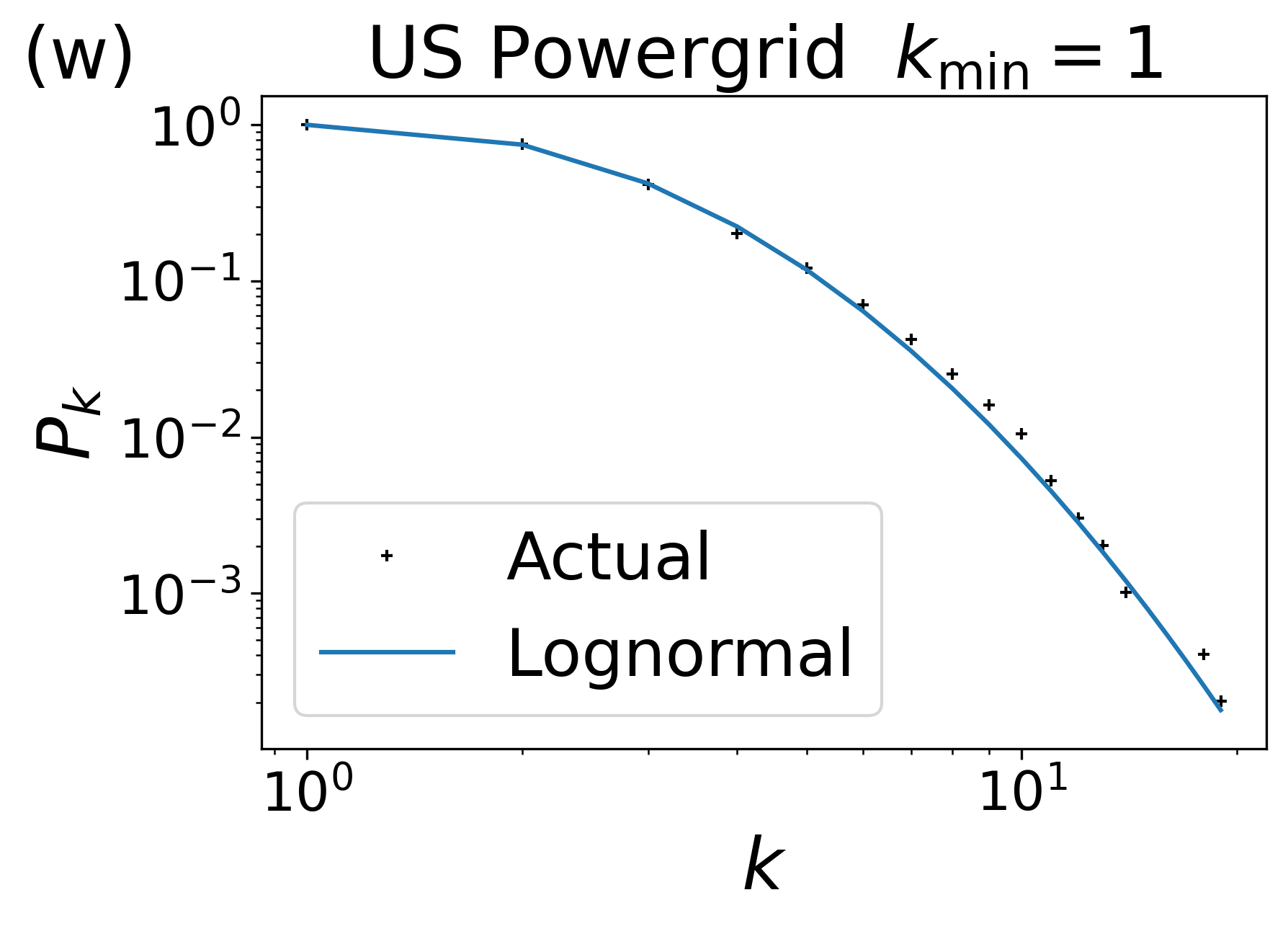}
\includegraphics[width=0.24\textwidth]{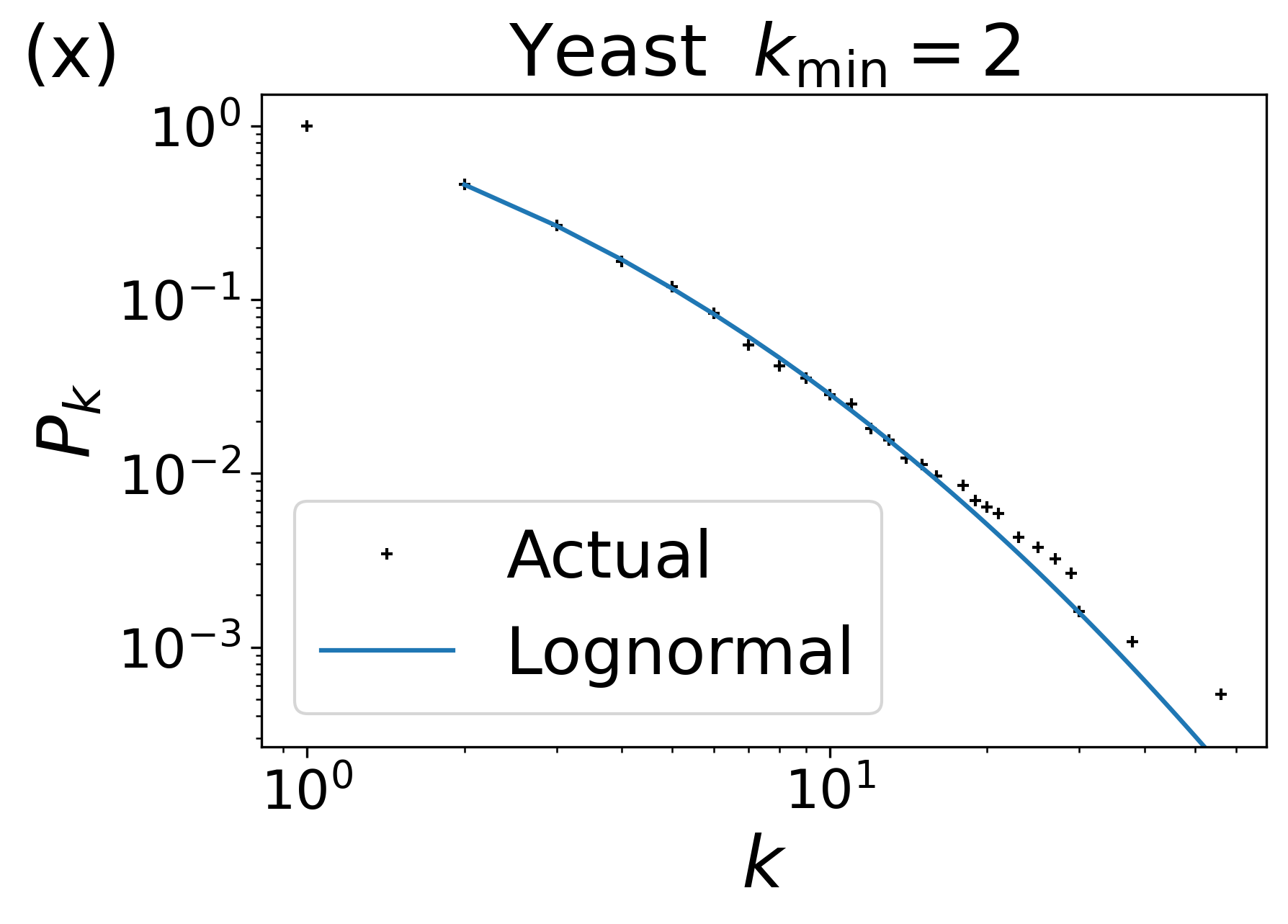}
\includegraphics[width=0.24\textwidth]{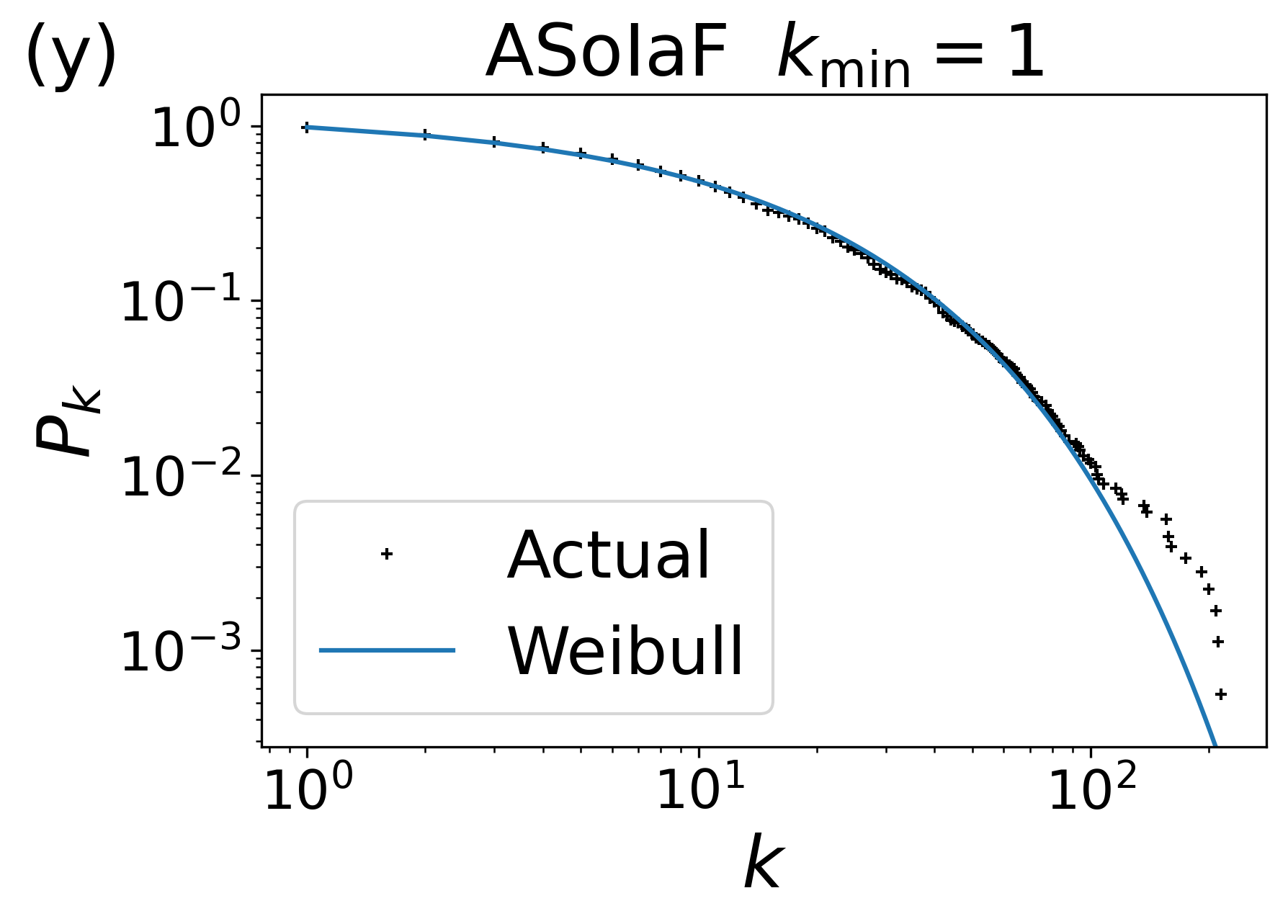}
\includegraphics[width=0.24\textwidth]{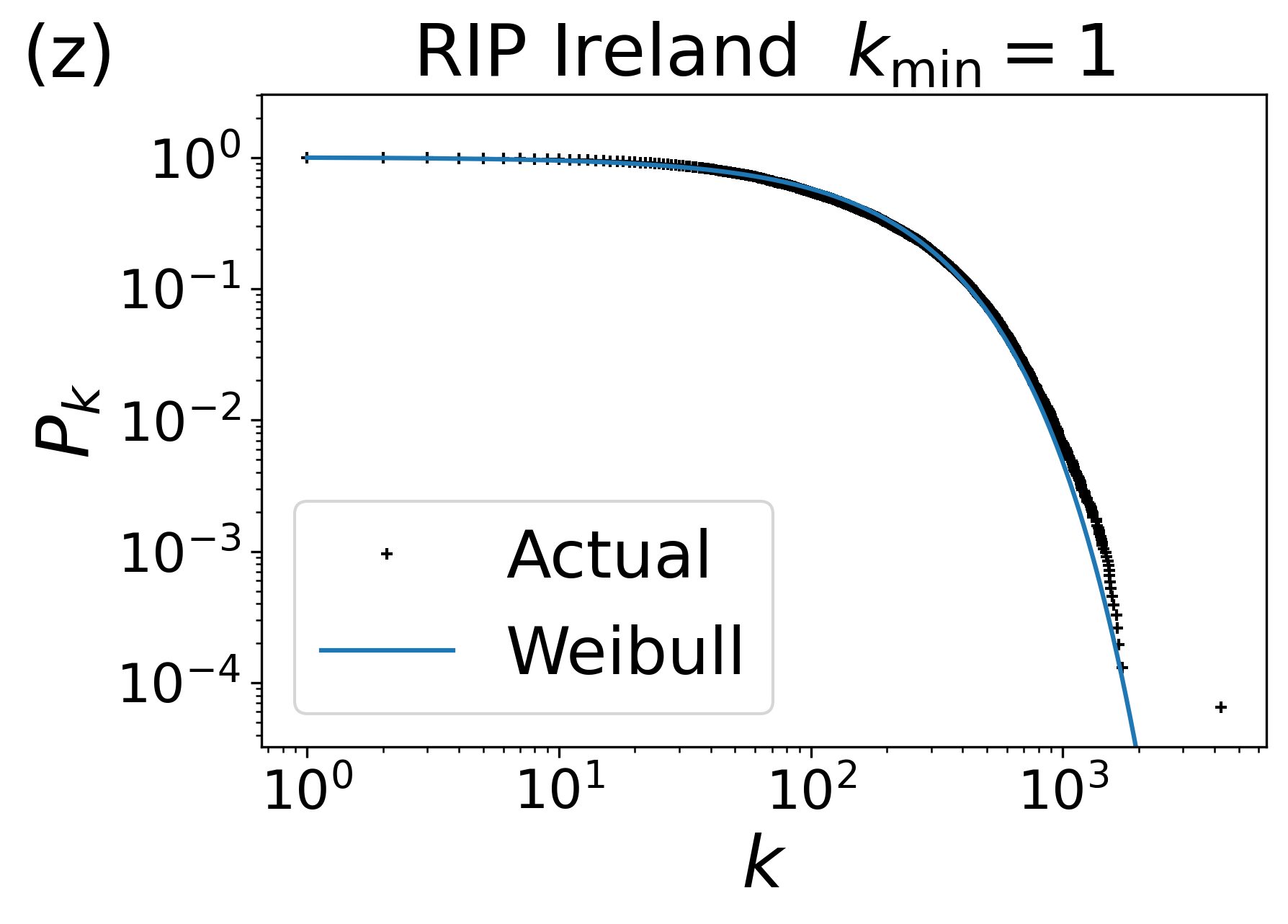}
\includegraphics[width=0.24\textwidth]{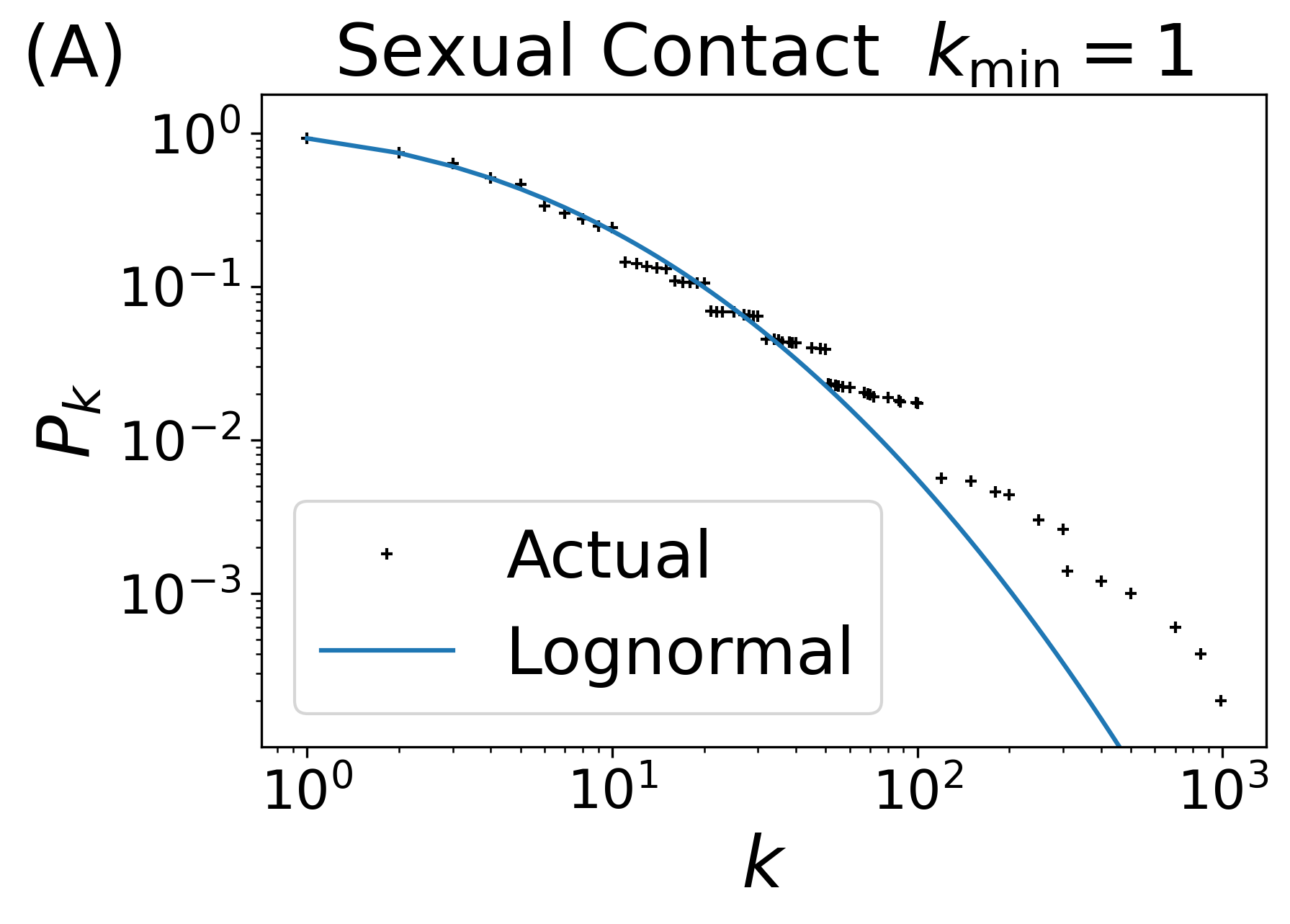}
\includegraphics[width=0.24\textwidth]{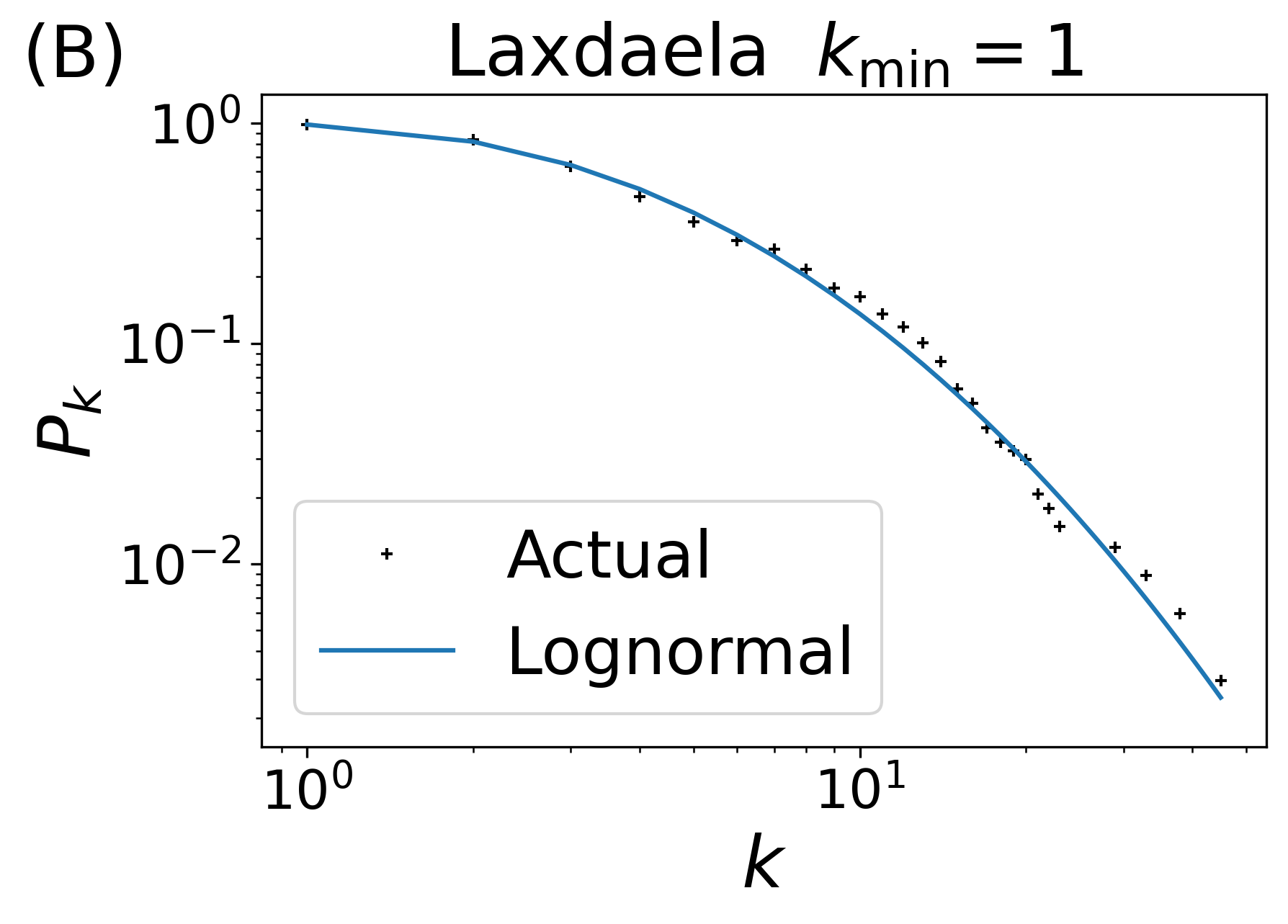}
\caption{Empirical and fitted CCDFs for all studied networks listed in table~\ref{tab:networkdata}. When assessing fit both PDFs and CCDFs should be considered.}
\label{fig:all_ccdfs}
\end{center}
\end{figure*}
\begin{figure*}[t!]
\begin{center}
\includegraphics[width=0.24\textwidth]{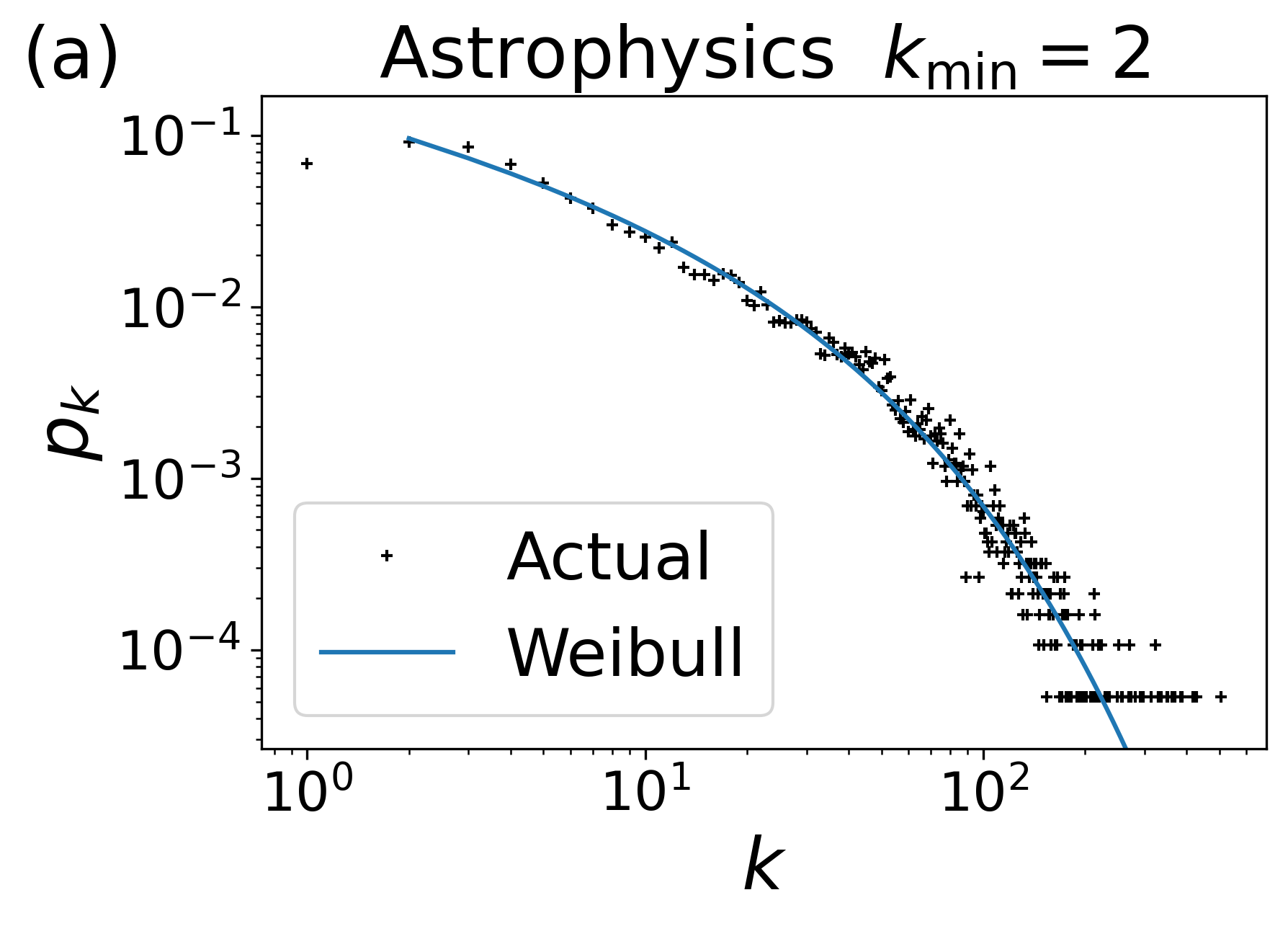}
\includegraphics[width=0.24\textwidth]{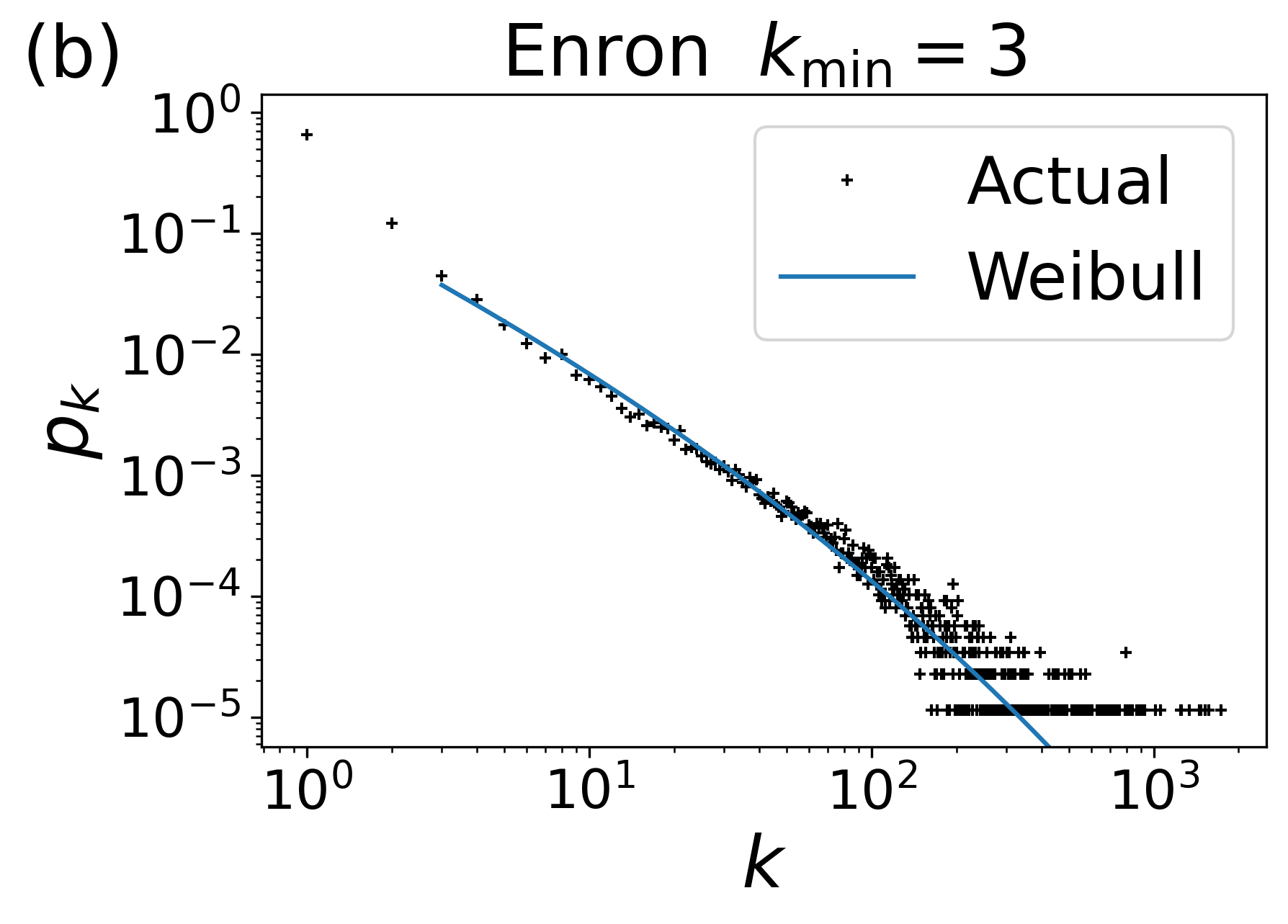}
\includegraphics[width=0.24\textwidth]{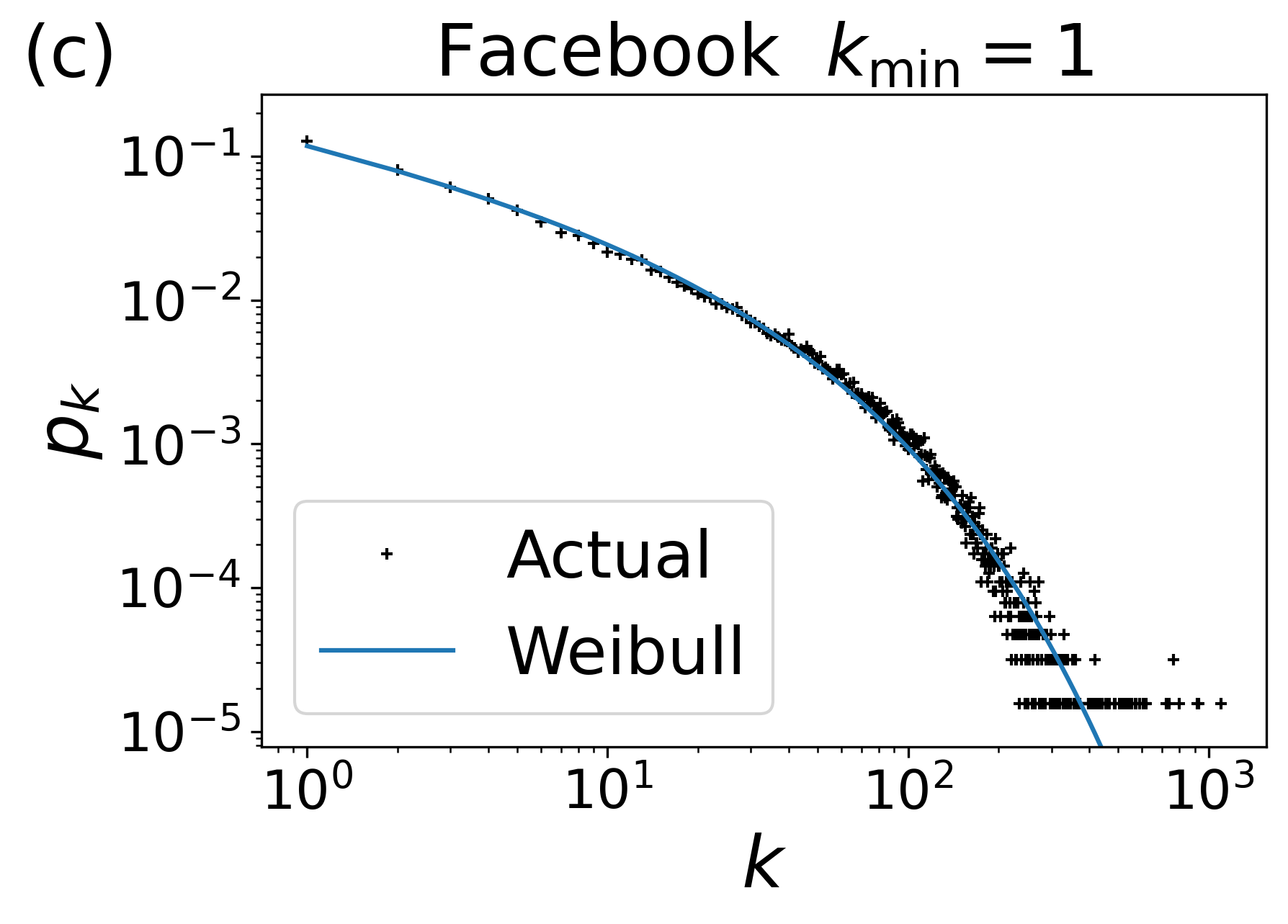}
\includegraphics[width=0.24\textwidth]{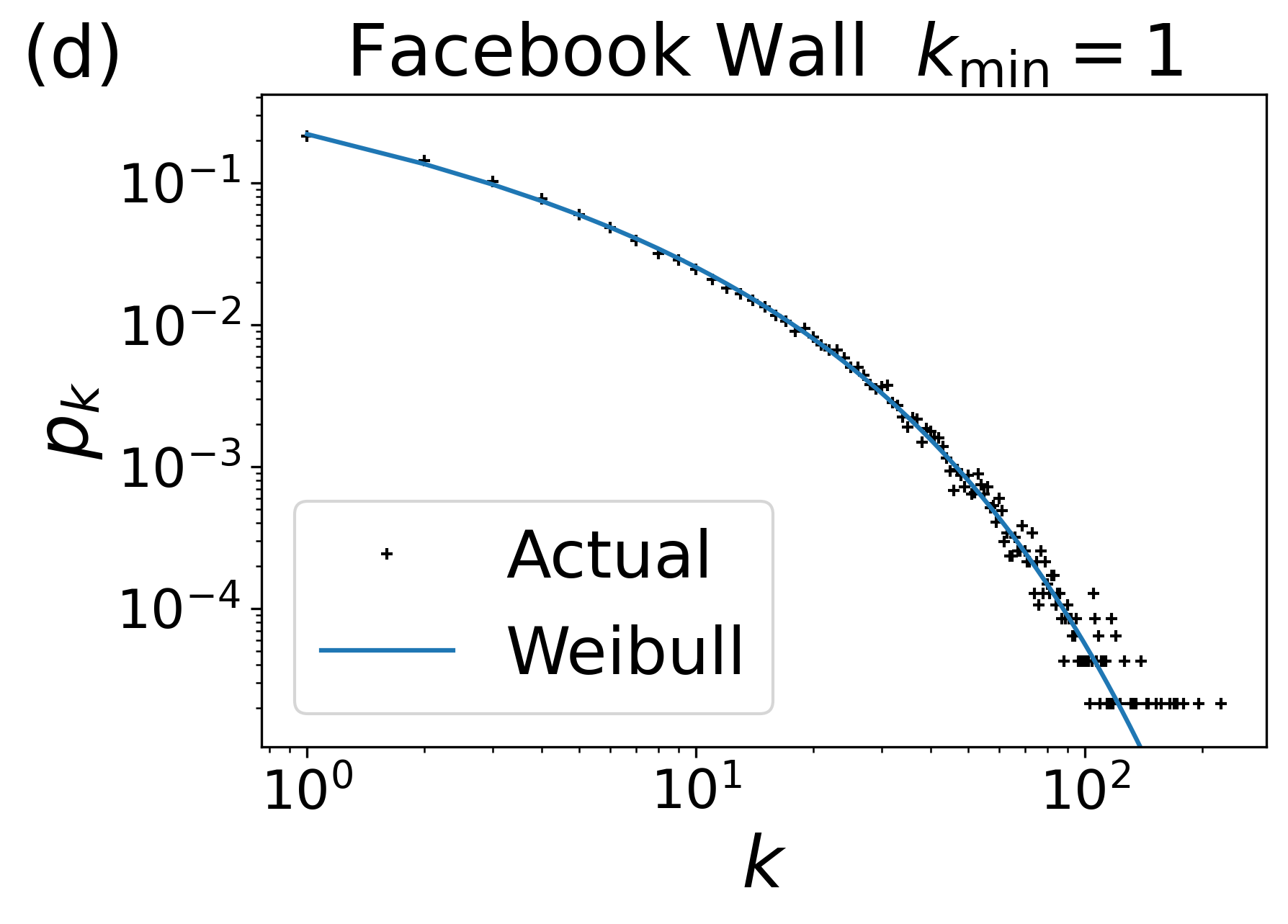}
\includegraphics[width=0.24\textwidth]{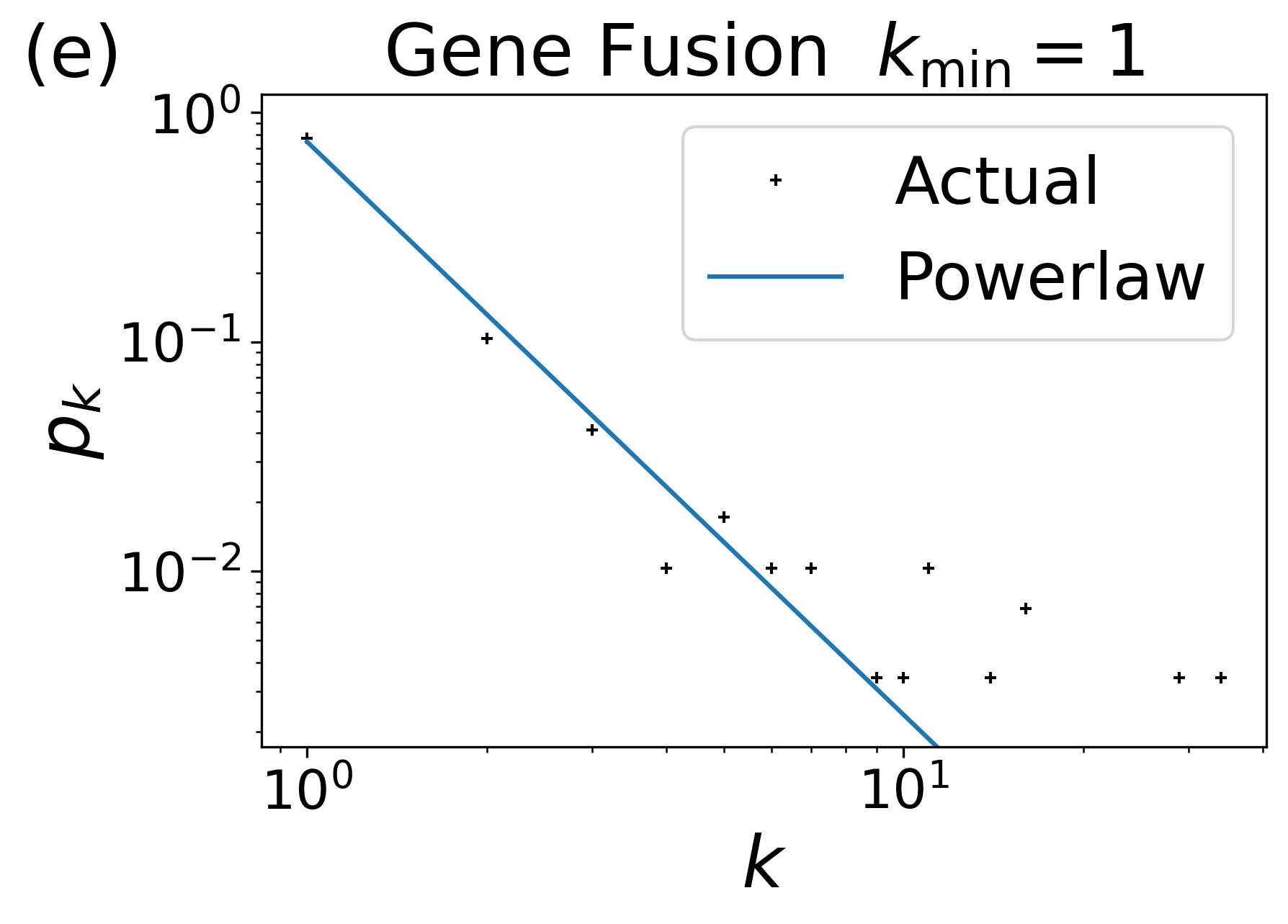}
\includegraphics[width=0.24\textwidth]{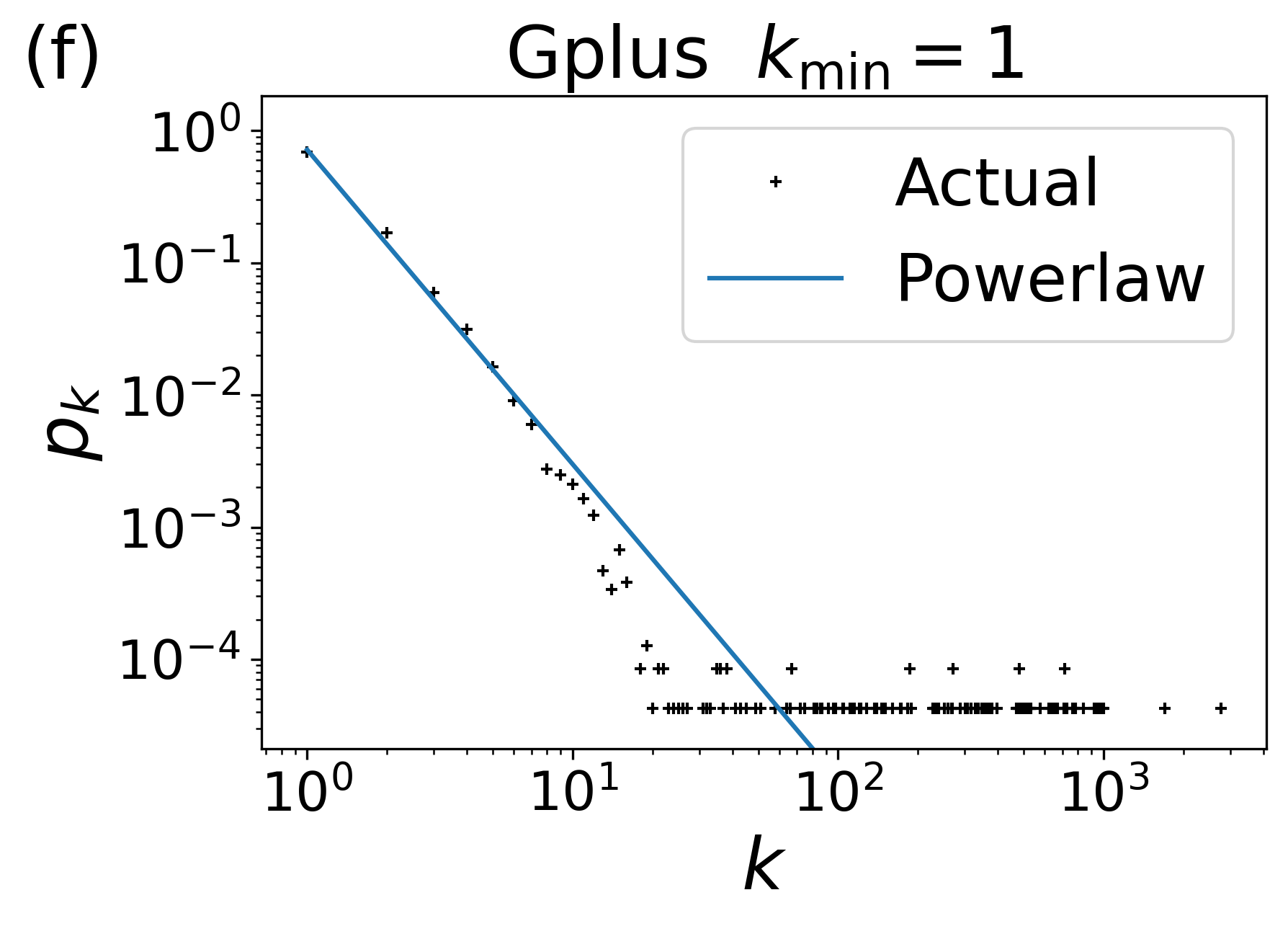}
\includegraphics[width=0.24\textwidth]{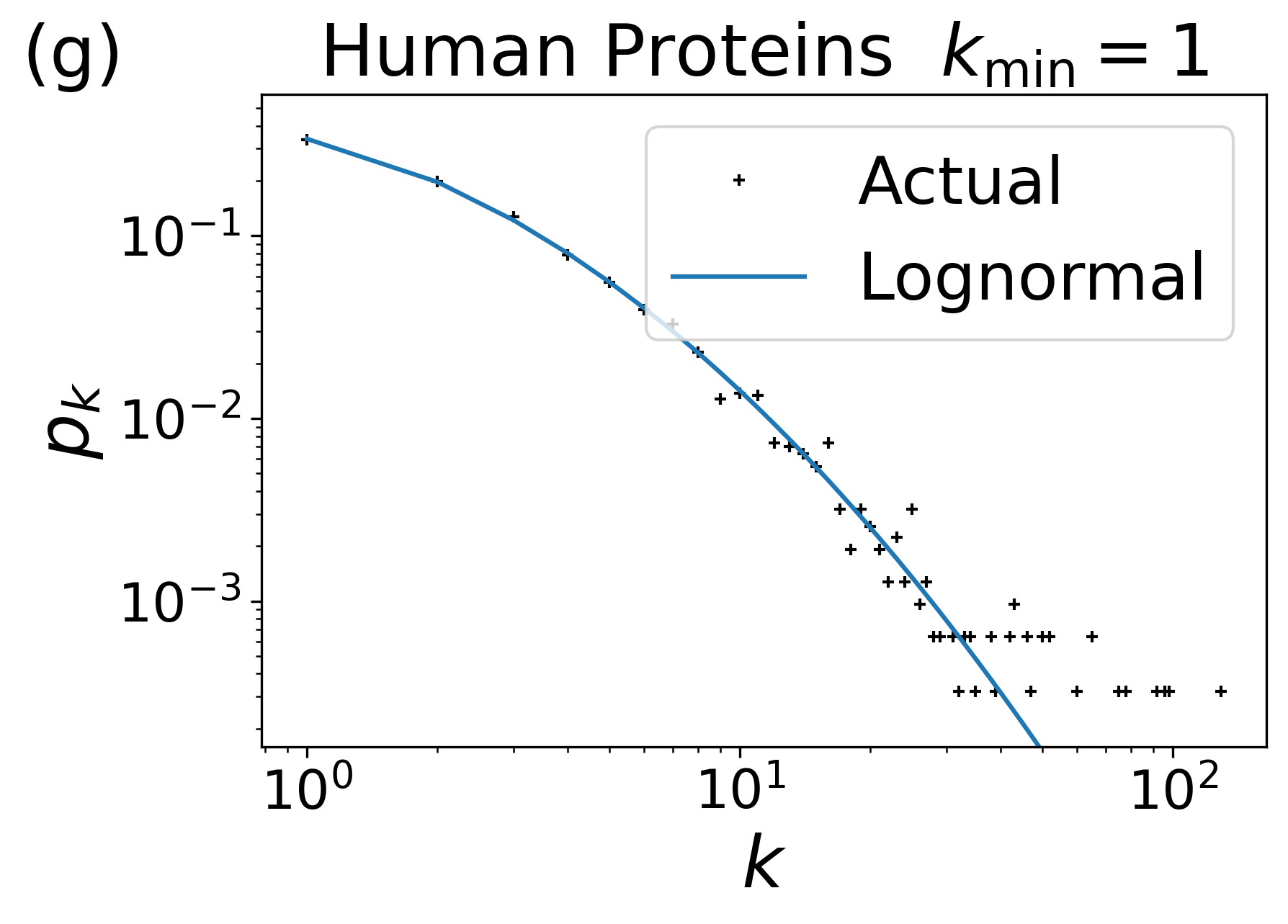}
\includegraphics[width=0.24\textwidth]{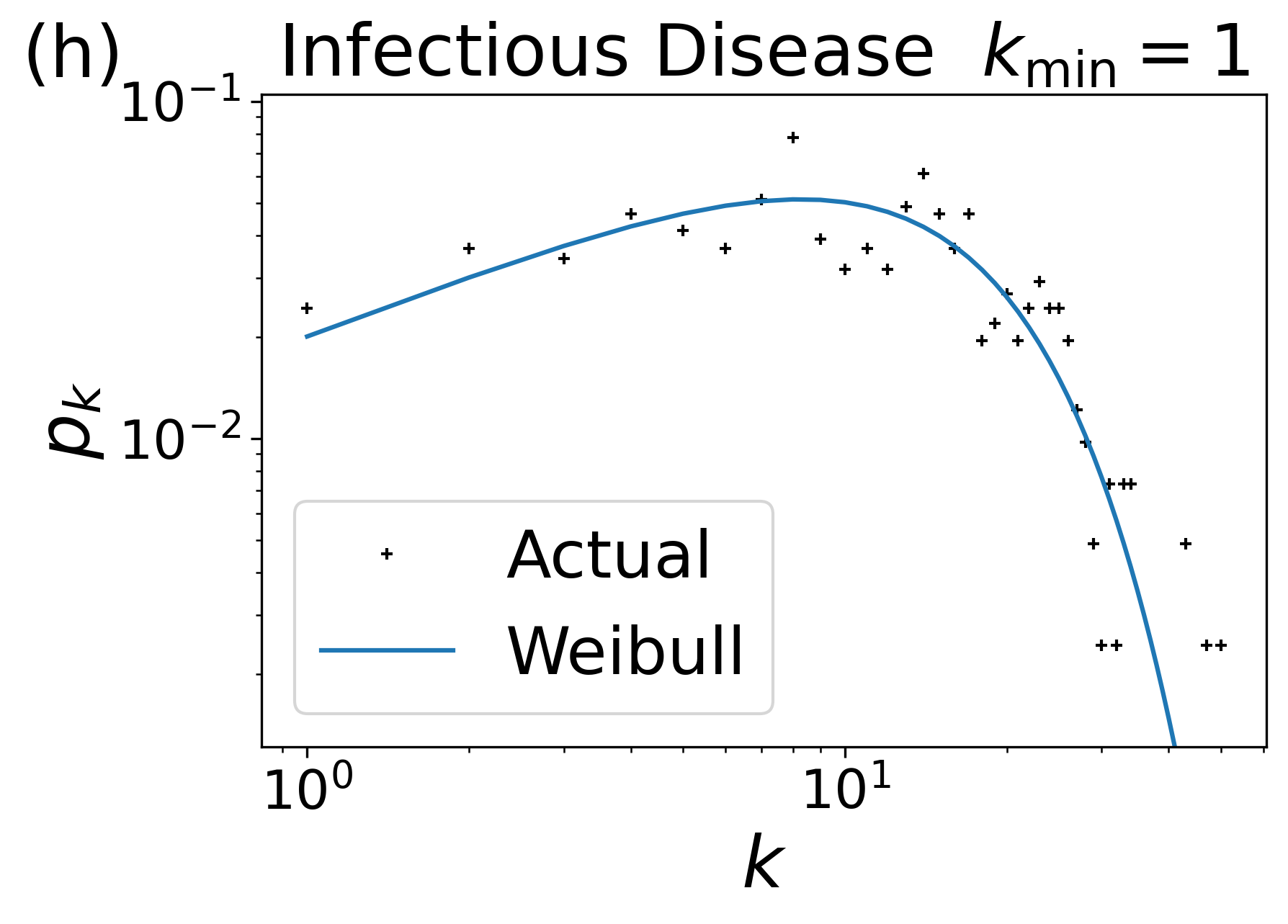}
\includegraphics[width=0.24\textwidth]{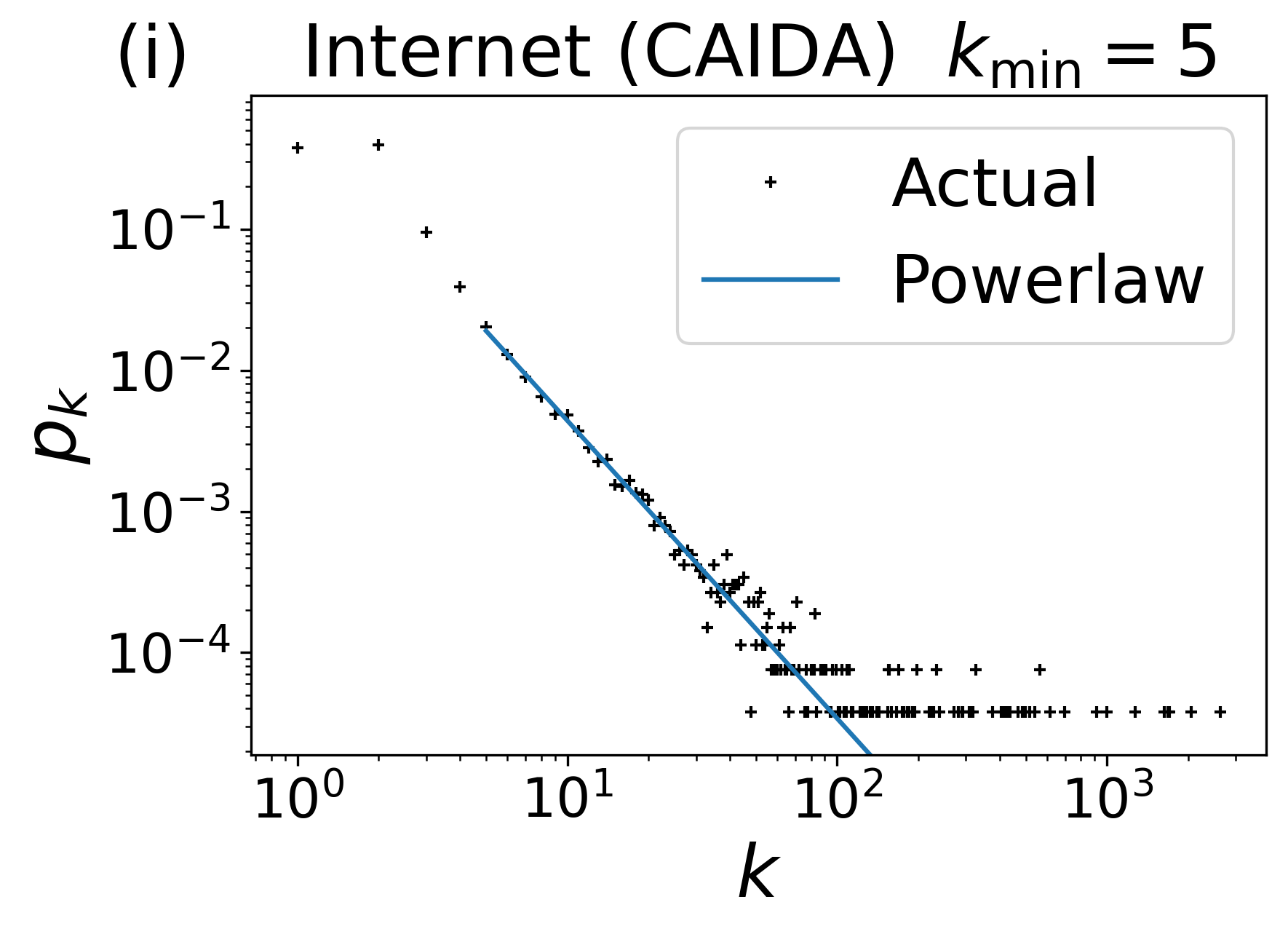}
\includegraphics[width=0.24\textwidth]{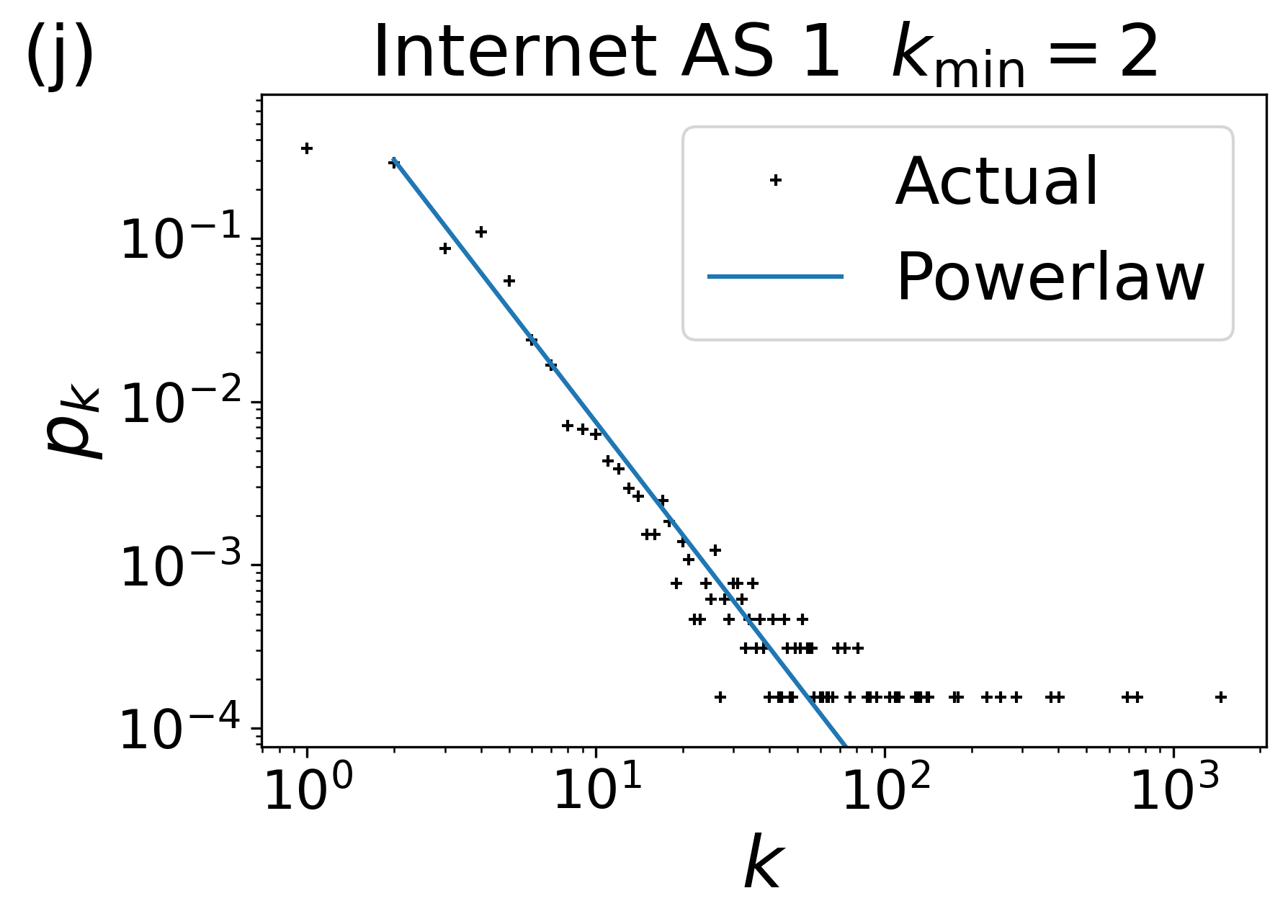}
\includegraphics[width=0.24\textwidth]{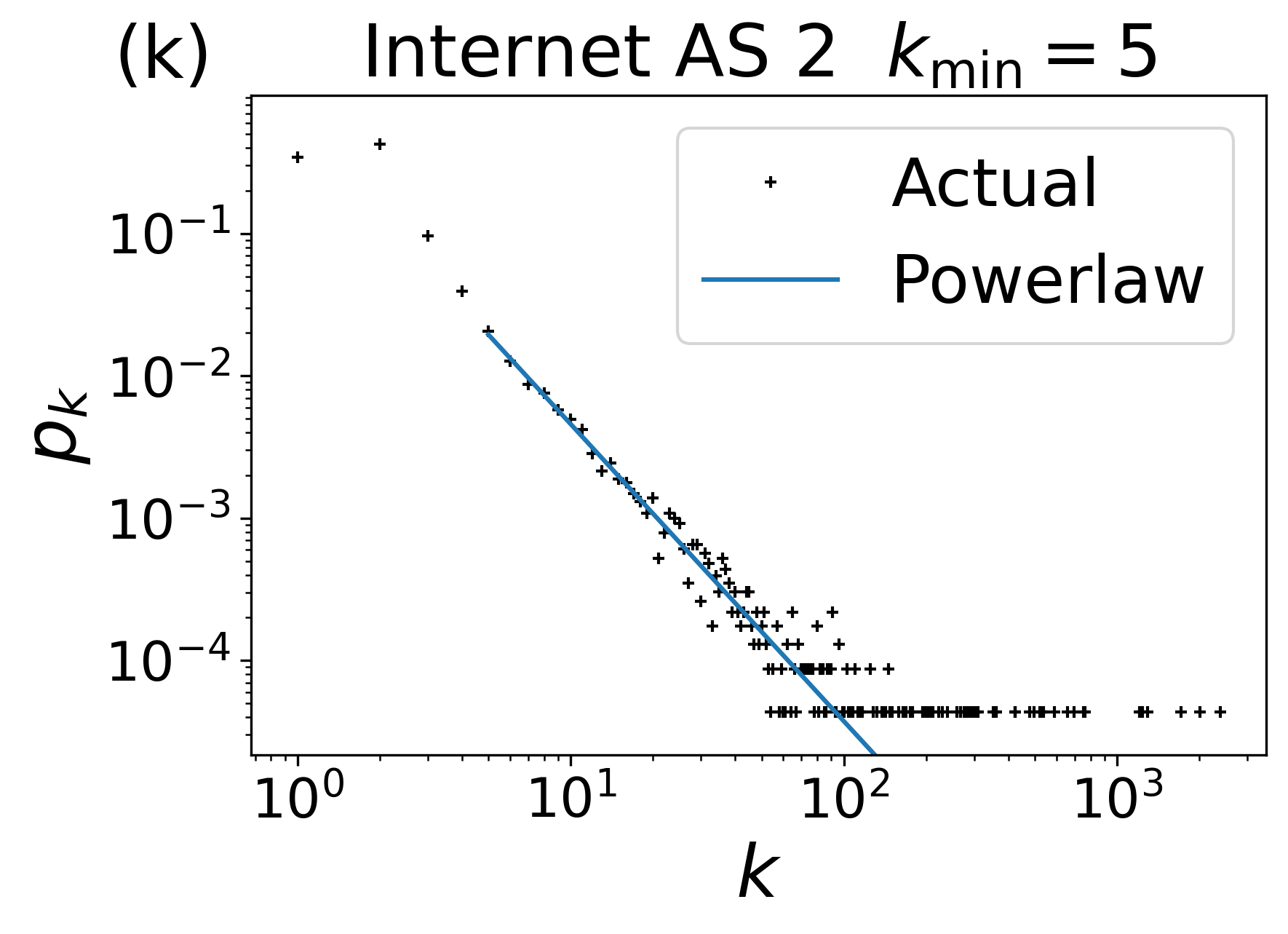}
\includegraphics[width=0.24\textwidth]{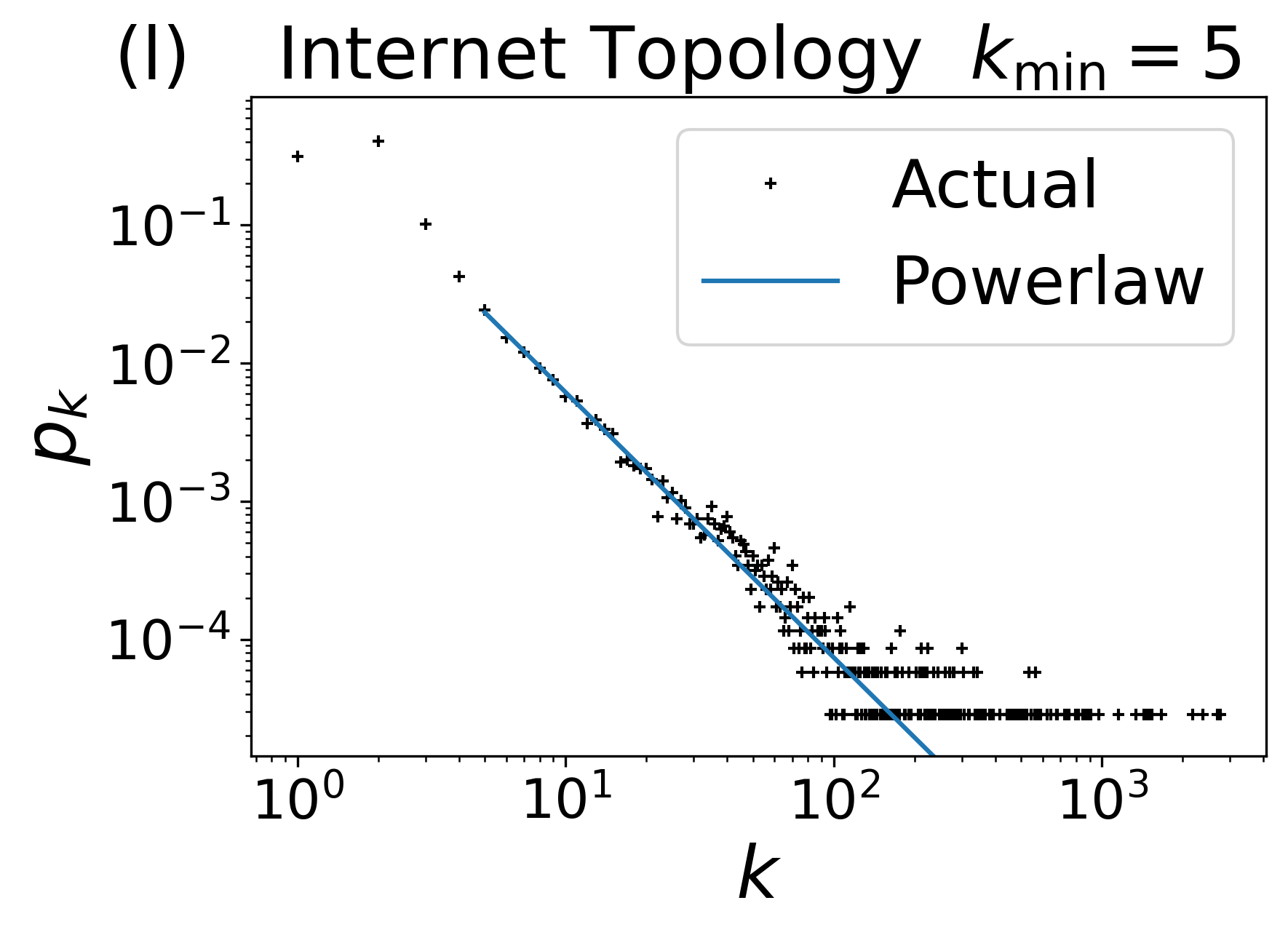}
\includegraphics[width=0.24\textwidth]{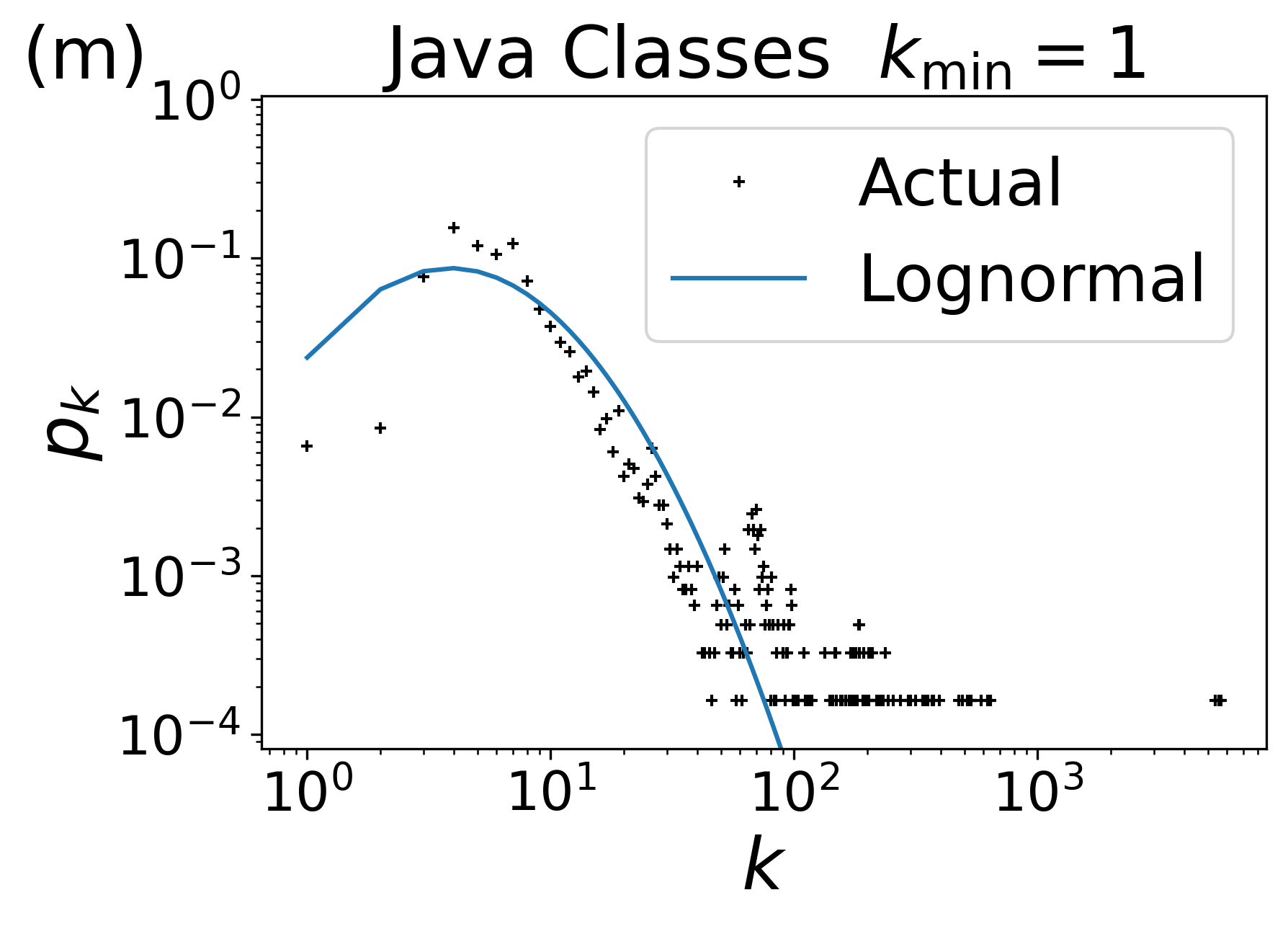}
\includegraphics[width=0.24\textwidth]{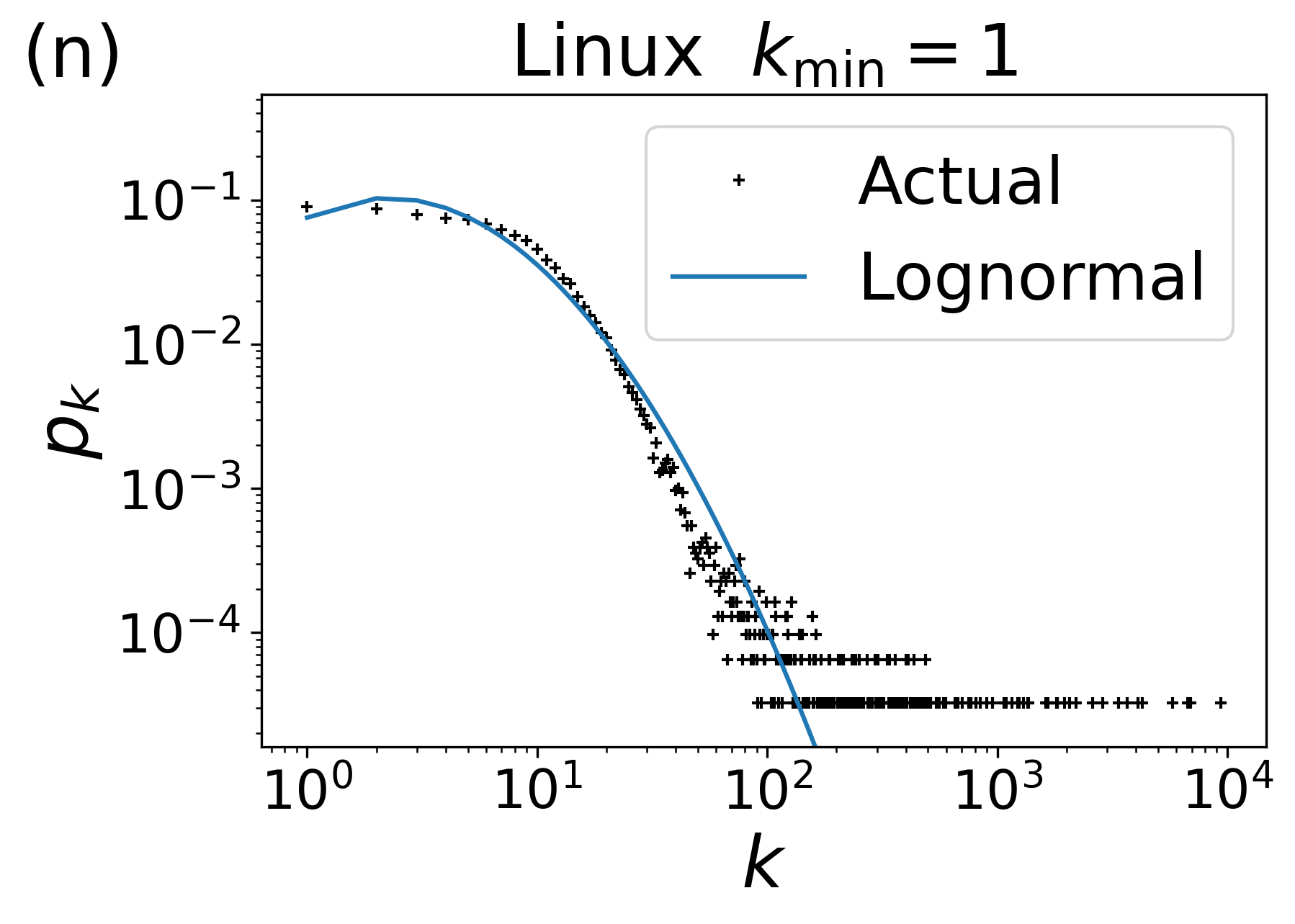}
\includegraphics[width=0.24\textwidth]{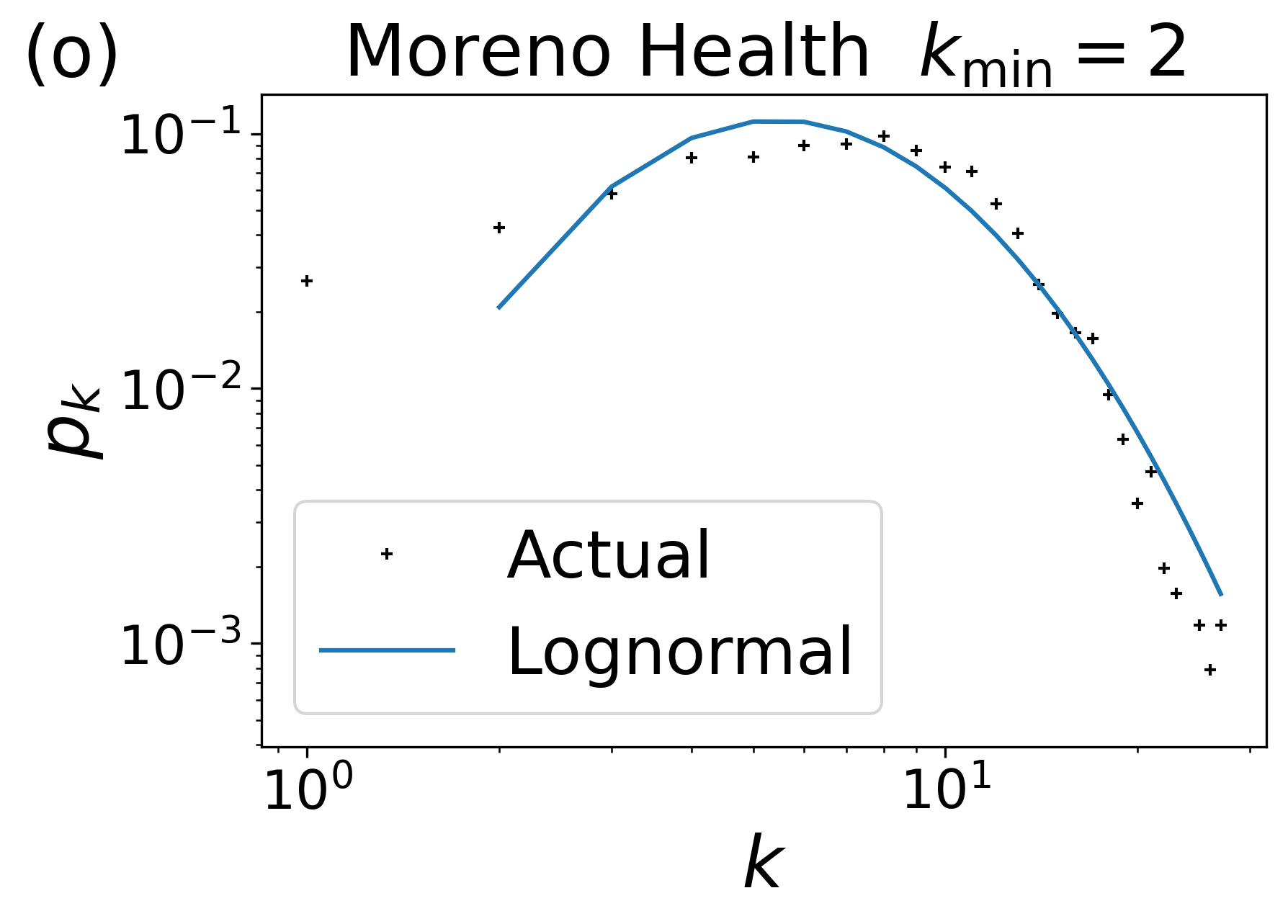}
\includegraphics[width=0.24\textwidth]{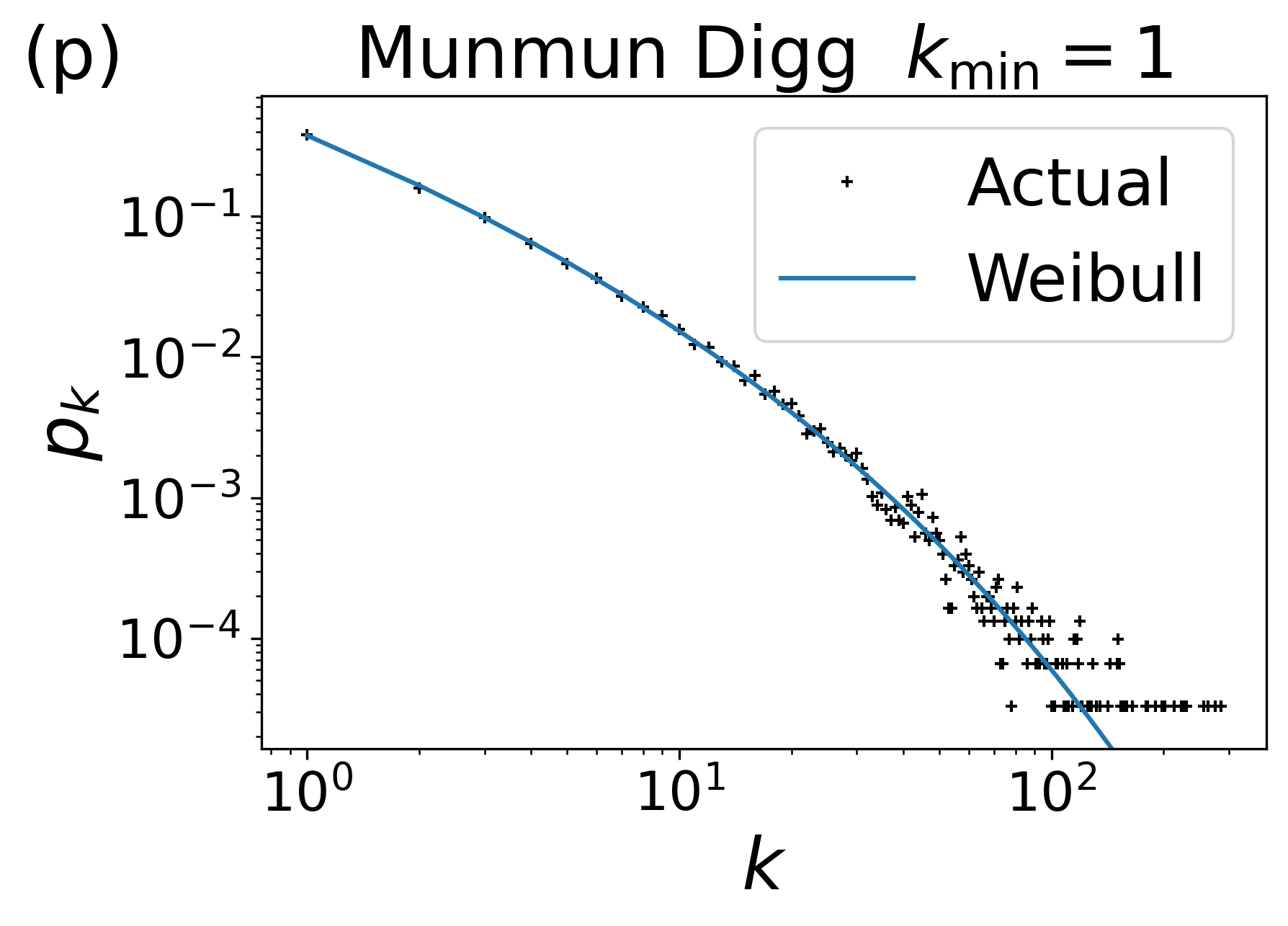}
\includegraphics[width=0.24\textwidth]{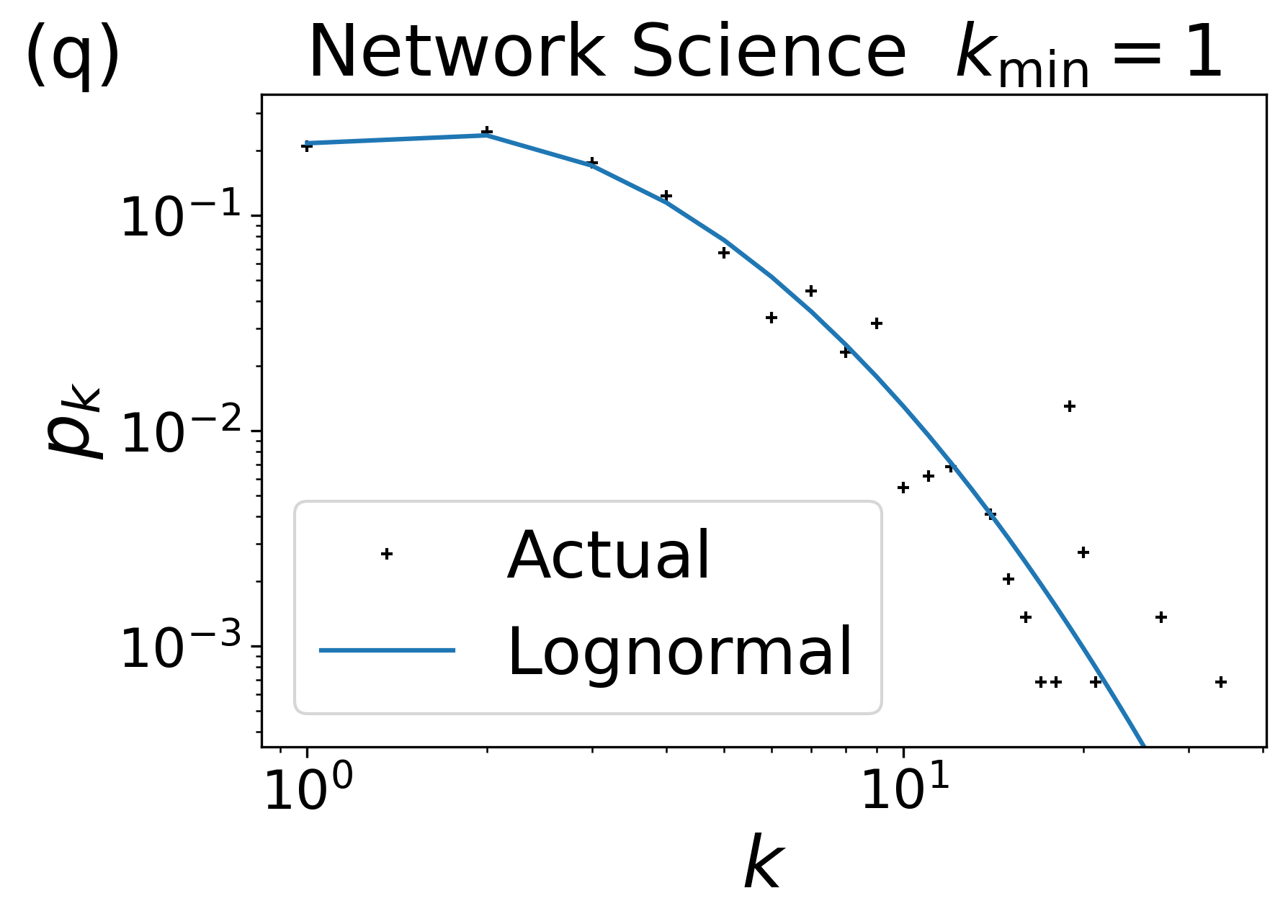}
\includegraphics[width=0.24\textwidth]{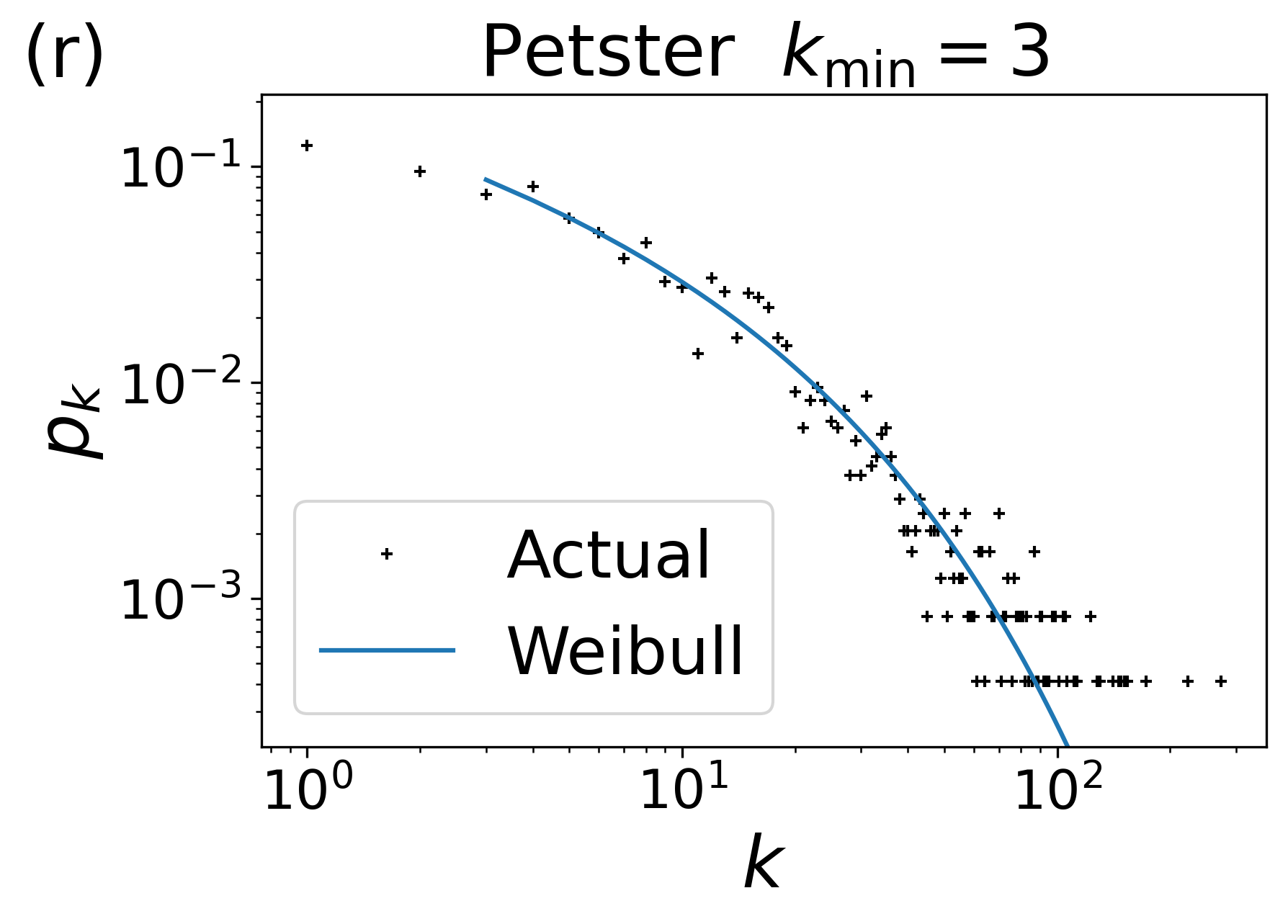}
\includegraphics[width=0.24\textwidth]{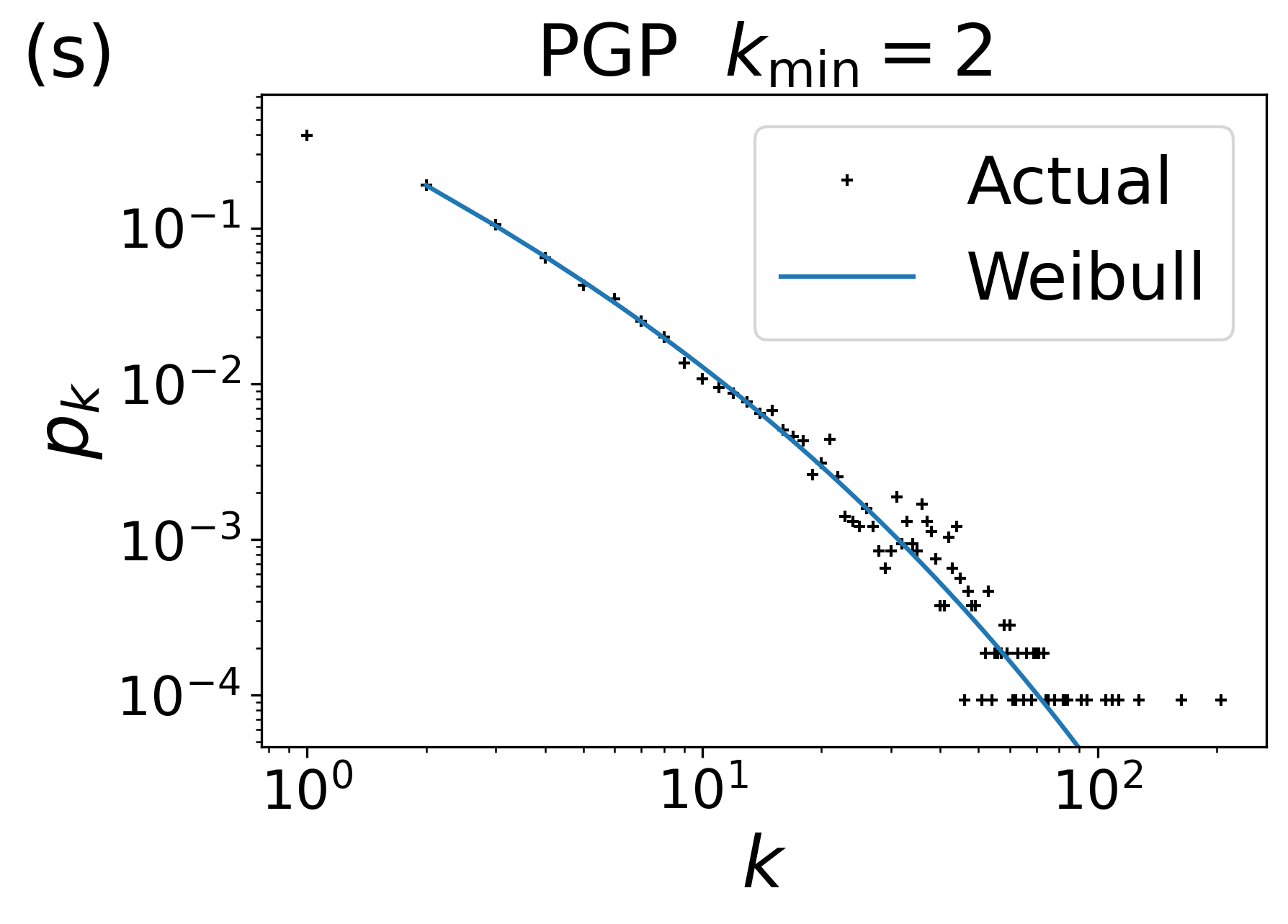}
\includegraphics[width=0.24\textwidth]{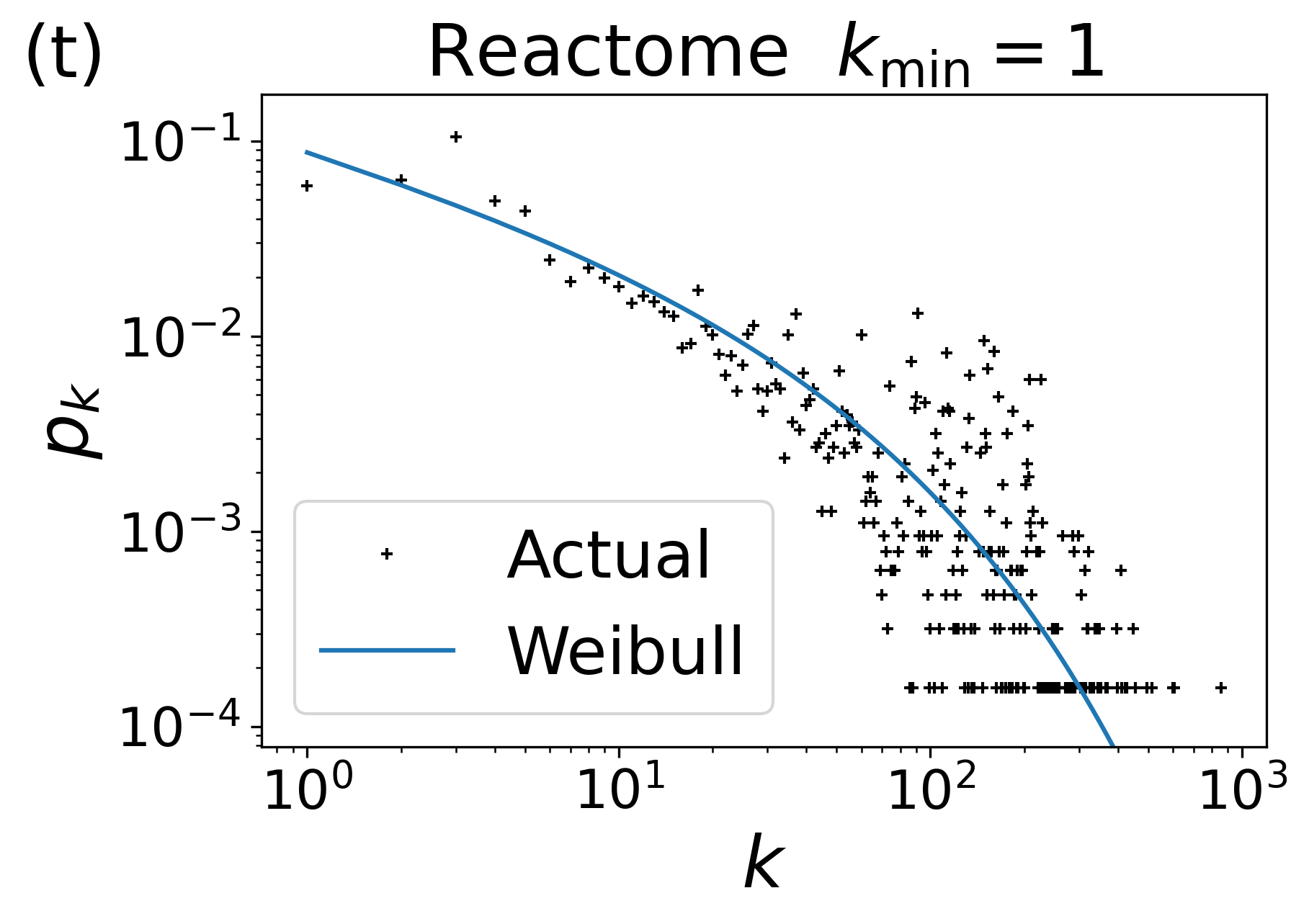}
\includegraphics[width=0.24\textwidth]{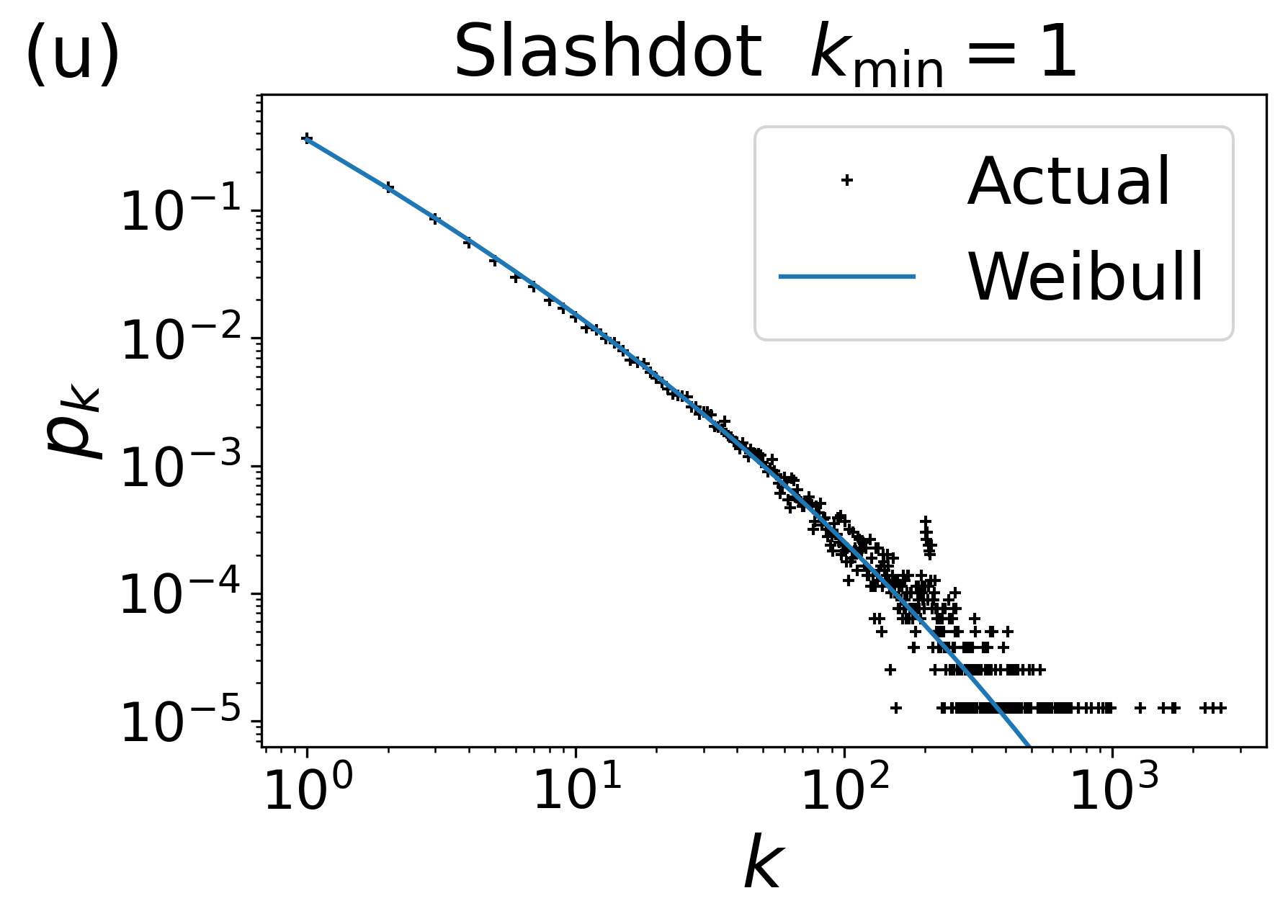}
\includegraphics[width=0.24\textwidth]{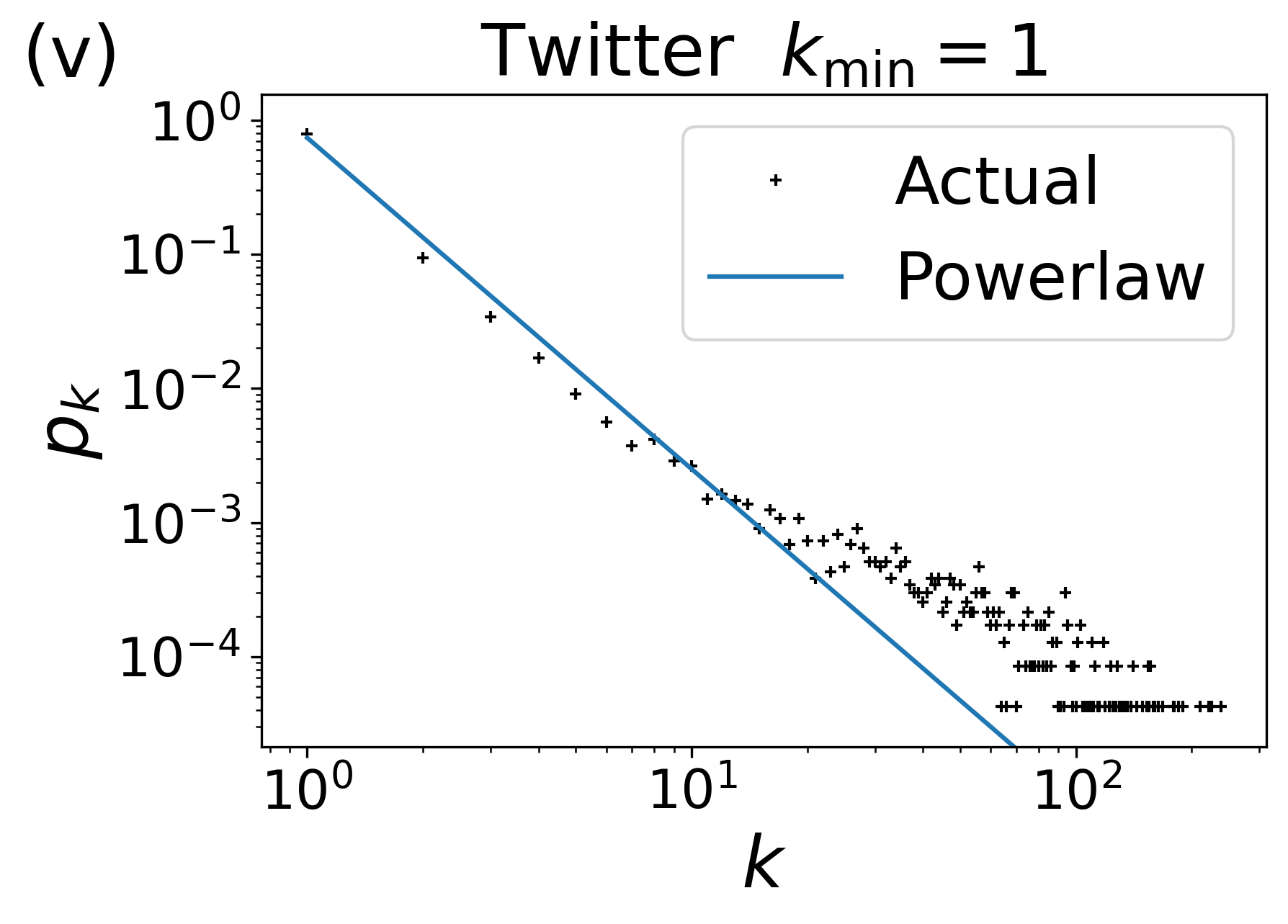}
\includegraphics[width=0.24\textwidth]{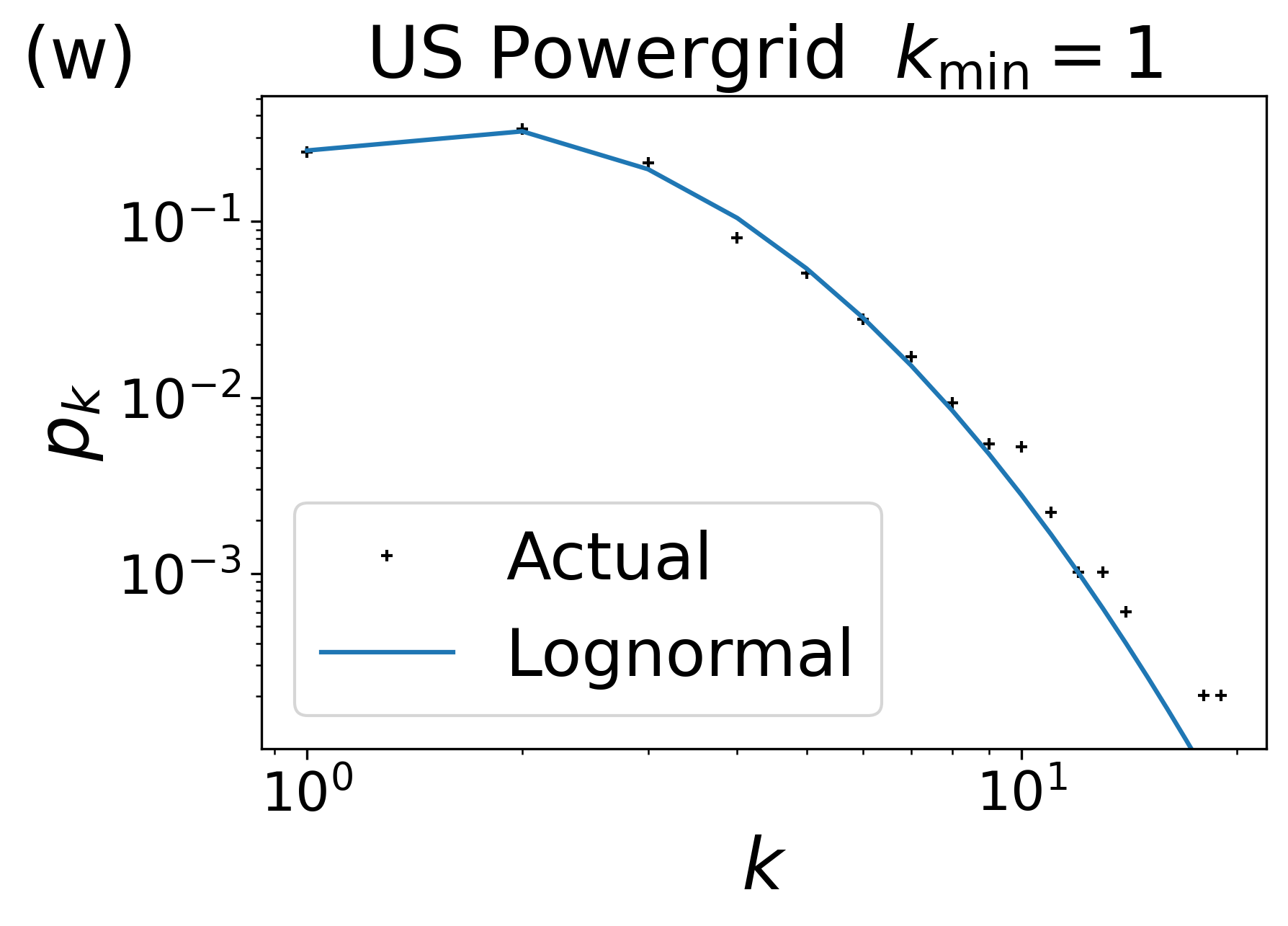}
\includegraphics[width=0.24\textwidth]{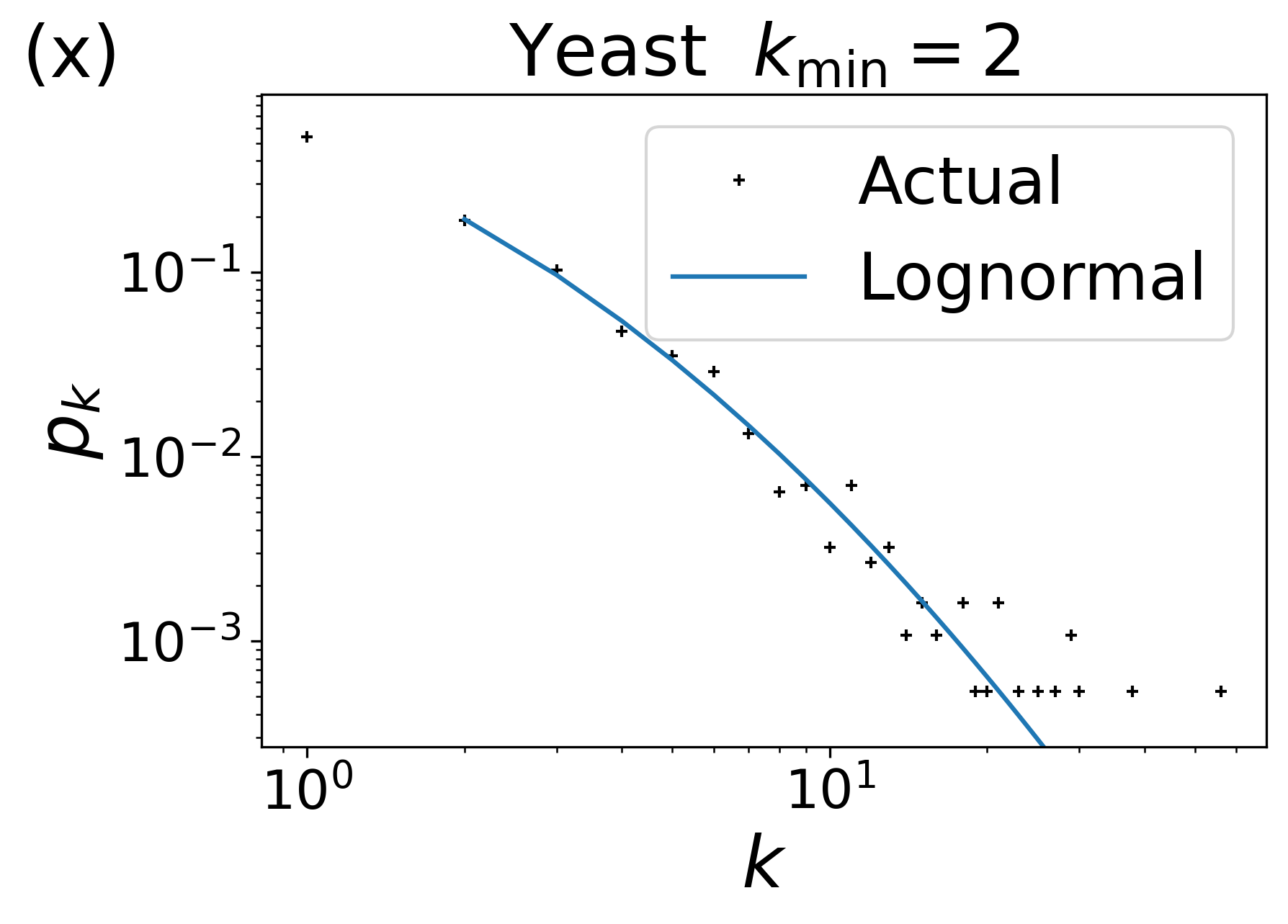}
\includegraphics[width=0.24\textwidth]{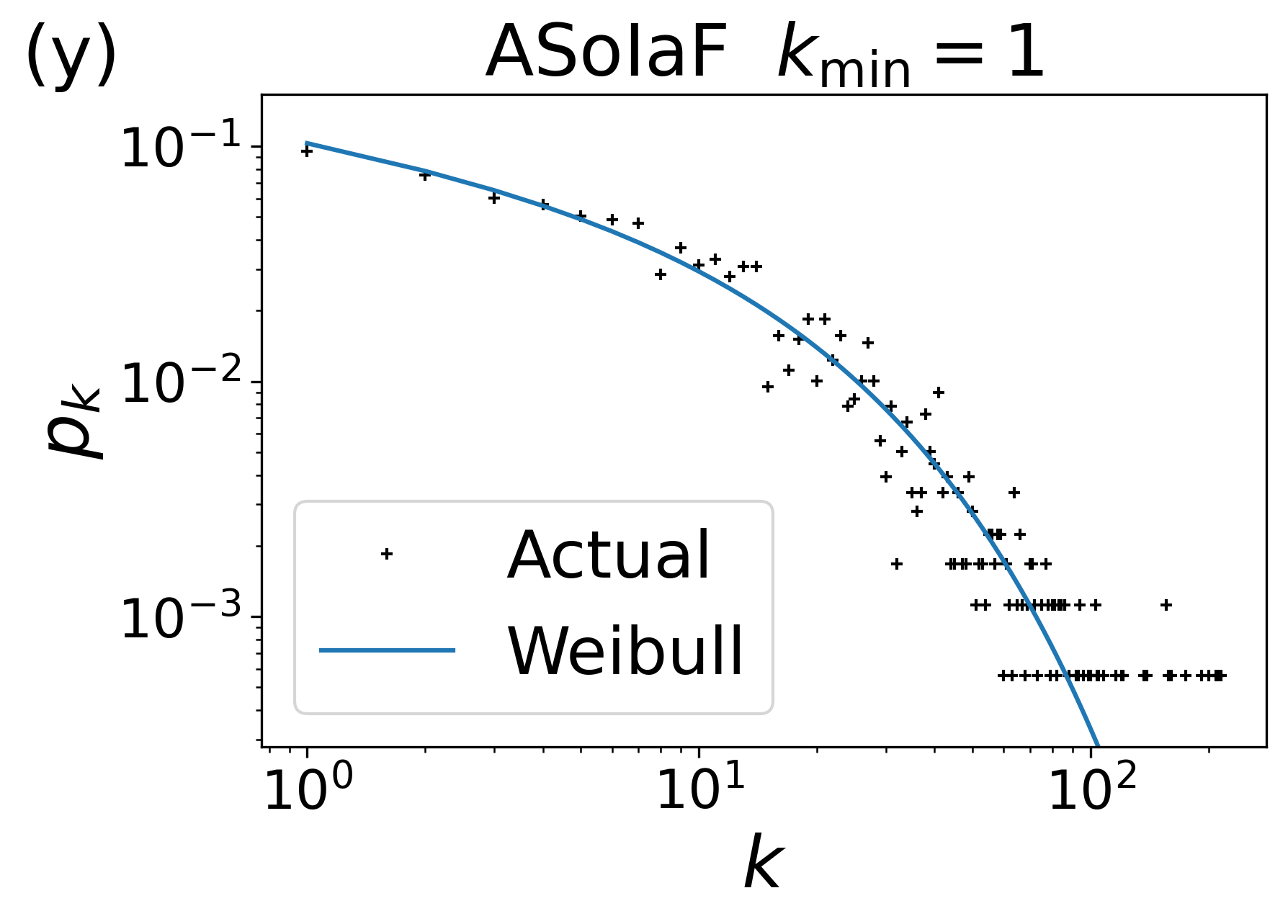}
\includegraphics[width=0.24\textwidth]{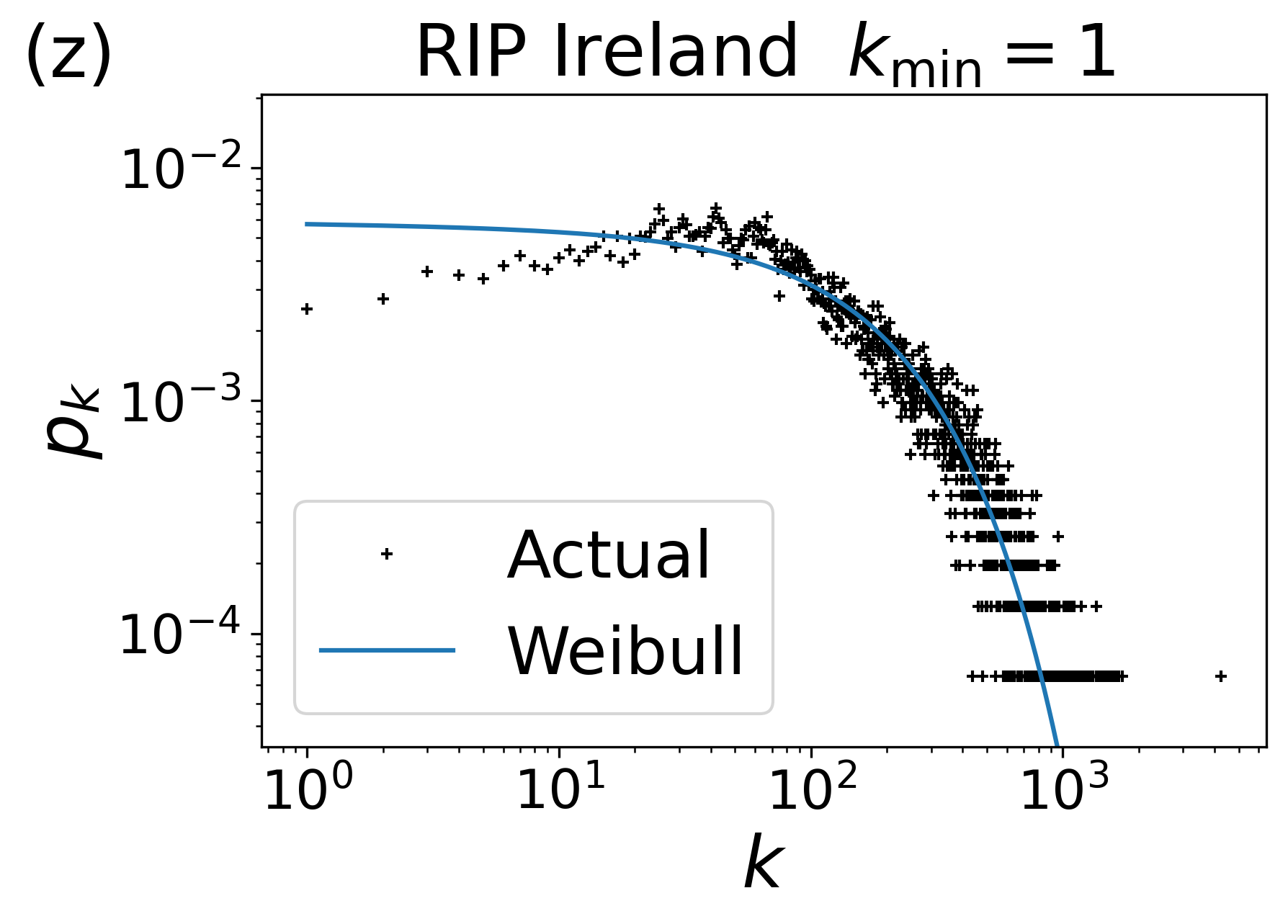}
\includegraphics[width=0.24\textwidth]{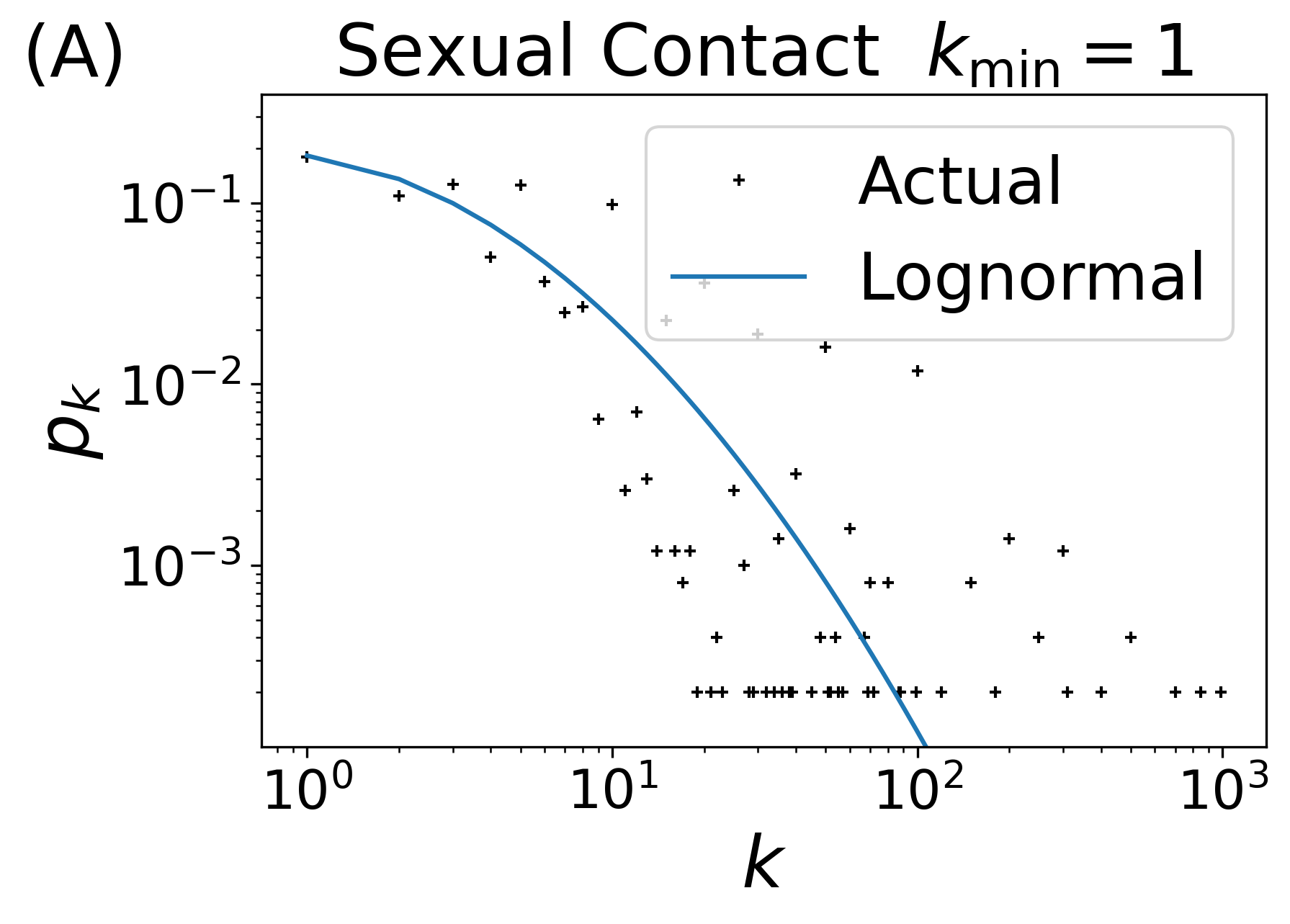}
\includegraphics[width=0.24\textwidth]{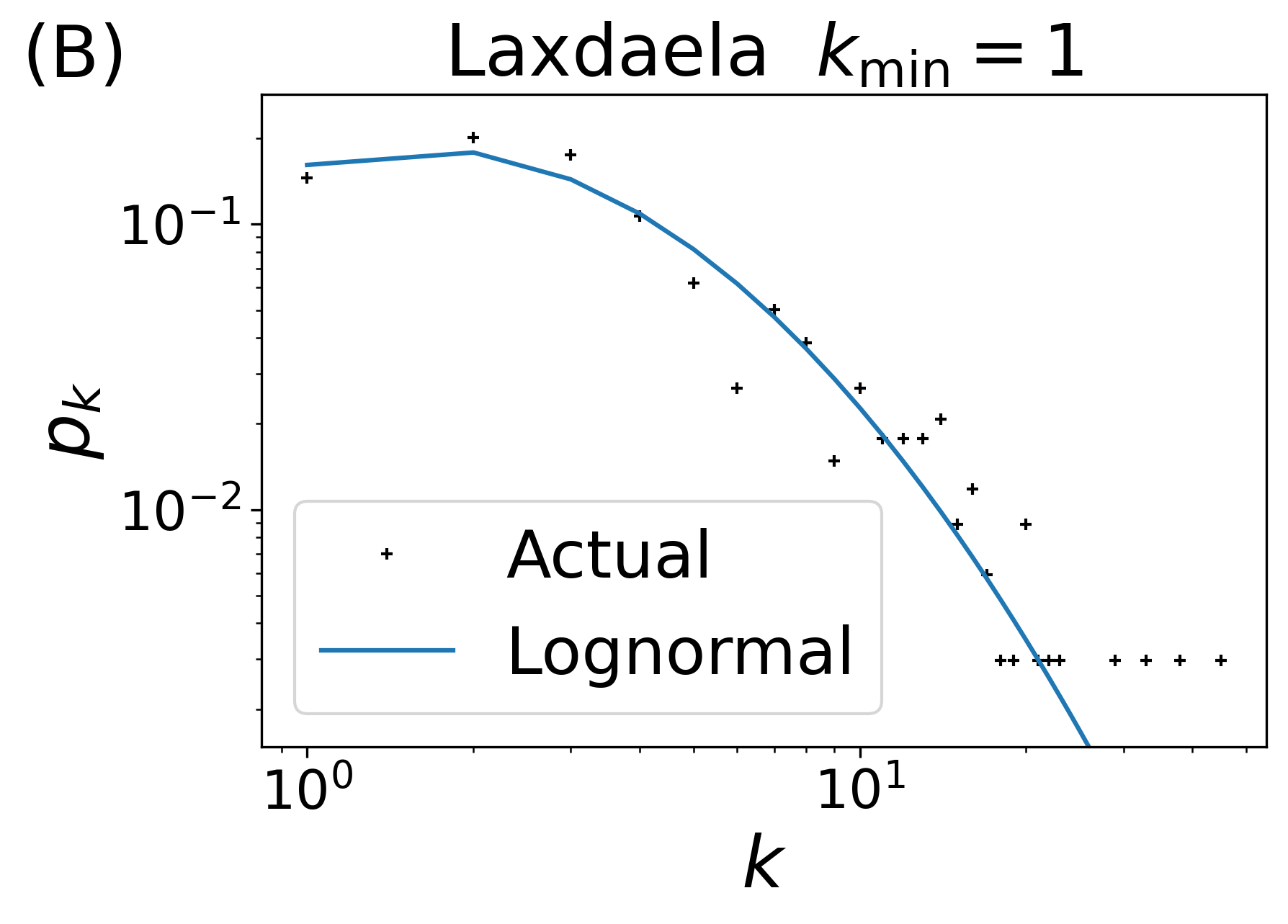}
\caption{Empirical and fitted PDFs for all studied networks listed in table~\ref{tab:networkdata}. When assessing fit both PDFs and CCDFs should be considered.}
\label{fig:all_pdfs}
\end{center}
\end{figure*}

\end{document}